\documentclass[english,british]{revtex4}
\usepackage[T1]{fontenc}
\usepackage[latin9]{inputenc}
\usepackage{babel}

\usepackage{units}
\usepackage{textcomp}
\usepackage{amsmath}
\usepackage{graphicx}
\usepackage{amssymb}
\usepackage{esint}

\makeatletter

\providecommand{\tabularnewline}{\\}

\@ifundefined{textcolor}{}
{%
 \definecolor{BLACK}{gray}{0}
 \definecolor{WHITE}{gray}{1}
 \definecolor{RED}{rgb}{1,0,0}
 \definecolor{GREEN}{rgb}{0,1,0}
 \definecolor{BLUE}{rgb}{0,0,1}
 \definecolor{CYAN}{cmyk}{1,0,0,0}
 \definecolor{MAGENTA}{cmyk}{0,1,0,0}
 \definecolor{YELLOW}{cmyk}{0,0,1,0}
 }

\makeatother

\begin{document}

\title{Theory of Attosecond Pulses from Relativistic Surface Plasmas}

\author{Daniel an der Brügge}

\email{dadb@tp1.uni-duesseldorf.de}

\author{Alexander Pukhov}

\affiliation{Institut für theoretische Physik, Heinrich-Heine-Universität Düsseldorf}
\begin{abstract}
High harmonic generation by relativistically intense laser pulses
from overdense plasma layers is surveyed. High harmonics are generated
in form of (sub-)attosecond pulses when the plasma surface rebounds
towards the observer with relativistic velocity. Different cases are
considered. The {}``relativistically oscillating mirror'' (ROM)
model, describing the most typical case, is analyzed in detail. The
resulting harmonic spectrum is usually a power law with the exponent
-8/3 \cite{baeva2006relativistic}, but possible exceptions due to
{}``higher order $\gamma$-spikes'' are considered. It is shown
that under certain conditions, ultra-dense electron nanobunches can
be formed at plasma surface that emit coherent synchrotron radiation.
The resulting spectrum is much flatter and leads to the formation
of a giant attosecond pulse in the reflected radiation. The harmonics
radiation is also considered in time domain, where they form a train
of attosecond pulses. It is characterized and a possibility to select
a single attosecond pulse via polarization gating is described. Further,
the line structure in relativistic harmonic spectra is analyzed. It
is shown that the harmonics have an intrinsic chirp and it can be
responsible for experimentally observed spectral modulations. Finally,
high harmonic generation is considered in realistic three-dimensional
geometry. It is shown that free space diffraction can act as a high
pass filter, altering the spectrum and the reflected field structure.
The high harmonics tend to be self-focused by the reflecting surface.
This leads to a natural angular divergence as well as to field boost
at the focal position. Coherently focusing the harmonics using an
optimized geometry may result in a significantly higher field than
the field of the driving laser.

\end{abstract}
\maketitle

\section{Introduction }

The tremendous progress in the femtosecond laser technology leads
us to the question, if it is possible to create even shorter pulses,
with durations in the attosecond or, may be, zeptosecond time range.

One, already well-established way to produce light pulses of attosecond
scale duration is the generation of high order harmonics by the non-linear
interaction of gas atoms with lasers of intensities close to the ionization
threshold \cite{corkum2007attosecond}. This mechanism is however
limited to not-too-high laser intensities and a relatively low efficiency.
Production of coherent attosecond pulses of higher energy could open
up the way to entirely new methods of attosecond research such as
XUV pump-probe spectroscopy \cite{ferenc2009attosecond,igor2007probing}.
The currently most promising way towards more intense attosecond pulses
is the generation of high order harmonics (HHG) at solid density plasma
surfaces.

For a complete understanding of this attosecond pulse generation scheme,
it is necessary to study three stages in the laser-plasma interaction
process:
\begin{description}
\item [{Plasma~Formation.}] Before the main laser pulse hits the solid
target surface, the pedestal of the pulse already ionizes it and turns
it into a plasma. The plasma then thermally expands and at the same
time is pushed inside by the laser ponderomotive potential. Depending
on the contrast ratio of the laser system and the exact structure
of the pre-pulse, this may yield very different surface density profiles.
These processes are well understood today and can reliably be simulated
by hydro-codes such as Multi-FS \cite{ramis1988multitextemdash}.
\item [{Harmonics~Generation.}] The second stage is the harmonics generation
itself. It happens during the interaction of the main laser pulse
with the pre-formed plasma density gradient. If the laser pulse duration
is in the order of just a few ten femtoseconds or below, the motion
of the ions during this period can be neglected, and the interaction
takes place between the laser electromagnetic fields and the plasma
electrons.
\item [{Diffraction.}] After the radiation has been emitted from the surface,
it will propagate through space. Due to the extremely broad spectrum
of the emitted radiation and its coherent phase properties, it is
well worth to take a closer look at its diffraction and focusing behaviour.
\end{description}
The paper is organized as follows. First of all, in section \ref{sec:history}
we provide a short historical overview of high harmonic generation
from overdense plasmas in relativistic regime.

In section~\ref{sec:Harmonics-Theory}, we examine the theory of
relativistic HHG. Different models are discussed and spectra are analytically
derived from the models via asymptotic analysis. This method has the
merit of yielding {}``universal'' spectra in the sense that these
are independent on details of the electron motion, but only on some
basic properties around the so-called $\gamma$-spikes. The most typical
case is the power law decay $I\propto\omega^{-8/3}$ \cite{bgp-theory2006},
runing up to a critical harmonic number $\omega_{r}/\omega_{0}\sim\gamma^{3}$,
where $\gamma$ is the relativistic $\gamma-$factor of the oscillating
plasma surface. At higher harmonic numbers, the spectrum rolls over
into an exponential decay. Although the spectrum \cite{bgp-theory2006}
holds in most cases, special and even more efficient regimes are possible.
We found that the most efficient regime of single pulse HHG is governed
by the formation of highly compressed electron nanobunches in front
of the surface and results in a slowly decaying spectral power law
with an exponent of 6/5 - instead of 8/3. The analytical results are
substantiated by numerical simulations.

In Section~\ref{sec:Relativistic-Harmonics-Radiation} we consider
properties of this novel radiation source. Particularly, we discuss
the structure of the attosecond pulses.

Section~\ref{sec:Fine-Structure-Harmonics} deals with the line structure
of the high harmonics spectra. It is demonstrated that the generated
{}``harmonic'' lines can appear severely modulated due to the unequal
spacing inside the attosecond pulse train. Thus, these spectral modulations
do not imply incoherent noise, but rather are the result of a natural
frequency chirp of the harmonics. We show that the line structure
contains information about the motion of the surface plasma on the
femtosecond timescale.

Finally, section~\ref{sec:3d_harmonics} treats the surface HHG in
a realistic 3D geometry. Here, diffraction takes on an important role
shaping the radiation spectra. With carefully designed surfaces or
laser pulses we may harness diffraction as a sort of \emph{spatial
spectral filter}. With a well designed focusing geometry, it should
even be possible to focus the harmonics coherently in both space and
time, yielding unprecedented intensities that exceed the intensity
of the laser itself by more than a thousand times.

\section{Historical overview }

\label{sec:history}

The basic idea for HHG at overdense plasma surfaces has been around
for almost thirty years now and endured several generations of high
power lasers. In this work, we focus on the most efficient, highly
relativistic regime. Before we move on to the actual study, it is
worth to have a brief look at the history of the topic.

The first observation of high harmonics from plasma surfaces was reported
from the Los Alamos Scientific Laboratory \cite{PhysRevLett.46.29,PhysRevA.24.2649}
in 1981. At that time, huge $\textrm{CO}_{2}$ lasers were used at
nanosecond pulse duration and the observed radiation was non-coherently
emitted into the whole half-space in front of the target. A theoretical
explanation for this first observation was given in Ref.~\cite{PhysRevLett.49.202}.
The spectrum extended up to a sharp cutoff, which was found to be
the plasma frequency corresponding to the maximum electron density.
Therefore, non-linear collective plasma behaviour could be identified
as the source of the harmonics. In the strongly inhomogeneous plasma,
laser light was resonantly converted to plasma oscillations, which
in turn produced harmonics by sum frequency mixing with the laser
light.

For some time then, it became silent around surface HHG, but interest
rose again, when the CPA technique, invented in 1985 by Strickland
and Mourou \cite{strickland1985compression}, revolutionized ultraintense
laser science in the 1990's. With the newly possible fs-duration,
multi-TW pulses, HHG entered an entirely new regime \cite{linde1995generation,kohlweyer1995harmonic}.
Because of the much shorter pulse duration, the plasma surface is
not destroyed by the pulse and the harmonics are cleanly emitted around
the specular direction along with the reflected fundamental \cite{tarasevitch2000generation}.

In the mid-nineties, there were first theoretical reports about a
novel HHG mechanism based on a non-linearity of purely relativistic
origin, providing a source for harmonics without the limitation of
a strict cutoff at the plasma shelf density \cite{bulanov1994interaction,lichters:3425,linde1996highorder,gibbon1996harmonic}.
The mechanism could roughly be described by a simple model, now commonly
termed the {}``relativistically oscillating mirror'' (ROM) \cite{lichters:3425,linde1996highorder}.
However, for the time being, lasers were still not strong enough to
unambiguously demonstrate the relativistic effect in distinction to
the non-relativistic plasma non-linearities.

In the first decade of the new millennium, theory of surface HHG made
further substantial advances. It was found, that for fs-laser systems
the harmonics due to the plasma non-linearity were much stronger than
could be expected from the old theory. This was attributed to so-called
Brunel electrons \cite{brunel1987notsoresonant} that trigger the
plasma oscillations instead of the evanescent laser field, leading
to {}``coherent wake emission'' (CWE) \cite{quere:125004}. The
ROM model was put on a solid basis by Baeva, Gordienko and Pukhov
(BGP) \cite{bgp-theory2006}, who managed to calculate a universal
spectral envelope ($I\propto\omega^{-8/3}$) for the model by means
of asymptotic analysis. The BGP theory takes fully into account the
surface acceleration, leading to a smooth spectral cutoff at a frequency
scaling as $\omega_{c}\propto\gamma^{3}$, comparable to synchrotron
radiation, and not just proportional to $\gamma^{2}$ like the Doppler
frequency upshift at a constantly moving mirror. Around the same time,
experiments were first able to unambiguously demonstrate the relativistic
mechanism and confirmed the spectrum obtained in the refined ROM model,
see Ref.~\cite{Dromey2006}.

In 2010, an der Brügge and Pukhov \cite{bru?gge2010enhanced} discovered
another mechanism based on the relativistic non-linearity. They found
out, that for certain combinations of parameters, extremely dense
and narrow electron bunches may form at the surface. In this exciting
regime, not even the basic boundary condition of the ROM model is
valid and the frequency upconversion process can be much more efficient
than predicted by the model. The radiation is then described as coherent
synchrotron emission (CSE) from the electron {}``nanobunches''.

\section{\label{sec:Harmonics-Theory}Generation Processes and Models}

We discuss the theory of HHG at surface plasmas, with a focus on the
highly relativistic regime $a_{0}\gg1$.

In subsection~\ref{sub:harm_theo-starting_point}, we start by summarizing
the theoretical framework all models of the interaction are based
on. Once having the equations written down, it is straightforward
to derive some selection rules for the parity and the polarization
of the generated harmonics.

In subsection~\ref{sub:harm_theo-ROM}, the ROM model is discussed.
We are going to see that this reputedly well-known model still bears
some surprises, with respect to both its foundation and the spectra
that can be derived from it.

After this, we consider a variation of the ROM boundary condition
that was suggested in Ref.~\cite{tarasevitch2009towards} in conjunction
with a two-pulse-scheme (subsection~\ref{sub:harm_theo-TROM}). We
show, that this condition represents a {}``totally reflecting oscillating
mirror'' (TROM) with negligible skin depth and rigorously derive
a spectral envelope from the model via asymptotic analysis.

Especially for p-polarized oblique incidence, the formation of highly
dense and narrow electron nanobunches in front of the surface is often
observed. If these bunches carry a considerable amount of charge,
they emit intense high frequency radiation that is not described within
the ROM model. In this case, we can derive the spectrum by calculating
the coherent synchrotron emission (CSE) from these bunches - as presented
in subsection~\ref{sub:harm_theo-nanobunching}.

\subsection{\label{sub:harm_theo-starting_point}Starting point of analysis}

The foundation of the theory of surface HHG is explained here.

In Sub.~\ref{sub:harmonics-Basic-Equations}, we demonstrate the
Green function solution of the inhomogeneous wave equation. This solution
provides a general starting point of HHG theory. Then (Sub.~\ref{sub:harmonics-Selection-Rules}),
we have a closer look at the source term to derive general selection
rules concerning the parity and polarization of the generated harmonics.
In Sub.~\ref{sub:harm_theo-sub-relativistic}, we briefly deal with
surface HHG in the sub-relativistic regime. Sub.~\ref{sub:S-similarity}
explains the meaning of the ultra-relativistic $S$-similarity group
for HHG.

\subsubsection{\label{sub:harmonics-Basic-Equations}Solution of the inhomogeneous
wave equation}

\global\long\def\jperp{\mathbf{j}_{\perp}}
Let us begin with the classical wave equation for the electromagnetic
potential. Since the basic structure of the physical mechanism can
best be understood in a one dimensional slab geometry, we let $\mathbf{A}$
depend on only one spatial coordinate $x$ and the time $t$. By this
we do not generally exclude oblique incidence, because it can be treated
in a Lorentz boosted frame wherein the laser is normally incident
(see App.~\ref{sec:Bourdier}). In this geometry, the wave equation
in Coulomb gauge ($\nabla\cdot\mathbf{A}=0$) can be written as: \begin{equation}
\frac{1}{c^{2}}\partial_{t}^{2}\mathbf{A}(t,x)-\partial_{x}^{2}\mathbf{A}(t,x)=\frac{4\pi}{c}\mathbf{j}_{\perp}(t,x).\label{eq:inhom_wave_eq}\end{equation}

Equation~\eqref{eq:inhom_wave_eq} can generally be solved with the
help of a Green function. We formally write down the solution as\begin{equation}
\mathbf{A}(t,x)=4\pi\iint\,\mathbf{j}_{\perp}(t',x')\, G(t,x,t',x')\, dt'dx'.\label{eq:inhom_wave_sol}\end{equation}

By the choice of $G$, the asymptotic behaviour of $\mathbf{A}$ can
be controlled. We choose the Green function $G$ in order to solve
Eq.~\eqref{eq:inhom_wave_eq} and additionally fulfil the boundary
condition $|\mathbf{A}(t,x)|\rightarrow0$ for $x\rightarrow+\infty$,
i.e. there is no light coming from the right and all radiation coming
from the left is fully reflected. We obtain:\begin{equation}
G(t,x,t',x')=\frac{1}{2}\left[\theta\left(t-t'-\frac{|x-x'|}{c}\right)-\theta\left(t-t'-\frac{x-x'}{c}\right)\right],\label{eq:green_total_ref}\end{equation}
where $\theta$ denotes the Heaviside step function.

It proves convenient to continue working with the transverse electric
field $\mathbf{E}_{\perp}=-c^{-1}\partial_{t}\mathbf{A}$ instead
of the vector potential here. Thus Eq.~\eqref{eq:inhom_wave_sol}
becomes:\begin{equation}
\mathbf{E}_{\perp}(t,x)=\frac{2\pi}{c}\int_{x}^{\infty}\left[\mathbf{j}_{\perp}\left(t-\frac{x-x'}{c},\, x'\right)-\mathbf{j}_{\perp}\left(t+\frac{x-x'}{c},\, x'\right)\right]\, dx'.\label{eq:inhom_wave_sol_E}\end{equation}

Let us define $x_{v}\equiv\sup\left\{ x:j(t,x')=0,\,\forall t,\,\forall x'<x\right\} $:
the leftmost point which is reached by any charge during the laser-plasma
interaction process. It can be seen that for all $x<x_{v}$ to the
left of the plasma, the first term in Eq.~\eqref{eq:inhom_wave_sol_E}
represents the incoming radiation, while the second term represents
the reflected one. To the right of the plasma both terms cancel, as
our choice of the Green function requested.

Due to the assumption of one-dimensionality, the radiation does not
change while propagating in vacuum, and the incoming and outgoing
fields $\mathbf{E}_{i}$ and $\mathbf{E}_{r}$ are each function of
only one variable $t\pm x/c$. We may therefore drop the argument
$x$ and identify\begin{eqnarray}
\mathbf{E}_{i}(t) & \equiv & \frac{2\pi}{c}\int_{-\infty}^{+\infty}\mathbf{j}_{\perp}\left(t+\frac{x'}{c},\, x'\right)\, dx'\label{eq:E_incoming}\\
\mathbf{E}_{r}(t) & \equiv & -\frac{2\pi}{c}\int_{-\infty}^{+\infty}\mathbf{j}_{\perp}\left(t-\frac{x'}{c},\, x'\right)\, dx',\label{eq:E_reflected}\end{eqnarray}
so that $\mathbf{E}_{\perp}(t,\, x)=\mathbf{E}_{i}(t-x/c)+\mathbf{E}_{r}(t+x/c)$
for $x<x_{v}$.

Eqs.~\eqref{eq:E_incoming} and \eqref{eq:E_reflected} provide a
powerful and general starting point for the theory of harmonics generation.
Whereas Eq.~\eqref{eq:E_reflected} tells us how to obtain the reflected
field $\mathbf{E}_{r}$ from a given current distribution, Eq.~\eqref{eq:E_incoming}
provides a condition on the current for a given incident field $\mathbf{E}_{i}$.
Note that this equation stems from our choice of the Green function
and physically represents the condition of total reflection. If instead
we had chosen the Green function in a way that all fields vanish for
$t\rightarrow-\infty$, then there would be no $\mathbf{E}_{i}$,
but the generated field to the left of the plasma would be the same
as $\mathbf{E}_{r}$ in Eq.~\eqref{eq:E_reflected}. Equation~\eqref{eq:E_incoming}
is of course not sufficient to explicitly calculate $\mathbf{j}_{\perp}$,
but it can be harnessed to obtain $\mathbf{j}_{\perp}$ and consequently
$\mathbf{E}_{r}$ in conjunction with some additional assumption.
This is a possible approach to derive boundary conditions for the
ROM and TROM models (subsections~\ref{sub:harm_theo-ROM} and \ref{sub:harm_theo-TROM}).

\subsubsection{\label{sub:harmonics-Selection-Rules}Selection Rules}

Before we move on to present these models, let us collect some general
facts about the source term $\mathbf{j}_{\perp}$, stemming from the
plasma response to the laser. Therefore, we consider the fluid equations
for a cold relativistic plasma. These equations do not account for
kinetic effects like trajectory crossing, but they are adequate to
derive some general properties of the physical process. For a relatively
short laser pulse, we can neglect the ion response, so the current
is given by:\begin{equation}
\mathbf{j}=-e\left(n\mathbf{v}-n_{0}\mathbf{v}_{0}\right),\end{equation}
where $e$ is the elementary charge, $n$ is the electron density
and $\mathbf{v}$ is the electron fluid velocity. Note that we consider
all magnitudes in the inertial frame in which the laser is normally
incident. In this frame, the electrons and ions possess some initial
velocity $\mathbf{v}_{0}$ parallel to the surface, and the initial
density $n_{0}$ is not necessarily identical to the initial density
in the laboratory frame. The velocity $\mathbf{v}$ is related to
the relativistic momentum $\mathbf{p}$ like $\mathbf{v}=\mathbf{p}/\gamma m_{e}$,
where $\gamma\equiv\sqrt{1+(\mathbf{p}/m_{e}c)^{2}}$. Due to the
conservation of the canonical momentum\cite{kruer}, the transverse
component can directly be connected to $\mathbf{A}$ in the presumed
1D geometry:\begin{equation}
\mathbf{p}_{\perp}=\mathbf{p}_{0}+e\mathbf{A}.\label{eq:canonical_mom}\end{equation}

The set of equations is completed by the equation of motion for the
longitudinal momentum component $p_{x}$, the continuity equation
and the Poisson equation for the electrostatic potential due to charge
separation:\begin{eqnarray}
\frac{dp_{x}}{dt} & = & e\left(\partial_{x}\Phi-\frac{\mathbf{v}_{\perp}}{c}\cdot\mathbf{A}\right),\\
\partial_{t}n & = & -\partial_{x}\left(nv_{x}\right),\label{eq:continuity}\\
\partial_{x}^{2}\Phi & = & 4\pi e\left(n-n_{0}\right),\end{eqnarray}
wherein $d/dt=\partial_{t}+v_{x}\partial_{x}$ denotes the absolute
time derivative.

Having a closer look at these equations, it is possible to derive
some {}``selection rules'' with respect to parity (even or odd harmonic
numbers) and polarization (see also Ref.~\cite{lichters:3425}).
\begin{enumerate}
\item Assuming normal incidence of linearly polarized light, we take $\mathbf{p}_{0}=0$,
$\mathbf{A}=A\mathbf{e}_{y}$. It is obvious then, that $j_{z}=0$,
and the polarization of the incident light is conserved. For the $y$-component
of the source term, we obtain $j_{y}\propto nA/\gamma$. About the
longitudinal momentum $p_{x}$, which enters both $n$ {[}through
$v_{x}$ in Eq.~\eqref{eq:continuity}{]} and $\gamma$, we know
that $\dot{p}_{x}=e\left(\partial_{x}\Phi+eA^{2}/\gamma\right)$.
Thus, the longitudinal momentum is driven by the square of the laser
field $A^{2}$ and therefore has the same periodicity as $A^{2}$.
Consequently, it possesses only even harmonics of the laser frequency.
The same holds true for $n$ and $\gamma$, as can be seen from Eq.~\eqref{eq:continuity}
and $\gamma=\sqrt{1+eA^{2}+p_{x}^{2}}$. Finally, $j_{y}$ is a product
of $A\sim\cos\omega_{0}t$ (in zeroth order) with quantities that
possess only even harmonics of the fundamental laser frequency. We
conclude, that $j_{y}$ and therefore $A$ purely consist of odd harmonics
of the fundamental.
\item For s-polarized oblique incidence, we may assume $\mathbf{p}_{0}=p_{0}\mathbf{e}_{z}$
and initially $\mathbf{A}=A\mathbf{e}_{y}$. In this case, $j_{y}\propto nA/\gamma$
as in the normal incidence case, but additionally there is a source
term in $z$-direction: $j_{z}\propto np_{0}/\gamma-n_{0}p_{0}/\gamma_{0}$.
Again, $p_{x}$, $n$ and $\gamma$ contain only even harmonics of
the laser frequency. Consequently, $j_{y}$ and $A_{y}$ contain only
odd harmonics and $j_{z}$ and $A_{z}$ contain only even harmonics
of the fundamental.
\item For p-polarized oblique incidence, we can take $\mathbf{p}_{0}=p_{0}\mathbf{e}_{y}$
and $\mathbf{A}=A\mathbf{e}_{y}$. We immediately see, that there
is no source term in $z$-direction ($j_{z}=0$), and $j_{y}\propto p_{0}\left(n/\gamma-n_{0}/\gamma_{0}\right)+enA/\gamma$
obviously contains both even and odd harmonics. Another interesting
fact is that $p_{x}$ is now also driven by a term that is linear
in $A$. This implies, that harmonics can be observed here at lower
intensities compared to s-polarized and normal incidence.
\end{enumerate}
\begin{table}
\centering{}\begin{tabular}{|c|c|c|}
\hline
incident light & odd harmonics & even harmonics\tabularnewline
\hline
\hline
normal (linear) & same as incident & -\tabularnewline
\hline
oblique (s) & s & p\tabularnewline
\hline
oblique (p) & p & p\tabularnewline
\hline
\end{tabular}\caption{\label{tab:Selection-rules}Selection rules for polarization (s, p)
and parity (even, odd) of harmonics at plasma surfaces depending on
the polarization and the angle of the incident laser.}
\end{table}

In table~\ref{tab:Selection-rules}, the rules just derived are summarized
for reference.

\subsubsection{\label{sub:harm_theo-sub-relativistic}Sub-relativistic plasma non-linearity}

This work deals with generation of harmonics due to relativistic mechanisms.
These have to be distinguished from harmonics generated by sub-relativistic
plasma non-linearity. Here, we explain the sub-relativistic mechanism
in brief.

It is found that for p-polarized oblique laser incidence, the threshold
for harmonics generation is much lower than for s-polarized or normal
incidence. This is due to plasma non-linearities, which are not of
relativistic origin and only occur for p-polarized incidence. Under
this condition, two effects may lead to the excitation of plasma oscillations
inside the inhomogeneous plasma-gradient:
\begin{enumerate}
\item Resonant absorption of the laser field, see e.g. the book by Kruer
\cite{kruer}.
\item Electron bunches that are separated from the main plasma and then
re-enter, see the famous work by Brunel \cite{brunel1987notsoresonant}.
\end{enumerate}
Due to the strong inhomogeneity of the plasma, these oscillations
couple back to electromagnetic modes via sum frequency generation,
leading to the emission of high harmonics. When the excitation happens
by means of Brunel electrons, the mechanism is commonly referred to
as {}``coherent wake emission'' (CWE) \cite{quere:125004,thaury2010highorder}.
CWE is the prevalent sub-relativistic generation process for femtosecond-scale
laser pulses.

According to their generation mechanism, the sub-relativistic harmonics
have a strict frequency limit, given by the plasma frequency $\omega_{p}$
corresponding to the maximum density \cite{PhysRevLett.46.29,quere:125004,PhysRevLett.49.202}.
The subsequently discussed relativistic harmonics are not subject
to this limitation and can therefore easily be distinguished from
the ones generated by the non-relativistic mechanism. The transition
between both regimes for moderately relativistic laser pulses was
discussed by Tarasevitch \emph{et al.} in Ref.~\cite{tarasevitch:103902}.

\subsubsection{\label{sub:S-similarity}$S$-similarity}

The $S$-similarity \cite{gordienko:043109} is a similarity group
that characterizes the interaction of higly relativistic ($a_{0}\gg1$)
lasers with plasmas. The major dimensionless parameter here the similarity
number $S\equiv N_{e}/a_{0}N_{c}$. If $S$ is kept constant, but
$a_{0}$ and $N_{e}$ are changed, the behaviour is similar, so that
many important physical quantities can be obtained by simple scaling
laws. The theory was first applied to laser - underdense plasma interaction,
where it delivered useful scaling laws for laser-wakefield acceleration
(LWFA).

Concerning the interaction of lasers with overdense plasmas, similarity
theory has to be applied with care, since in the skin layer with its
extreme field gradients, ultra-relativistic and weakly or non-relativistic
electron motion happens in directly adjacent regions. Even if the
laser is highly relativistic in the sense $a_{0}\gg1$, most of the
electrons deeper inside the skin layer move with only moderately relativistic
velocities as long as $S>1$. For $S<1$, relativistic transparency
sets in. Nevertheless, some signatures of $S$-similarity can be observed
here.

According to similarity theory, the momenta of the highly relativistic
electrons can be written as \begin{equation}
\mathbf{p}_{i}(a_{0},N_{e};\, t)=a_{0}\,\hat{\mathbf{p}}_{i}(S;\, t),\end{equation}
where $\hat{\mathbf{p}}_{i}$ is a characteristic function that describes
the motion of the plasma electrons. Without knowing the details, we
can assume $|\hat{\mathbf{p}}_{i}|\sim1$ close to the surface, i.e.
a considerable fraction of the laser field accelerates the electrons
at the surface. This is certainly the case in parameter ranges where
harmonics are efficiently generated.

Let us now consider the $x$-component of the velocity. Dropping the
particle index $i$ for simplicity, we write $v_{x}=\hat{p}_{x}(a_{0}^{-2}+\hat{p}_{x}^{2}+\hat{\mathbf{p}}_{\perp}^{2})^{-1/2}$
and consequently\begin{eqnarray}
\gamma_{x} & \equiv & \frac{1}{\sqrt{1-v_{x}^{2}}}=\sqrt{\frac{a_{0}^{-2}+\hat{p}_{x}^{2}+\hat{\mathbf{p}}_{\perp}^{2}}{a_{0}^{-2}+\hat{\mathbf{p}}_{\perp}^{2}}}\nonumber \\
 & \approx & \left\{ \begin{array}{cc}
a_{0}\hat{p}_{x} & (\textrm{when}\;\hat{\mathbf{p}}_{\perp}^{2}=0)\\
\sqrt{1+\hat{p}_{x}^{2}/\hat{\mathbf{p}}_{\perp}^{2}} & (\textrm{otherwise})\end{array}.\right.\end{eqnarray}

It is evident that $\gamma_{x}$ grows very large - up to the order
of $a_{0}$ - around the times, when the transverse momentum components
vanish. Further, we see that $\gamma_{x}$ must possess a distinct
maximum here - a kind of spike - since otherwise $\gamma_{x}$ remains
in the order of one and $a_{0}\gg1$. Later we show, that the generation
of high order harmonics happens basically due to these {}``$\gamma$-spikes''.

Note, that this also implies that the generation of relativistic harmonics
is generally much more efficient for p-polarized and normally incident
light than it is for s-polarized light. In the case of s-polarized
oblique incidence, the momentum space is three-dimensional and there
are two generally non-vanishing transverse momentum components. Consequently,
the trajectory of $\mathbf{p}$ in momentum space does not necessarily
cross the $p_{x}$-axis and $\gamma_{x}$ probably remains in the
order of $1$ during the entire interaction process. For p-polarized
or normally incident light, one of the transverse momentum components
vanishes for symmetry reasons, $p_{z}=0$. Therefore, the momentum
space is two-dimensional and $\mathbf{p}$ has to cross the $p_{x}$-axis
every time $p_{y}$ changes sign.

\subsection{\label{sub:harm_theo-ROM}The relativistically oscillating mirror
(ROM) model}

\global\long\def\arp{\text{ARP}}
\global\long\def\xrom{x_{\textrm{surf}}}
\global\long\def\rom{\textrm{surf}}
\global\long\def\gai{\textrm{gAi}_{n}}

Due to its descriptive nature, the term {}``relativistically oscillating
mirror'' (ROM) is in common use. However, its usage varies among
authors and there has been no accurate and generally accepted definition
so far. In the frame of this work, we define the ROM model as the
model based on the boundary condition\begin{equation}
E_{i}\left(t-\frac{x_{\arp}(t)}{c}\right)+E_{r}\left(t+\frac{x_{\arp}(t)}{c}\right)=0,\label{eq: ARP_ansatz}\end{equation}
wherein $x_{\arp}$ denotes the coordinate of the {}``apparent reflection
point'' (ARP). In Ref.~\cite{lichters:3425}, where the term {}``oscillating
mirror'' was first used in the context of relativistic laser-plasma
interaction, it was applied to a model based on an oscillating step-like
plasma boundary. We are going to see soon that the above boundary
condition is closely related to that model. Further, the ARP is intuitively
understood as a sort of \emph{mirror}, which \emph{oscillates} at
\emph{relativistic} velocities.

This subsection consists of three parts. At first (Sub.~\ref{sub:harm_theo-ARP_foundation}),
we investigate the foundation of Eq.~\eqref{eq: ARP_ansatz}, trying
to clarify, under which conditions it is applicable. Then (Sub.~\ref{sub:harm_theo-ARP-verify}),
we demonstrate a simple way to check the validity of the model within
a simulation. Finally (Sub.~\ref{sub:spectrum-ARP}), we derive some
very general properties of the spectrum that follows from Eq.~\eqref{eq: ARP_ansatz}.

\subsubsection{\label{sub:harm_theo-ARP_foundation}Foundation of the ARP boundary
condition}

Here, the applicability of the boundary condition Eq.~\eqref{eq: ARP_ansatz}
is analyzed. To do this, we consider two possible ways to arrive at
the condition. The first approach was introduced by Gordienko \emph{et
al.} \cite{gordienko:043109} in 2004. It is based on the Taylor expansion
of the current distribution. The alternative approach connects the
ARP boundary condition to the assumption of a moving step-like electron
density profile. The correlation between the shape of the electron
density profile and the resulting radiation can be confirmed within
PIC simulations.

We begin with the approach from Ref.~\cite{gordienko:043109,bgp-theory2006}.
It is based on the Taylor expansion of the current distribution with
respect to time: $\jperp(t_{0}+h,\, x)\approx\mathbf{j}_{\perp}(t_{0},\, x)+h\,\partial_{t}\mathbf{j}_{\perp}(t_{0},\, x)+\mathcal{O}(h^{2})$.
This is inserted into Eq.~\eqref{eq:inhom_wave_sol_E}. Expanding
around $t_{0}=t-t'$ and using $h=(x-x')/c$, we find that the zeroth
order term vanishes immediately. Keeping the first order terms and
neglecting the second and higher orders we get \begin{equation}
\mathbf{E}_{\perp}\approx\frac{4\pi}{c}\int_{x}^{\infty}\frac{x-x'}{c}\partial_{t}\jperp(t-t',\, x')\, dx'.\label{eq:inhom_wave_taylor}\end{equation}

Now, the time derivative $\partial_{t}\jperp$ is estimated by the
current divided by the {}``skin layer evolution time'' $\tau\equiv\min\left(\jperp/\partial_{t}\mathbf{j}_{\perp}\right)$.
In this way, for $x=x_{\textrm{surf}}$ at the plasma surface, $\mathbf{E}_{\perp}$
can be approximated as: \begin{equation}
\mathbf{E}_{\perp}(x=x_{\textrm{surf}})\sim\frac{4\pi\delta}{c\tau}\mathbf{J}_{\perp},\end{equation}
where $\delta$ refers to the skin length and $\mathbf{J}_{\perp}$to
the instantaneous net current. The characteristic time $\tau$ can
be estimated by the inverse laser pulse frequency, $\tau\sim\omega_{0}^{-1}.$
In the linear approximation, the skin length is given by the plasma
frequency, $\delta=c/\omega_{p}$. Thus, for highly overdense plasmas,
we expect the field being small at the plasma surface and may consequently
apply Eq.~\eqref{eq: ARP_ansatz}.

In order to further clarify the conditions which lead to the fulfilment
of Eq.~\eqref{eq: ARP_ansatz}, one can also consider an alternative
derivation of the boundary condition \eqref{eq: ARP_ansatz}. This
derivation is based on the model of an oscillating, step-like boundary.
We start with an arbitrary polarization component of the wave equation
\eqref{eq:inhom_wave_eq}. The equation is then adapted to the step-like
density profile and normal incidence. Further, we make use of the
canonical momentum conservation \eqref{eq:canonical_mom} and switch
to relativistically normalized units ($ct\rightarrow t$, ...) for
convenience:\begin{equation}
\left(\partial_{x}^{2}-\frac{1}{c^{2}}\partial_{t}^{2}\right)A=\theta\left(x-\xrom(t)\right)\frac{\omega_{p}^{2}}{c^{2}\gamma}A,\label{eq:wave_eq_step_normal}\end{equation}
where $\omega_{p}$ is the electron plasma frequency and $\gamma$
is the electron $\gamma$-factor. We make the complex ansatz:

\begin{equation}
A(t,x)=\left\{ \begin{array}{lc}
A_{i}(t-x/c)+A_{r}(t+x/c)\quad & \left(x<\xrom(t)\right)\\
A_{s}(t+i\kappa x/c) & \left(x>\xrom(t)\right)\end{array}\right.,\label{eq:ROM_ansatz}\end{equation}
wherein $\kappa=\sqrt{\omega_{p}^{2}/\left(\gamma\omega_{0}^{2}\right)-1}$
is a real number, as the plasma is overdense. To take account for
the relativistic non-linearities, we allow for general functions instead
of strictly assuming $A_{r},\: A_{s}\propto\exp\left(i\omega_{0}t\right)$.
Note, that the vacuum part ($x<\xrom$) of Eq.~\eqref{eq:ROM_ansatz}
is an exact solution of Eq.~\eqref{eq:wave_eq_step_normal}. The
skin layer part is an exact solution for the fundamental mode, $A_{s}\propto\exp\left(i\omega_{0}t\right)$.
Taking into account that in many cases ($\omega^{-8/3}$ - spectrum)
the biggest share of energy is still contained in the laser fundamental
mode, we consider this approximation being reasonable enough within
our simple model.

Now the function $A$ as well as its first partial spatial derivative
must be continuous at the point $x_{\rom}(t)$ at every time $t$.
Defining $a_{i}(t)\equiv eA_{i}(t-x_{\rom}(t))/mc,\: a_{r}(t)\equiv eA_{r}(t+x_{\rom}(t))/mc,\: a_{s}(t)\equiv eA_{s}(t+i\kappa x_{\rom}(t))mc$,
we get:

\begin{eqnarray}
a_{i}+a_{r} & = & a_{s}\label{eq: contin 1}\\
\frac{1}{\dot{x}_{\rom}/c-1}\dot{a_{i}}+\frac{1}{\dot{x}_{\rom}/c+1}\dot{a_{r}} & = & \frac{1}{\dot{x}_{\rom}/c-i/\kappa}\dot{a}_{s}.\label{eq: contin 2}\end{eqnarray}

Solving for $\dot{a}_{s}$ yields

\begin{equation}
\dot{a}_{s}=2\frac{\dot{x}_{\rom}/c-i/\kappa}{1+i/\kappa}\, E_{i}(t-x_{\rom}(t)/c),\label{eq: A_s}\end{equation}
where $\dot{a}_{i}=(\dot{x}_{\rom}/c-1)\, E_{i}$ was used. Inserting
this back into Eq.~\eqref{eq: contin 2}, we obtain

\begin{equation}
E_{r}(t+x_{\rom}(t)/c)+\frac{\kappa-i}{\kappa+i}E_{i}(t-x_{\rom}(t)/c)=0.\label{eq: E_r}\end{equation}

Since $|(\kappa-i)/(\kappa+i)|=1$ for $\kappa\in\mathbb{R}$, it
is now seen that Eq.~\eqref{eq: E_r} agrees with Eq.~\eqref{eq: ARP_ansatz}
except for a phase term. This phase can be included in the function
$x_{\arp}(t)$, setting \begin{equation}
x_{\arp}(t)=x_{\rom}(t)+\frac{1}{\omega_{0}}\arccos\left(1-\frac{2\omega_{0}^{2}\gamma}{\omega_{p}^{2}}\right).\end{equation}

We have shown here that the ARP boundary condition \eqref{eq: ARP_ansatz}
is valid under three main assumptions: a step-like electron density
profile, normal incidence and the interaction is dominated by the
laser fundamental. In this case, the ansatz \eqref{eq:ROM_ansatz}
is reasonable. If the density inside the plasma is not exactly constant,
but there is a sharp rising edge behind which comparatively weak fluctuations
follow, the ansatz \eqref{eq:ROM_ansatz} might still be useful, as
the precise behaviour of the field deep inside the skin layer has
no strong influence on the reflection. The new calculation equips
us with a rough idea of when Eq.~\eqref{eq: ARP_ansatz} can be expected
to be useful.

\subsubsection{\label{sub:harm_theo-ARP-verify}Verifying the ARP boundary condition}

Equation~\eqref{eq: ARP_ansatz} has a simple interpretation that
allows us to verify within simulation data, whether it is fulfilled
or not. For Eq.~\eqref{eq: ARP_ansatz} to have any useful physical
meaning, it is required that $|\dot{x}_{\arp}(t)|<c$ at all time.
Otherwise, the fields would be un- or overdetermined, leading to contradictions
or useless tautologies.

It is obvious then, that the reflected field $E_{r}$ is nothing but
a phase modulation of the negative of the incident one $(-E_{i})$.
In a PIC simulation, we can easily check this by looking at the fields
in the time domain. If and only if Eq.~\eqref{eq: ARP_ansatz} is
fulfilled, then both functions possess the same sequence of extrema
and monotonic intervals.

\begin{figure}
\centering{}\includegraphics[width=6in]{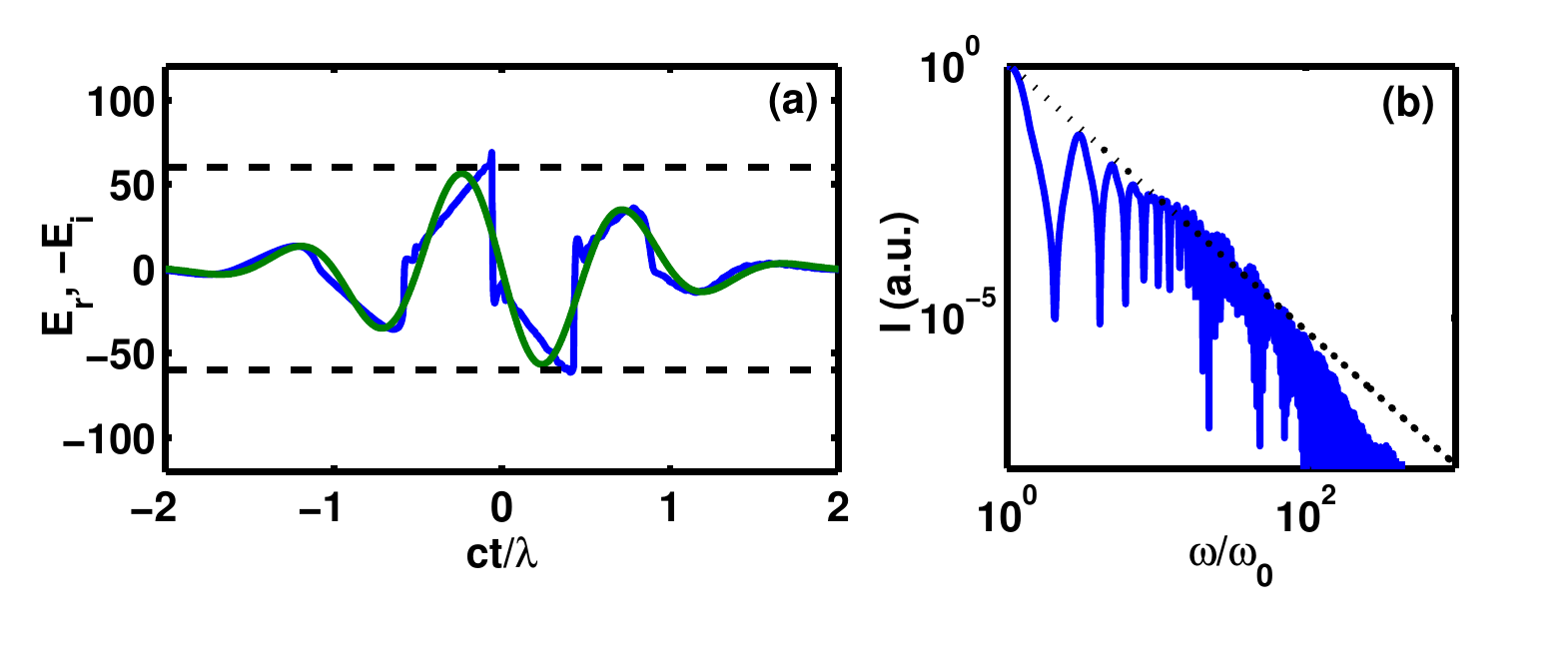}\caption{\label{fig:ROM-new_model_time_spectra}Radiation in time (a) and spectral
(b) domain in the ROM regime. Simulation parameters are: normal incidence,
plasma density $N_{e}=250\, N_{c}$, laser field amplitude is $a_{0}=60$.
In both frames, the reflected field is represented by a blue line.
In (a), the green line represents the field of the incident laser
and the black dashed lines mark the maximum field of it. In (b), the
dotted black line represents an $8/3$ power law.}
\end{figure}

Fig.~\ref{fig:ROM-new_model_time_spectra}(a) shows an example, where
Eq.~\eqref{eq: ARP_ansatz} is fulfilled to a good approximation.
This example was obtained by normal incidence on a sharply defined
plasma, therefore confirming the considerations presented above. However,
compare this to Fig.~\ref{fig:CSE-new_model_time_spectra}(a) to
see that Eq.~\eqref{eq: ARP_ansatz} cannot always be applied. In
Sec.~\ref{sub:harm_theo-nanobunching}, we are going to discuss this
interesting case in more detail. Now we continue with the case, where
the ROM model works.

\subsubsection{\label{sub:spectrum-ARP}Analytical derivation of the spectrum}

Given the validity Eq.~\eqref{eq: ARP_ansatz}, it is possible to
calculate the general form of the spectral envelope with only a few
straightforward assumptions.

We start by writing down the Fourier transform of $E_{r}(t)$ from
Eq.~\eqref{eq: ARP_ansatz}, taking into account the retarded time:
\begin{equation}
E_{r}(\omega)=-\int E_{i}\left(t-\frac{x_{\arp}}{c}\right)\, e^{i\omega(t+x/c)}\,\left(1+\frac{\dot{x}_{\arp}}{c}\right)\, dt.\end{equation}

The incoming laser pulse is described by an envelope approximation
$E_{i}(t)=g(t)\left[\exp\left(i(\omega_{0}t+\phi_{0})\right)-\exp\left(-i(\omega_{0}t+\phi_{0}\right)\right]/2$,
where $g(t)$ is a slowly varying function. We arrive at\begin{eqnarray}
E_{r}(\omega) & = & e^{i\phi_{0}}E_{+}-e^{-i\phi_{0}}E_{-}\nonumber \\
E_{\pm} & = & -\int g\left(t-\frac{x_{\arp}}{c}\right)\nonumber \\
 &  & \times\exp\left[i\left(\omega\left(t+\frac{x_{\arp}}{c}\right)\pm\omega_{0}\left(t-\frac{x_{\arp}}{c}\right)\right)\right]\nonumber \\
 &  & \times\left(1+\frac{\dot{x}_{\arp}}{c}\right)\, dt.\label{eq:ARP_integral}\end{eqnarray}

Now note that for high $\omega$, the exponential term leads to a
rapid oscillation of the integrand during most of the time. Because
of this oscillation, most contributions cancel, except for those where
the phase of the integrand stands still. This means that the integral
can be handled by the method of stationary phase. The somewhat more
technical details of this calculation are shifted to appendix~\ref{sec:Stationary-Phase-Method},
but before presenting the final result, we would like to remark two
interesting points:

First, the stationary phase points correspond to the instants when
the ARP moves towards the observer with maximum velocity. These moments
are crucial for the generation of high order harmonics. The corresponding
ARP gamma factor $\gamma_{\arp}=\left(1-\dot{x}_{\arp}^{2}/c^{2}\right)^{-1/2}$
possesses a sharp spike at these instants, which is the reason why
we also call them \emph{$\gamma$-spikes} \cite{bgp-theory2006}.

Second, the spectrum depends on the exact behaviour of the ARP in
the neighbourhood of these points. In Ref.~\cite{bgp-theory2006}
it was presumed, that the derivative of the ARP acceleration is different
from zero at the $\gamma$-spike. Other cases are imaginable however,
and it is intriguing to see, what difference they make. We here consider
the most general case, in which $\ddot{x}_{\arp}$ has a zero of order
$2n-1$ at the $\gamma$-spike, meaning that $d^{k}x_{\arp}/dt^{k}=0$
for all $2\leq k\leq2n$. We will subsequently refer to $n$ as the
\emph{order of the $\gamma$-spike}.

After the calculations in appendix~\ref{sec:Stationary-Phase-Method},
the spectrum can be written as:

\begin{eqnarray}
I_{n}(\omega) & \sim & \omega^{-\frac{4n+4}{2n+1}}\left[\sum_{\sigma\in\{-1,1\}}e^{\sigma i\phi_{0}}\,\gai\left(\frac{\omega\gamma^{-2}-\sigma4\mbox{\ensuremath{\omega}}_{0}}{2\left(\alpha\omega\right)^{1/(2n+1)}}\right)\right]^{2},\label{eq:general_ARP_spectrum}\end{eqnarray}
wherein $\gamma$ refers to the peak value of $\gamma_{\arp}(t)$
and $\alpha$ is a constant related to the behaviour of the ARP trajectory
close to the $\gamma$-spike. As the ARP motion happens on the timescale
of the laser period, we may estimate $\alpha^{2n}\sim\omega_{0}$
and consequently write $\alpha\equiv\tilde{\alpha}\omega_{0}^{1/2n}$,
where $\tilde{\alpha}$ is a numeric constant on the order of one.
$\phi_{0}$ is connected to the phase, at which the $\gamma$-spikes
occur. As we are going to see later, it has no signifcant influence
on the spectrum, unless $\phi_{0}\approx(n+1/2)\pi$. $\gai$ is a
generalized Airy-function as defined in appendix~\ref{sec:num-integration-gen-airy}.
These functions are not commonly available in general purpose numerical
function libraries. With a small trick, they are however not hard
to compute. The details of the numerical calculation of the integral
are explained in App.~\ref{sec:num-integration-gen-airy}. We now
regard the common case $n=1$ and the special case $n>1$, corresponding
to instantaneously vanishing acceleration at the $\gamma$-spike,
separately.

\noindent The most typical case $n=1$ has been investigated by Baeva,
Gordienko and Pukhov (BGP) in Ref.~\cite{bgp-theory2006}. In this
case, Eq.~\eqref{eq:general_ARP_spectrum} can be written with a
conventional Airy function:

\begin{eqnarray}
I_{n}(\omega) & \sim & \omega^{-\frac{8}{3}}\left[\sum_{\sigma\in\{-1,1\}}e^{\sigma i\phi_{0}}\,\mathrm{Ai}\left(\frac{\omega\gamma^{-2}-\sigma4\mbox{\ensuremath{\omega}}_{0}}{2\left(\alpha\omega\right)^{1/3}}\right)\right]^{2},\label{eq:BGP_spec}\end{eqnarray}

\begin{figure}
\centering{}\includegraphics[width=3in]{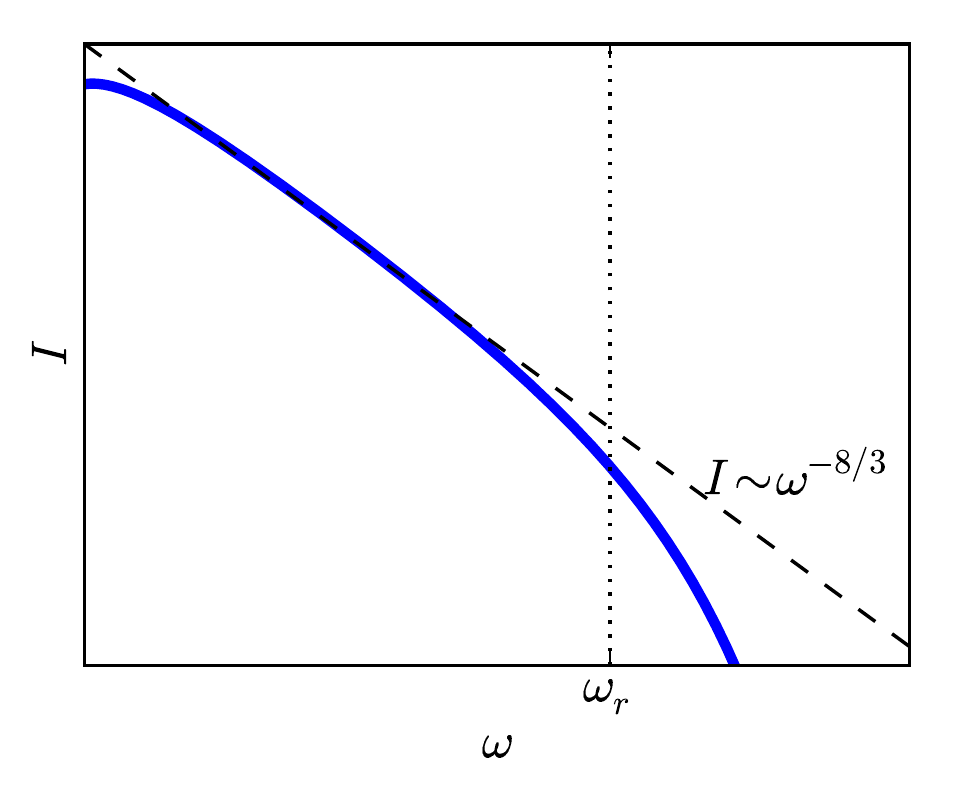}\caption{\label{fig:BGP-vs-83-simple}The BGP-spectrum Eq.~\eqref{eq:BGP_spec}
in a double logarithmic representation. The dashed black line denotes
an 8/3-power law and the dotted line represents the roll-over frequency
$\omega_{r}=\sqrt{8\tilde{\alpha}}\,\omega_{0}\gamma^{3}$. For the
picture, we took $\tilde{\alpha}=1$, $\gamma=5$ and $\phi_{0}=0$,
but the resulting spectra are largely independent of these variables,
provided the $\omega$-axis is scaled appropriately.}
\end{figure}

The spectrum is presented in Fig.~\ref{fig:BGP-vs-83-simple}. We
see that they consist of a power law which {}``rolls over'' into
an exponential decay at a frequency, which depends on $\gamma$. By
comparing the argument of the Airy function to one, we find this characteristic
roll-over frequency to be \begin{equation}
\omega_{r}=\sqrt{8\tilde{\alpha}}\,\omega_{0}\gamma^{3}.\end{equation}
Note the favourable scaling of this characteristic frequency $\omega_{r}$
with $\gamma^{3}$, in contrast to the Doppler shift from the reflection
at a constantly moving mirror, which produces a frequency upshift
by a factor of only $4\gamma^{2}$.

Below this frequency, the airy functions are almost constant and the
spectrum can be approximated by the famous power law \begin{equation}
I_{{\rm BGP}}(\omega\ll\omega_{r})\sim\frac{1}{\omega^{8/3}}.\label{eq:BGP_low}\end{equation}

This scaling can nicely be seen in Fig.~\ref{fig:BGP-vs-83-simple}.
Technically, there is an exception to this scaling: the special case
$\phi_{0}\approx(n+1/2)\pi$. In that case, the leading orders of
the Airy functions cancel each other out, yielding a slightly steeper
power law decay: $I_{\mathrm{BGP}}\sim\omega^{-10/3}$, see also Fig.~\ref{fig:ARP_spectra}(b).
This special case seems to be of not much physical importance however.

For frequencies much larger than $\omega_{r}$, the Airy function
dominates and the decay becomes exponential:\begin{equation}
I_{{\rm BGP}}(\omega\gg\omega_{r})\sim\left(\frac{\omega_{r}}{\omega}\right)^{-3}\exp\left(-\frac{4}{3}\frac{\omega}{\omega_{r}}\right).\label{eq:BGP_high}\end{equation}

\begin{figure}
\centering{}\includegraphics[width=3in]{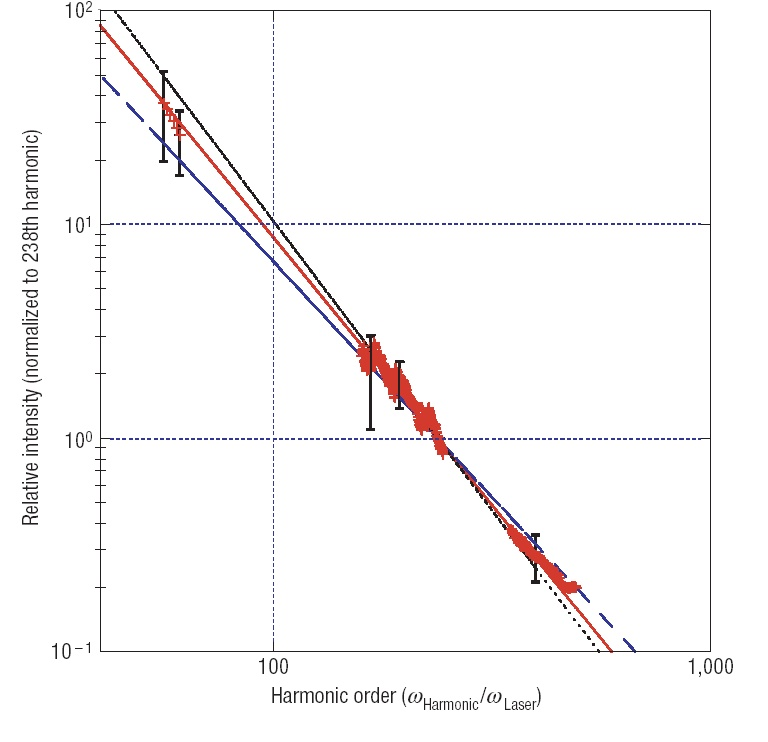}\caption{\label{fig:dromey_spectrum_widerange}Experimental measurement of
the relativistic harmonics spectrum by Dromey et al., taken from Ref.~\cite{Dromey2006}.
The dots indicate measured harmonics normalized to the 238th harmonic,
the lines correspond to power law fits $I\propto I^{-q}$ with $q=2.5$
for the best fit (red line), and $q=2.2$ (blue) respectively $q=2.7$
(black), close to the theoretical value $q=8/3$ of the BGP spectrum
Eq.~\eqref{eq:BGP_spec}. }
\end{figure}

\begin{figure}
\centering{}\includegraphics[width=3in]{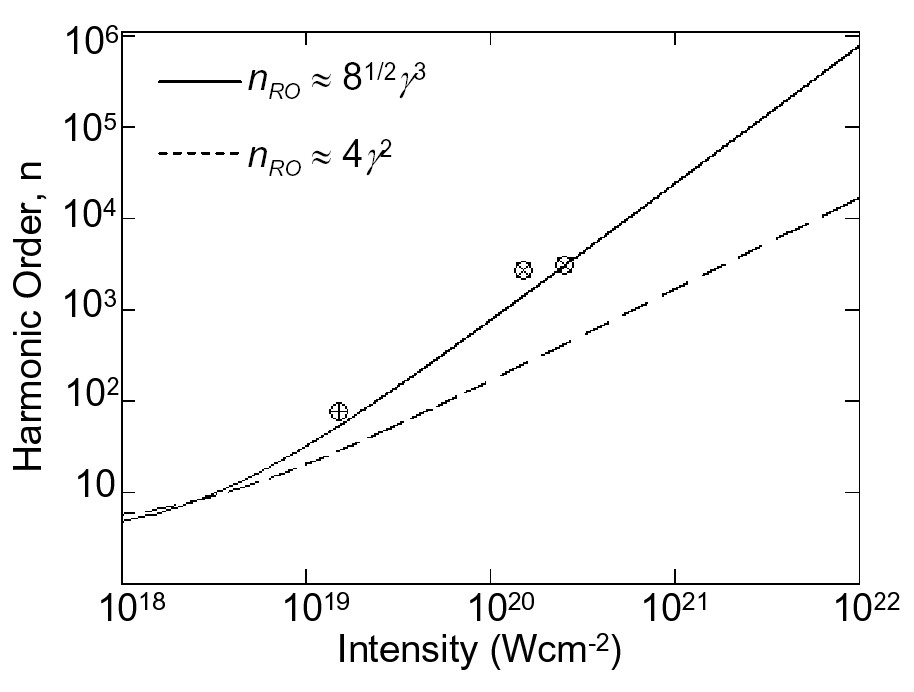}\caption{\label{fig:dromey_cutoff_dependency}Measurement of the dependence
of the harmonics spectrum cut-off on the laser intensity, taken from
Ref.~\cite{Dromey2006}.}
\end{figure}

\noindent An example for this sort of spectrum from a PIC simulation
can be seen in Fig.~\ref{fig:ROM-new_model_time_spectra}(b). An
experimental spectrum, measured over a wide range of frequencies,
is shown in Fig.~\ref{fig:dromey_spectrum_widerange}. It is found
to be a power law, and the exponent $q\in[2.2;\,2.7]$ agrees with
the BGP spectrum \eqref{eq:BGP_spec}. Further, the scaling of the
cut-off frequency was found to be $\sim\gamma^{3}$, also in agreement
with the theoretical predictions of the discussed model.

\begin{figure}
\centering{}\includegraphics[width=6in]{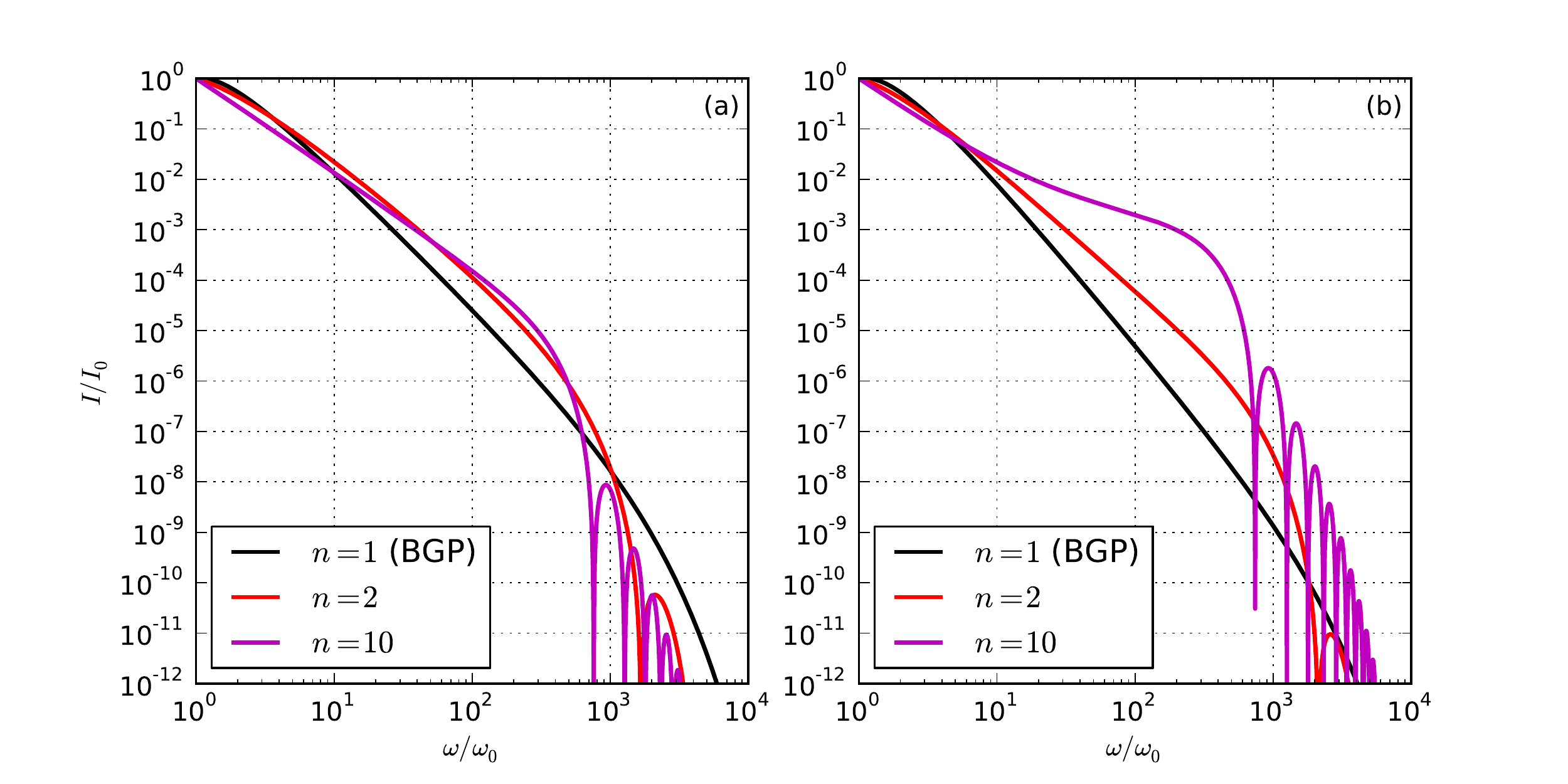}\caption{\label{fig:ARP_spectra}Logarithmic plot of the spectra Eq.~\eqref{eq:general_ARP_spectrum}
following from the ARP boundary condition, for orders $n=1,\,2,\,10$
of the $\gamma$-spike. (a) shows the case $\phi_{0}=0$, (b) shows
the special case $\phi_{0}=\pi/2$. Unless $\phi_{0}\approx(n+1/2)\pi$,
the observed spectra resemble (a) rather than (b). All spectra have
been normalized to $I_{0}=I(\omega_{0})$, and $\alpha=1,\,\gamma=8$
was used throughout.}
\end{figure}

\noindent Let us now move on to the cases $n>1$. Fig.~\ref{fig:ARP_spectra}
shows the spectra Eq.~\eqref{eq:general_ARP_spectrum} for different
orders $n$ of the $\gamma$-spike. To the best of our knowledge,
the cases $n>1$ have not been investigated before.

In Fig.~\ref{fig:ARP_spectra}(a), the typical case $\phi_{0}=0$
is displayed, Fig.~\ref{fig:ARP_spectra}(b) shows the case $\phi_{0}=\pi/2$,
where the first order term of the generalized Airy-functions just
cancel out. It makes sense to also consider this special case here,
as there is reason to believe that the occurence of higher-order $\gamma$-spikes
may be related to the occurence of the spikes at a certain phase.
Comparing the spectra for the higher order $\gamma$-spikes $n>1$
to the BGP case we note the following differences:
\begin{enumerate}
\item The power law part of the spectra decays slightly slower with increasing
$n$. In addition, particularly in the case $\phi_{0}=\pi/2$, the
non-power law part gains influence for increasing $n$ at low frequencies
already and leads to even slower decaying spectra. This is possibly
favourable for the efficient production of attosecond pulses.
\item Because of the oscillatory behaviour of the generalized Airy functions
$\gai(x)$ at positive $x$ and for $n>1$, the spectra become strongly
modulated at frequencies $\omega>\omega_{r}$, compare Fig.~\ref{fig:ARP_spectra}.
Also, this effect is more pronounced for $\phi_{0}=\pi/2$. This might
explain some of spectral modulations observed in numerical and real
experiments before, see e.g. Ref.~\cite{boyd:125004}.
\item The roll-off frequency, which scales as $\gamma^{3}$ in the BGP case,
approaches a $\gamma^{2}$ scaling in the limit $n\rightarrow\infty$,
reminiscent of the Doppler effect from a mirror moving with constant
velocity. This seems reasonable, since for higher order $\gamma$-spikes,
the acceleration is very small in the neighbourhood of the stationary
phase point. Therefore, its influence on the spectrum decreases with
$n$.
\end{enumerate}
Using the estimates from Sec.~\ref{sub:S-similarity} based on the
$S$-similarity theory, we can further assume that for fixed $S$-parameter
$\gamma\propto a_{0}$. In the BGP-case $n=1$ this means $\omega_{r}\propto a_{0}^{3}$,
which is in agreement with experimental observations reported in Ref.~\cite{Dromey2006}.

To sum up this subsection, we have reviewed the popular {}``relativistically
oscillating mirror'' (ROM) model for the relativistic generation
of harmonics at overdense plasma surfaces, based on Eq.~\eqref{eq: ARP_ansatz}.
We have found, that it is applicable for normal incidence and step-like
electron plasma boundaries. Because of its simplicity - it reduces
the whole complex interaction physics to one simple function $x_{\arp}(t)$
- the model helps us to gain insight into the basic mechanism that
leads to the generation of high harmonics. Further, we have analytically
calculated the possible spectra in the relativistic limit with the
help of asymptotic analysis. Here we noticed, that even within the
model, spectra that deviate considerably from the well known BGP $8/3$-power
law are in principle possible.

\subsection{\label{sub:harm_theo-TROM}Totally reflecting oscillating mirror
(TROM) and $\omega^{-2/3}$ spectrum}

As we see e.g. from Fig.~\ref{fig:CSE-new_model_time_spectra}, the
ROM model based on Eq.~\eqref{eq: ARP_ansatz} as it was used in
Ref.~\cite{bgp-theory2006}, is not always valid in the highly relativistic
regime - not even as an approximation. It is thus worth looking for
alternatives.

This subsection is about another model that one might intuitively
associate with the name {}``relativistically oscillating mirror'',
we call it the {}``totally reflecting oscillating mirror'' (TROM).
The model is rigorously based on the assumption of total reflection
from a perfectly localized current layer.

First (Sub.~\ref{sub:Derivation-TROM}), we derive the corresponding
boundary condition. Then (Sub.~\ref{sub:Spectrum-TROM}), we demonstrate
the spectral properties that follow from this boundary condition,
utilizing asymptotic analysis once again. Finally (Sub.~\ref{sub:Feasibilty-TROM})
we give some remarks about the possible physical realization of the
model.

\subsubsection{\label{sub:Derivation-TROM}Foundation of the TROM boundary condition}

\global\long\def\trom{\textrm{TROM}}

The TROM model is particularly interesting because of its mathematical
lucidity. It can be rigorously derived from only two straightforward
assumptions. These assumptions are:
\begin{enumerate}
\item There is total reflection, no light passes through the mirror. Therefore,
we can relate the plasma current to the incident radiation via Eq.~\eqref{eq:E_incoming}.
\item The skin layer of the reflecting plasma is infinitely thin. Therefore,
the current can completely be described by $j(t,x)=\bar{j}(t)\delta(x-x_{\trom}(t))$.
\end{enumerate}
Inserting the current profile into Eqs.~\eqref{eq:E_incoming} and
\eqref{eq:E_reflected}, we obtain: \begin{eqnarray}
E_{i}(t) & = & -2\pi\frac{\bar{j}(t-x_{\trom}(t)/c)}{c+\dot{x}_{\trom}(t-x_{\trom}(t)/c)}\label{eq:incident_TROM}\\
E_{r}(t) & = & 2\pi\frac{\bar{j}(t+x_{\trom}(t)/c)}{c-\dot{x}_{\trom}(t+x_{\trom}(t)/c)}.\label{eq:reflected_TROM}\end{eqnarray}

Now, the assumption of total reflection is exploited by using Eq.~\eqref{eq:inhom_wave_sol_E}.
We eliminate $\bar{j}$ and arrive at the boundary condition:\begin{equation}
E_{r}\left(t+\frac{x_{\trom}(t)}{c}\right)+\frac{1-\dot{x}_{\trom}/c}{1+\dot{x}_{\trom}/c}\, E_{i}\left(t-\frac{x_{\trom}(t)}{c}\right)=0.\label{eq:TROM_boundary}\end{equation}

Compare this to Eq.~\eqref{eq: ARP_ansatz}, which represents the
ROM model. The difference lies in the pre-factor of $E_{i}$, which
amplifies the reflected field at times, when the mirror moves towards
the observer. Since these are the regions which are responsible for
high frequency radiation, we expect a flatter spectrum here compared
to the ROM model.

Further note, that Eq.~\eqref{eq:TROM_boundary} is always the correct
boundary condition for a totally reflecting mirror in the limit of
constant velocity. In this case, Eq.~\eqref{eq:TROM_boundary} could
simply be derived by a Lorentz transformation to the system, where
the mirror is at rest. For a strongly accelerated mirror however,
we need the additional assumption of a perfectly localized skin layer
to obtain Eq.~\eqref{eq:TROM_boundary}.

\subsubsection{\label{sub:Spectrum-TROM}Analytical derivation of the TROM spectrum}

We now derive the spectrum corresponding to Eq.~\eqref{eq:TROM_boundary}.
The beginning of the calculation is analogue to the calculation in
subsection~\ref{sub:spectrum-ARP}, and we arrive at

\begin{eqnarray}
E_{\pm} & = & -\int g\left(t-\frac{x_{\trom}}{c}\right)\exp\left[i\left(\omega\left(t+\frac{x_{\trom}}{c}\right)\pm\omega_{0}\left(t-\frac{x_{\trom}}{c}\right)\right)\right]\nonumber \\
 &  & \times\left(1-\frac{\dot{x}_{\trom}}{c}\right)\, dt.\label{eq:TROM_integral}\end{eqnarray}

Compare this to Eq.~\eqref{eq:ARP_integral}. The difference lies
in the last factor: Whereas in Eq.~\eqref{eq:ARP_integral} it is
$1+\dot{x}_{\arp}/c$, we have $1-\dot{x}_{\trom}/c$ here. This difference
is crucial, since at the stationary phase points, where $\dot{x}_{\trom/\arp}\approx-c$,
the term in Eq.~\eqref{eq:ARP_integral} becomes very small, whereas
the term in Eq.~\eqref{eq:TROM_integral} even has a maximum.

Again, we can analytically calculate the corresponding spectrum, as
shown in App.~\ref{sec:Stationary-Phase-Method}. In general, for
a $\gamma$-spike of the order $n$ we obtain: \begin{eqnarray}
I_{\trom}^{n}(\omega) & \propto & \omega^{-\frac{2}{2n+1}}\left[\sum_{\sigma\in\{-1,1\}}\sigma\,\gai\left(\frac{\omega\gamma^{-2}-\sigma4\mbox{\ensuremath{\omega}}_{0}}{2\left(\alpha\omega\right)^{1/(2n+1)}}\right)\right]^{2}.\label{eq:general_TROM_spectrum}\end{eqnarray}

This is the same as the ROM spectrum Eq.~\eqref{eq:general_ARP_spectrum},
except for the different exponent in the power law. The TROM spectrum
is much flatter. For high order $\gamma$-spikes, the power law part
even tends to $\omega^{0}$, so that the spectrum is merely determined
by the generalized Airy functions.

In the more likely case $n=1$, Eq.~\eqref{eq:general_TROM_spectrum}
can to a good approximation be simplified:

\begin{equation}
I_{\trom}^{1}(\omega)\propto\frac{1}{\omega^{2/3}}\left[\textrm{Ai}\left(\left(\frac{\omega}{\omega_{r}}\right)^{2/3}\right)\right]^{2}.\end{equation}

Compared to the $\omega^{-8/3}$ decay predicted for the ROM model
{[}Eq.~\eqref{eq:BGP_spec}{]}, we obtain a slowly decaying $\omega^{-2/3}$
power law here.

\subsubsection{\label{sub:Feasibilty-TROM}Physical Feasibility of the TROM model}

As we have seen, the TROM model yields a distinctly flatter spectrum
than the ROM one. Therefore, if there were a physical system that
behaves according to the TROM model, it could be much more efficient
in the production of attosecond pulses. Let us try to answer (a) why
this is difficult and (b) how it might still be possible.

The difficulty can readily be seen from Eq.~\eqref{eq:incident_TROM}.
We notice that the current $\bar{j}$ does not necessarily vanish
at the instant when the surface moves at maximum velocity. This is
in contrast to the normal behaviour of an ultra-relativistic plasma.
The transverse current is the product of the transverse fluid velocity
component $v_{y}$ and the charge density $\rho$. Since the transverse
velocity component becomes very small at the instant of maximum longitudinal
velocity, a finite $\bar{j}$ implies a huge plasma density. But very
dense plasmas are hard to drive to relativistic motion.

For single pulse schemes, the realization is probably impossible.
The behaviour of ultra-relativistic plasmas is governed by the $S$-parameter
$S\equiv N_{e}/a_{0}N_{c}$. If the $S$-parameter is too low, it
leads to an extended skin layer in contradiction to the assumption
of a perfectly localized current layer. If the $S$-parameter is too
high, the plasma is not driven to relativistic motion at all.

In Ref.~\cite{tarasevitch2009towards}, Tarasevitch \emph{et al.}
propose the realization of the boundary condition \eqref{eq:TROM_boundary}
via a two pulse scheme. In the scheme, the first, relativistically
strong pulse drives the plasma surface to oscillation. The second
pulse is much weaker and has a polarization orthogonal to the first
one. It is used as a probe and the spectrum in the direction of its
polarization is recorded. Indeed, for a certain set of parameters
it was possible to observe the generation of harmonics according to
Eq.~\eqref{eq:TROM_boundary}. Thereby, they heuristically also find
a $2/3$-power law spectrum, confirming the above calculations.

In this case, the probe pulse {}``harvests'' the harmonics generated
by the much stronger driver pulse. Thus, the scheme is not appropriate
to increase the overall efficiency of frequency conversion or attosecond
pulse production. In the following section, we are going to look at
a physical mechanism, where the overall efficiency is indeed increased
considerably in comparison to the ROM case.

\subsection{\label{sub:harm_theo-nanobunching}Coherent synchrotron emission
(CSE) from electron nanobunches}

Cases where the ARP boundary condition \eqref{eq: ARP_ansatz} does
not apply are studied here. We find, that the radiation can be described
as coherent synchrotron emission (CSE) from extremely compressed electron
{}``nanobunches'' that form in front of the surface.

At first (Sub.~\ref{sub:nanobunching-process}), the generation process
is investigated by close examination of PIC data. Then, the spectrum
is calculated analytically (Sub.~\ref{sub:nanobunch-spectrum}).
Finally, we analyze the sensitivity of the process to changes in the
laser-plasma parameters (Sub.~\ref{sub:nanobunch-sensitivity}).

\subsubsection{\label{sub:nanobunching-process}Electron nanobunching process}

\begin{figure}
\centering{}\includegraphics[width=6in]{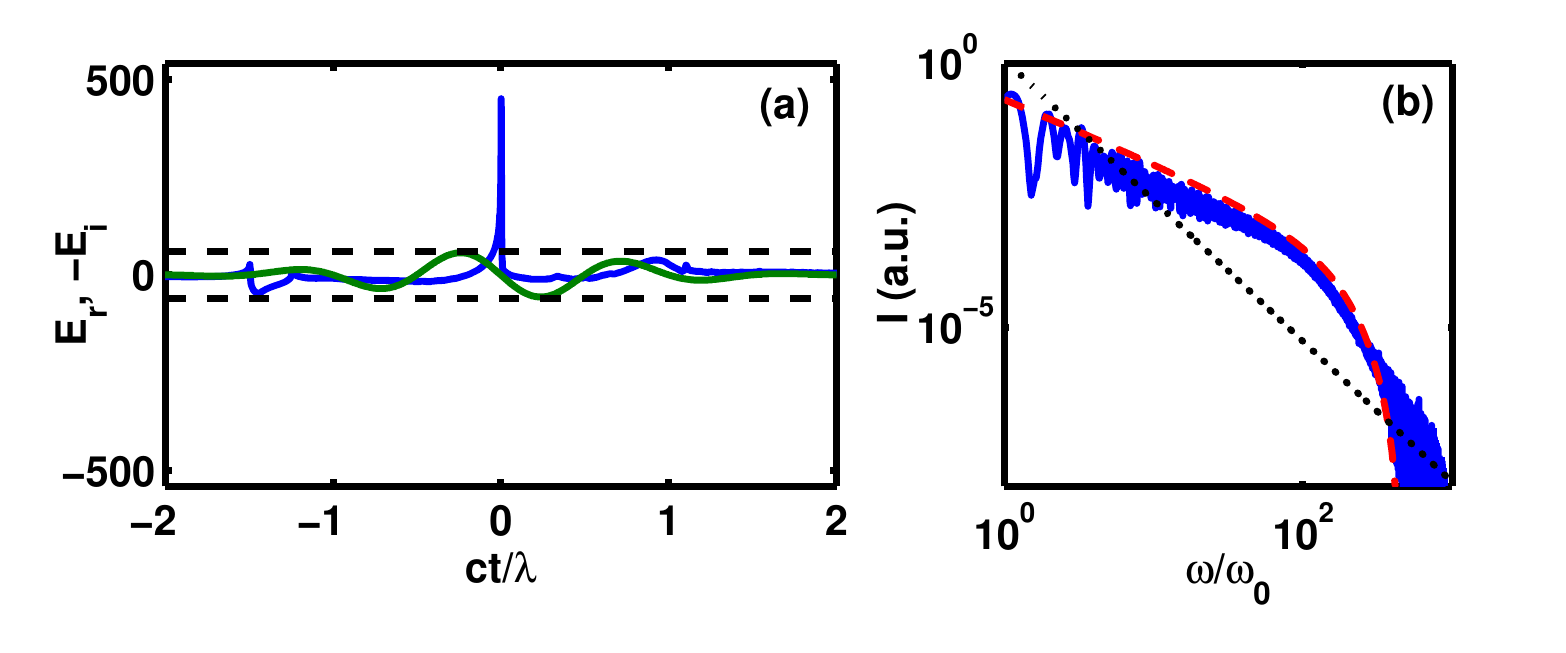}\caption{\label{fig:CSE-new_model_time_spectra}Radiation in time (a) and spectral
(b) domain for simulation in the {}``nanobunching'' regime. Compare
this to Fig.~\ref{fig:ROM-new_model_time_spectra}. Simulation parameters
are: plasma density ramp $\propto\exp(x/(0.33\,\lambda))$ up to a
maximum density of $N_{e}=95\, N_{c}$ (lab frame), oblique incidence
at $63^{\circ}$ angle (p-polarized), $a_{0}=60$. Again, the reflected
field is represented by a blue line, the green line represents the
field of the incident laser and the black dashed lines mark the maximum
field of it. In (b), the dotted black line represents an $8/3$ power
law and the red dashed line corresponds to the analytical nanobunch
CSE spectrum given by Eqs.~\eqref{eq:synch_spec_2nd-1} and \eqref{eq:gaussian_shape},
with $\omega_{rs}=800\,\omega_{0}$ and $\omega_{rf}=225\,\omega_{0}$. }
\end{figure}

Let us now have a fresh look at Fig.~\ref{fig:CSE-new_model_time_spectra}.
It is evident, that the maximum of the reflected field reaches out
about an order of magnitude higher than the amplitude of the incident
laser. The reflected radiation can clearly \textit{not} be obtained
from the incident one by phase modulation and the ARP boundary condition
Eq.~\eqref{eq: ARP_ansatz} fails.

Consequently, the spectrum deviates from the $8/3$-power law, compare
Fig.~\ref{fig:CSE-new_model_time_spectra}(b). Indeed, the efficiency
of harmonics generation is much higher than estimated by the calculations
in Ref.~\cite{bgp-theory2006}: about two orders of magnitude at
the hundredth harmonic. Also, we can securely exclude coherent wake
emission (CWE) as the responsible mechanism, since this would request
a cut-off around $\omega=10\omega_{0}$. The radiation has to be attributed
to a new sort of mechanism.

\begin{figure}
\begin{centering}
\includegraphics[width=4in]{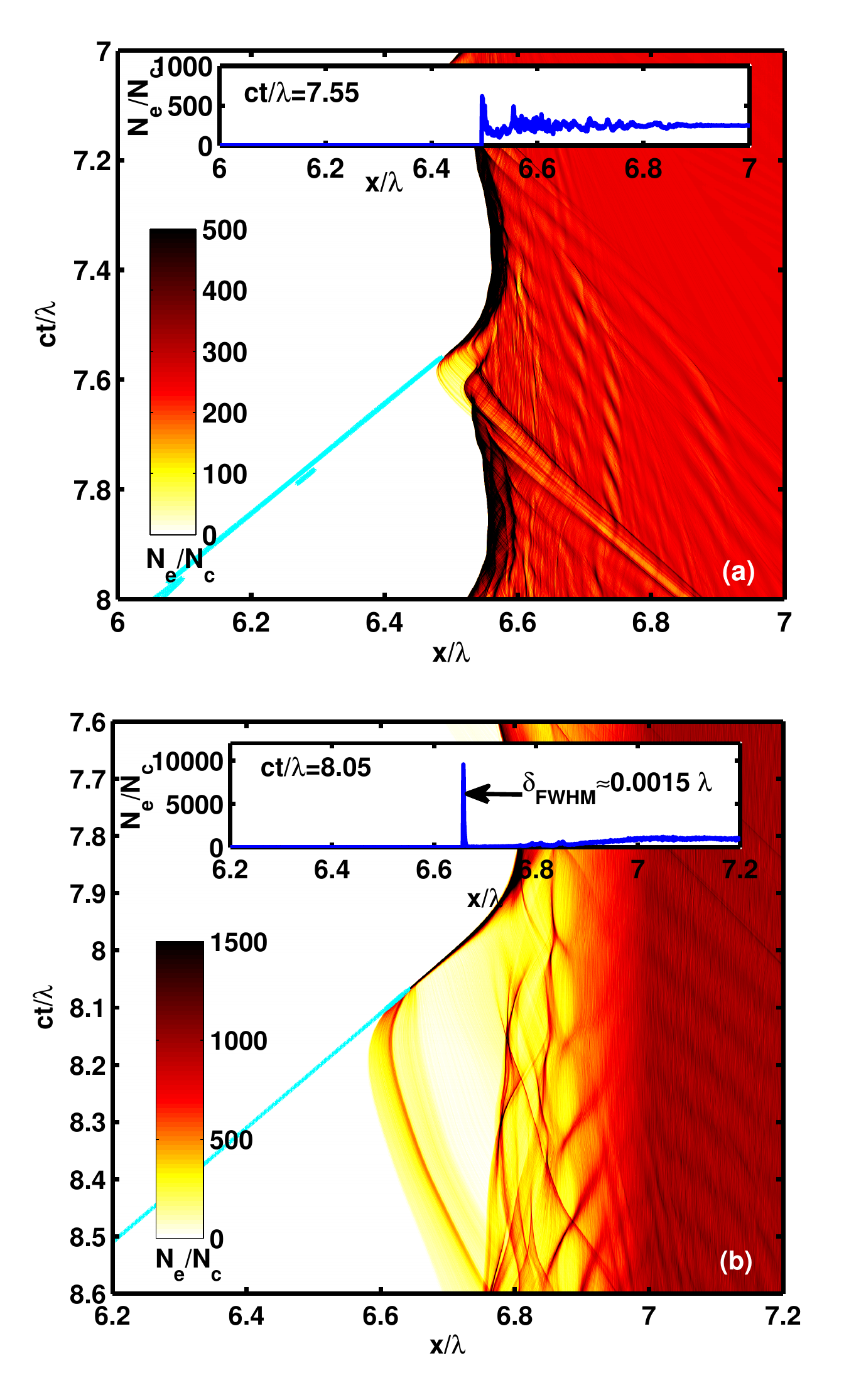}
\par\end{centering}

\caption{\selectlanguage{english}%
\label{fig:cse-generation}\foreignlanguage{british}{The electron
density and contour lines (cyan) of the emitted harmonics radiation
for $\omega/\omega_{0}>4.5$, in (a) the ROM and (b) the nanobunching
regime. The small windows inside the main figures show the detailed
density profile at the instant of harmonic generation. All magnitudes
are taken in the simulation frame. The simulation parameters in panel
(a) are the same as those in Fig.~\ref{fig:ROM-new_model_time_spectra}
and (b) corresponds to Fig.~\ref{fig:CSE-new_model_time_spectra}.}\selectlanguage{british}
}
\end{figure}

To get a picture of the physics behind, let us have a look at the
motion of the plasma electrons that generate the radiation. Figure~\ref{fig:cse-generation}
shows the evolution of the electron density corresponding to both
sample cases from Figs.~\ref{fig:ROM-new_model_time_spectra} and
\ref{fig:CSE-new_model_time_spectra}. In addition to the density,
contour lines of the spectrally filtered reflected radiation are plotted.
These lines illustrate where the main part of the high frequency radiation
emerges.

We observe that in both cases the main part of the harmonics is generated
at the point, when the electrons move towards the observer. This shows
again that in both cases the radiation does not stem from CWE. For
CWE harmonics, the radiation is generated inside the plasma, at the
instant when the Brunel electrons re-enter the plasma \cite{quere:125004}.

Apart from that mutuality, the two presented cases are very different.
Figure~\ref{fig:cse-generation}(a) corresponds to the ROM case.
It can be seen that the density profile remains roughly step-like
during the whole interaction process and the plasma skin layer radiates
as a whole. This explains why the ROM model works well here, as we
have seen before in Fig.~\ref{fig:ROM-new_model_time_spectra}.

Figure~\ref{fig:cse-generation}(b) looks clearly different. The
density distribution at the moment of harmonics generation is far
from being step-like, but possesses a highly dense (up to $\sim10000\, N_{c}$
density) and very narrow $\delta$-like peak, with a width of only
a few nanometres. This electron {}``nanobunch'' emits synchrotron
radiation coherently.

\begin{figure}
\begin{centering}
\includegraphics[width=4in]{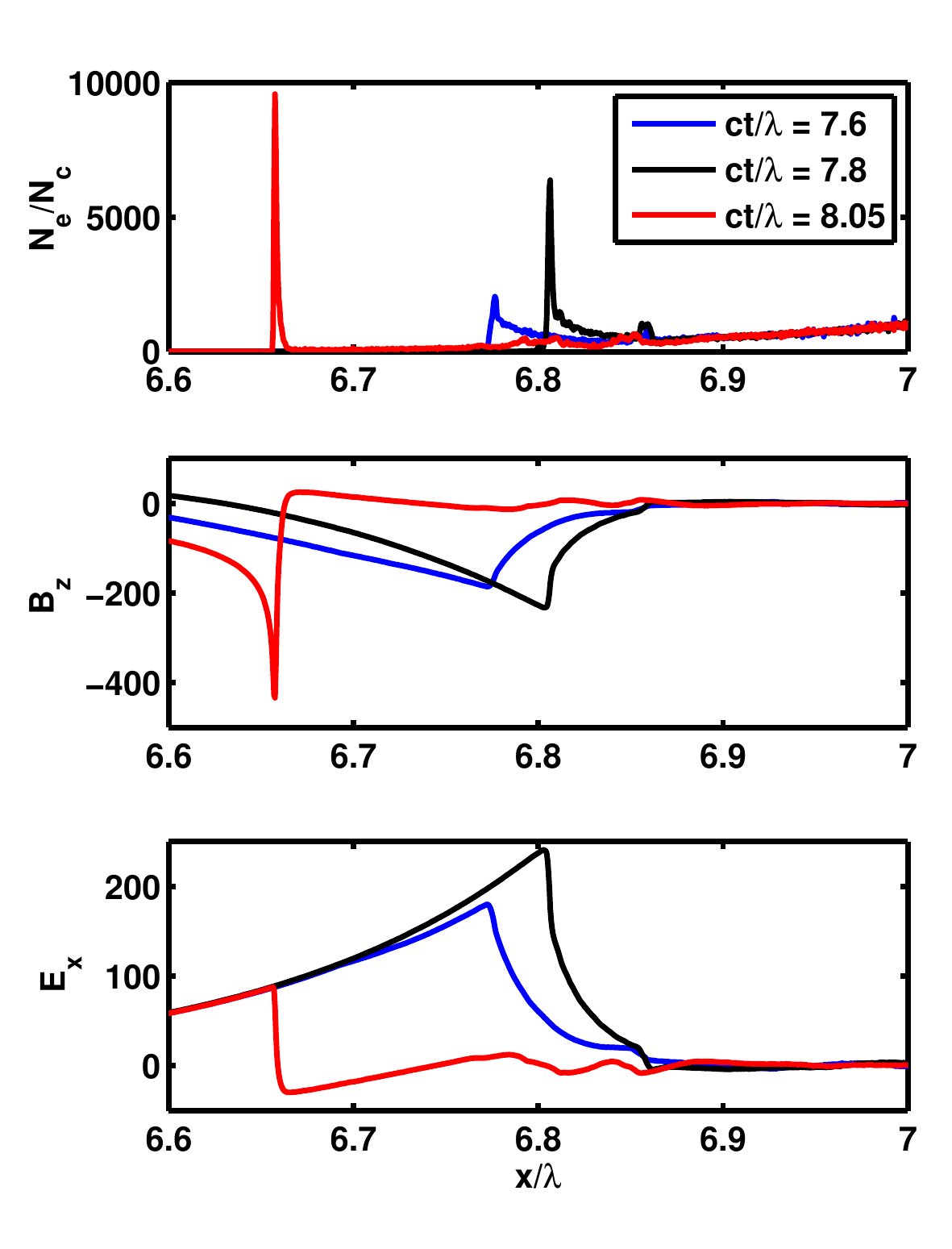}
\par\end{centering}

\caption{\selectlanguage{english}%
\label{fig:formation}\foreignlanguage{british}{Formation of the nanobunch
in the simulation corresponding to Fig.~\ref{fig:CSE-new_model_time_spectra}
and \ref{fig:cse-generation}(b). We depict the electron density $N_{e}$
in units of the critical density $N_{c}$, the transverse magnetic
field component $B_{z}$ and the longitudinal electric field component
$E_{x}$ in relativistically normalized units.}\selectlanguage{british}
}
\end{figure}

The high frequency radiation is emitted by a highly compressed electron
bunch moving \emph{away} from the plasma. However, the electrons first
become compressed by the relativistic ponderomotive force of the laser
that is directed into the plasma, compare the blue lines in Fig.~\ref{fig:formation}.
During that phase, the longitudinal electric field component grows
until the electrostatic force turns around the bunch, compare the
green lines in Fig.~\ref{fig:formation}. Normally, the bunch will
loose its compression in that instant, but in some cases, as in the
one considered here, the fields and the bunch current match in a way
that the bunch maintains or even increases its compression. The final
stage is depicted by the red lines in Fig.~\ref{fig:formation}.

We emphasize, that such extreme nanobunching does not occur in every
case of p-polarized oblique incidence of a highly relativistic laser
on an overdense plasma surface. On the contrary, it turns out that
the process is highly sensitive to changes in the plasma density profile,
laser pulse amplitude, pulse duration, angle of incidence and even
the carrier envelope phase of the laser. For a longer pulse, we may
even observe the case, that nanobunching is present in some optical
cycles but not in others. The parameters in the example were selected
in a way to demonstrate the new effect unambiguously, i.e. the nanobunch
is well formed and emits a spectrum that clearly differs from the
BGP one. The dependence of the effect on some parameters is discussed
in subsection~\ref{sub:nanobunch-sensitivity}.

Because of the one dimensional slab geometry, the spectrum is not
the same as the well known synchrotron spectrum \cite{Jackson} of
a point particle. We now calculate the spectrum analytically.

\subsubsection{\label{sub:nanobunch-spectrum}Analytical derivation of the nanobunch
1D CSE spectrum}

\global\long\def\cse{\textrm{CSE}}

The calculation of the spectrum is based on two assumptions:
\begin{enumerate}
\item As in the TROM model, the radiation is generated by a narrow bunch
of electrons. Optimal coherency for high frequencies will certainly
be achieved, if the current layer is infinitely narrow: $j(t,x)=j(t)\delta(x-x_{el}(t))$.
To include more realistic cases, we allow in our calculations for
a narrow, but finite electron distribution: \begin{equation}
j(t,x)=j(t)f(x-x_{el}(t))\label{eq:current-1}\end{equation}
 with variable current $j(t)$ and position $x_{el}(t)$, but fixed
shape $f(x)$.
\item In contrast to the TROM model, we give up on calculating $j(t)$ directly
from the incident radiation by the assumption of total reflection.
This means, that although we know that the bunch itself is not capable
of totally reflecting the incoming radiation and consequently there
are some additional currents inside the plasma, we do not care for
them as their contribution to the high frequency spectrum are small
compared to the contribution by the highly compressed bunch. \\
However, to get some kind of result, an assumption about the functions
$j(t)$ is required. Since we are dealing with the ultrarelativistic
regime $a_{0}\gg1$, it is reasonable to assume that changes in the
velocity components are governed by changes in the direction of motion
rather than by changes in the absolute velocity, which is constantly
very close to the speed of light $c$. We are going to see, that this
assumption is enough to obtain the spectrum.
\end{enumerate}
Following Eq.~\eqref{eq:E_reflected}, the radiation field is expressed
as $E_{\cse}\left(t,x\right)=2\pi c^{-1}$$\int j\left(t+(x-x')/c,\, x'\right)\, dx'$.
We take the Fourier transform, thereby considering the retarded time,
and arrive at the integral \begin{equation}
\tilde{E}_{\cse}(\omega)=\frac{2\pi}{c}\,\tilde{f}(\omega)\int_{-\infty}^{+\infty}j(t)\,\exp\left[-i\omega\left(t+\frac{x_{el}(t)}{c}\right)\right]\, dt,\end{equation}
wherein $\tilde{f}(\omega)$ denotes the Fourier transform of the
shape function.

In analogy to the standard synchrotron radiation by a point particle,
the integral can be solved with the method of stationary phase. Therefore,
we Taylor expand the current $j(t)$ and the electron bunch coordinate
$x_{el}(t)$ around the instant, where $\dot{x}_{el}$ is closest
to $-c$. Due to the ultrarelativistic behaviour, the current vanishes
at these instants and we write: $j(t)=\alpha_{0}t^{n}$. After the
calculations shown in appendix~\ref{sec:Stationary-Phase-Method},
the result can be expressed as \begin{equation}
\tilde{E}_{\cse}(\omega)=\tilde{f}(\omega)\,\frac{-4\pi^{2}\alpha_{0}i^{n}}{c(\alpha_{1}\omega)^{\nicefrac{n+1}{2n+1}}}\,\frac{d^{n}\gai(\xi)}{d\xi^{n}},\end{equation}

\noindent where $\gai(\xi)$ refers to a generalized Airy function,
defined in Eq.~\eqref{eq:gen_airy} and $\xi=\omega^{2n/2n+1}\delta/\alpha_{1}^{1/(2n+1)}$.

\noindent Anyway, note that high order $\gamma$-spikes ($n\gg1$)
imply, that the nanobunch remains for a comparatively long time at
low transverse currents. This appears to be unlikely here, as a static
nanobunch would not stay together for long time without magnetic fields
that can counteract the Coulomb explosion. Therefore, we go on to
discuss only the two most likely special cases $n=1$ and $n=2$:
\begin{enumerate}
\item The current changes sign at the stationary phase point and we can
Taylor expand $j(t)=\alpha_{0}\, t$. Consequently, $x_{el}(t)=-v_{0}t+\alpha_{1}t^{3}/3$.
The spectral envelope can now be written as:\begin{equation}
I(\omega)\propto|\tilde{f}(\omega)|^{2}\,\omega^{-4/3}\,\left[\textrm{Ai}'\left(\left(\frac{\omega}{\omega_{rs}}\right)^{2/3}\right)\right]^{2},\label{eq:synch_spec_1st-1}\end{equation}
where $\textrm{Ai}'$ is the Airy function derivative, $\omega_{rs}\approx2^{3/2}\sqrt{\alpha_{1}}\gamma_{0}^{3}$,
and $\gamma_{0}=(1-v_{0}^{2})^{-1/2}$ is the relativistic $\gamma$-factor
of the electron bunch at the instant when the bunch moves towards
the observer. As in the ROM models, the spectral envelope \eqref{eq:synch_spec_1st-1}
does not depend on all details of the electron bunch motion $x_{el}$,
but only on its behaviour close to the stationary points, i.e. the
$\gamma$-spikes.
\item \noindent In the case, when the current does not change sign at the
stationary phase point, we Taylor expand $j(t)=\alpha_{0}t^{2}$ and
$x_{el}(t)=-v_{0}t+\alpha_{1}t^{5}/5$. This yields to the spectral
envelope \begin{equation}
I(\omega)\propto|\tilde{f}(\omega)|^{2}\,\omega^{-6/5}\,\left[\textrm{S}''\left(\left(\frac{\omega}{\omega_{rs}}\right)^{4/5}\right)\right]^{2},\label{eq:synch_spec_2nd-1}\end{equation}
 with $\textrm{S}''$ being the second derivative of $\textrm{S}(x)\equiv\textrm{gAi}_{2}(x)=(2\pi)^{-1}$\\
$\times\int\exp\left[i\left(xt+t^{5}/5\right)\right]\, dt$, a
special case of the canonical swallowtail integral \cite{1984Swallowtail}.
For the characteristic frequency $\omega_{rs}$ we now obtain $\omega_{rs}\approx2^{5/4}\sqrt[4]{\alpha_{1}}\gamma_{0}^{2.5}$.
Because now even the derivative of $\ddot{x}_{el}$ is zero at the
stationary phase point, the influence of acceleration on the spectrum
decreases and the characteristic frequency scaling is closer to the
$\gamma^{2}$-scaling for a mirror moving with constant velocity.
\end{enumerate}
\begin{figure}
\begin{centering}
\includegraphics[width=3in]{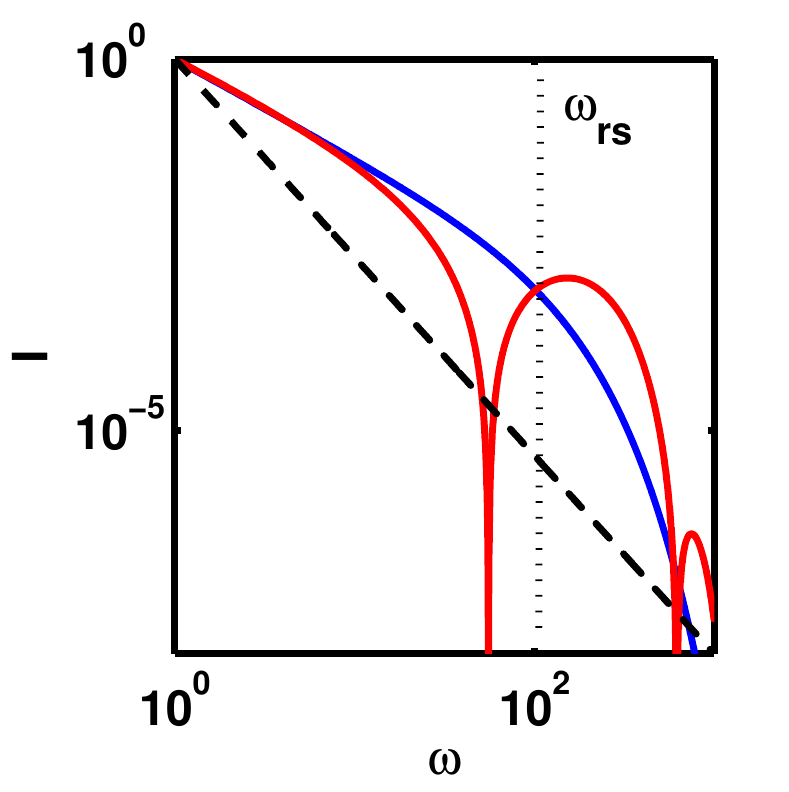}
\par\end{centering}

\caption{\selectlanguage{english}%
\label{fig:CSE_spectra-1}\foreignlanguage{british}{Coherent 1D synchrotron
spectra for an infinitely thin electron layer $\tilde{f}(\omega)\equiv1$
and $\omega_{rs}=100$. The blue line corresponds to Eq.~\eqref{eq:synch_spec_1st-1}
and the red line to Eq.~\eqref{eq:synch_spec_2nd-1}. For comparison,
the dashed black line denotes the BGP $8/3$-power law.}\selectlanguage{british}
}
\end{figure}

In Fig.~\ref{fig:CSE_spectra-1} the CSE spectra of the synchrotron
radiation from the electron sheets are depicted. Comparing them to
the $8/3$-power law from the BGP-case, we notice that, because of
the smaller exponents of their power law part, the CSE spectra are
much flatter. E.g., around the 100th harmonic we win more than two
orders of magnitude. Note that, as in the case of higher order $\gamma$-spikes
in the ROM model, side maxima are found in the spectrum \eqref{eq:synch_spec_2nd-1}.
This might provide an explanation for modulations that are occasionally
observed in harmonics spectra, compare Ref.~\cite{boyd:125004,pukhov2009relativistic2}.

\begin{figure}
\begin{centering}
\includegraphics[width=4in]{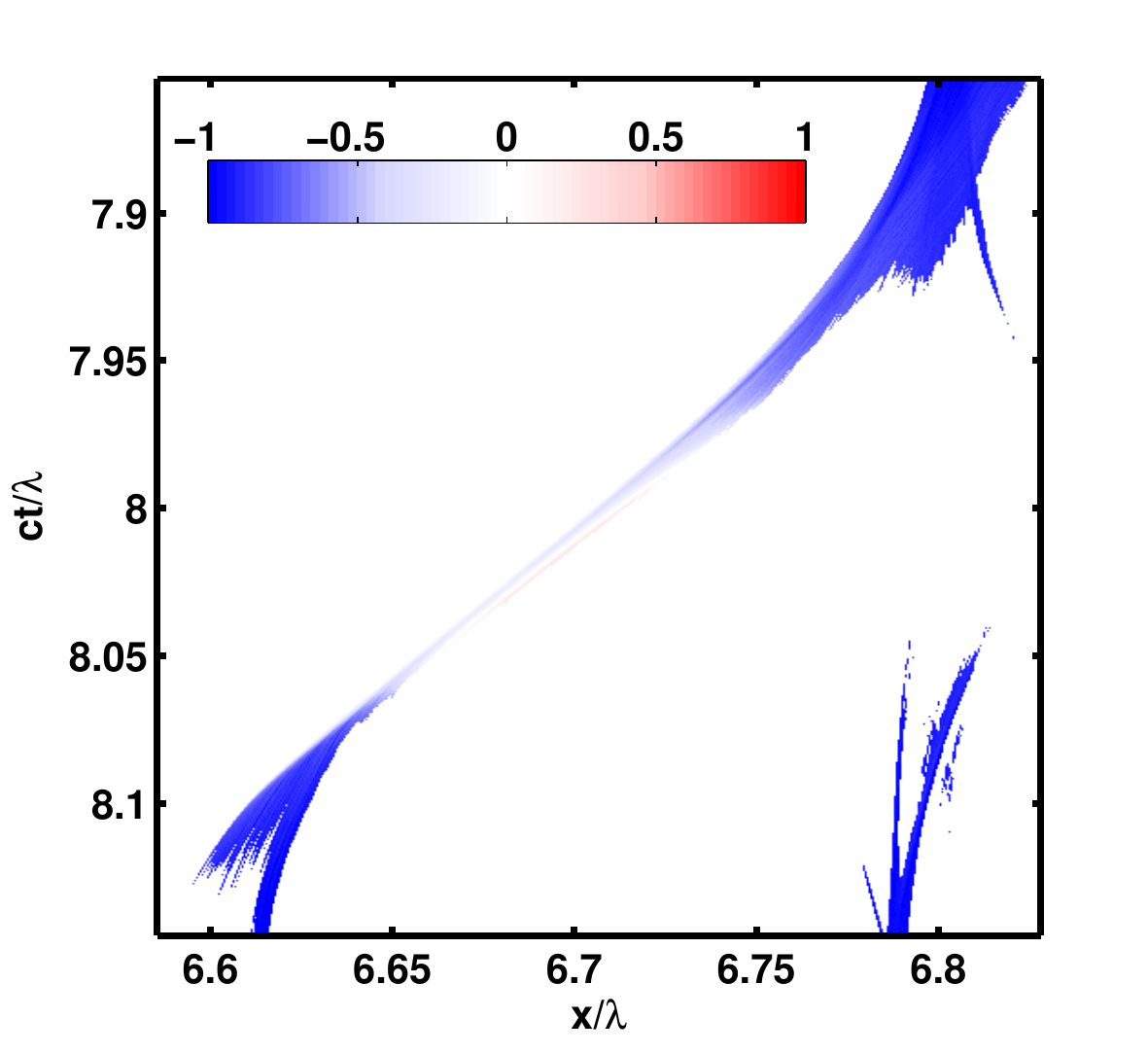}
\par\end{centering}

\caption{\selectlanguage{english}%
\label{fig:vy}\foreignlanguage{british}{Normalized transverse fluid
velocity $v_{y}/c$ of the electron nanobunch. Parts of the plasma
with a density below $500\, N_{c}$ are filtered out.}\selectlanguage{british}
}
\end{figure}

To compare the analytically obtained spectrum with the PIC result,
the finite size of the electron bunch must be taken into account.
We assume a Gaussian density profile which leads us to

\begin{equation}
|\tilde{f}(\omega)|^{2}=\exp\left[-\left(\frac{\omega}{\omega_{rf}}\right)^{2}\right].\label{eq:gaussian_shape}\end{equation}

Thus the spectral cut-off is determined either by $\omega_{rs}$,
corresponding to the relativistic $\gamma$-factor of the electrons,
or by $\omega_{rf}$ corresponding to the bunch width. A look at the
motion of the electron nanobunch in the PIC simulation (Fig.~\ref{fig:vy})
tells us that there is no change in sign of the transverse velocity
at the stationary phase point, consequently we use Eq.~\eqref{eq:synch_spec_2nd-1}.
We choose $\omega_{rf}=225\,\omega_{0}$ and $\omega_{rs}=800\,\omega_{0}$
to fit the PIC spectrum, corresponding to a Gaussian electron bunch
$f(x)=\exp\left[-(x/\delta)^{2}\right]$ with a width of $\delta=10^{-3}\lambda$
and an energy of $\gamma\sim10$. This matches well with the measured
electron bunch width $\delta_{\text{FWHM}}=0.0015\,\lambda$ {[}see
Fig.~\ref{fig:cse-generation}(b){]} and the laser amplitude $a_{0}=60$,
since we expect $\gamma$ to be smaller but in the same order of magnitude
as $a_{0}$. In this case $\omega_{rf}<\omega_{rs}$, so the cut-off
is dominated by the finite bunch width. Still, both values are in
the same order of magnitude, so that the factor coming from the Swallowtail-function
cannot be neglected and actually contributes to the shape of the cut-off.
The modulations that appear in Fig.~\ref{fig:CSE_spectra-1} for
frequencies around $\omega_{rs}$ and above cannot be seen in the
spectra, because it is suppressed by the Gauss-function Eq.~\eqref{eq:gaussian_shape}.
The analytical synchrotron spectrum agrees excellently with the PIC
result, as the reader may verify in Fig.~\ref{fig:CSE-new_model_time_spectra}(b).

\subsubsection{\label{sub:nanobunch-sensitivity}Sensitivity of the nanobunching
process to parametric changes}

Now, we have a look at the dependence of the harmonics radiation in
and close to the nanobunching regime on the laser and plasma parameters.
Exemplary, the laser intensity and the pre-plasma scale length are
varied here. The pulse duration however will be left constantly short,
so that we can simply focus our interest on the main optical cycle.
For longer pulses, the extent of nanobunching may vary from one optical
cycle to another, which makes a parametric study more difficult. We
are going to examine two dimensionless key quantities: the intensity
boost $\eta\equiv\max(E_{r}^{2})/\max(E_{i}^{2})$ and the pulse compression
$\Gamma\equiv(\omega_{0}\tau)^{-1}$. It is straightforward to extract
both magnitudes from the PIC data, and both are quite telling. The
intensity boost $\eta$ is a sign of the mechanism of harmonics generation.
If the ARP boundary condition Eq.~\eqref{eq: ARP_ansatz} is approximately
valid, we must of course have $\eta\approx1$. Then again, if the
radiation is generated by nanobunches, we expect to see strongly pronounced
attosecond peaks {[}see Eq.~\eqref{eq:atto_peak}{]} in the reflected
radiation and therefore $\eta\gg1$. The pulse compression $\Gamma$
is defined as the inverse of the attosecond pulse duration. In the
nanobunching regime, we expect it to be roughly proportional to $\eta$,
as the total efficiency of the attosecond pulse generation remains
$\eta_{\text{atto}}\lesssim1$, compare Eq.~\eqref{eq:cse_eff}.
In the BGP regime, there are no attosecond pulses observed without
spectral filtering. So the FWHM of the intensity peak is on the order
of a quarter laser period, and we expect $\Gamma\sim1$.

\begin{figure}
\begin{centering}
\includegraphics[width=4in]{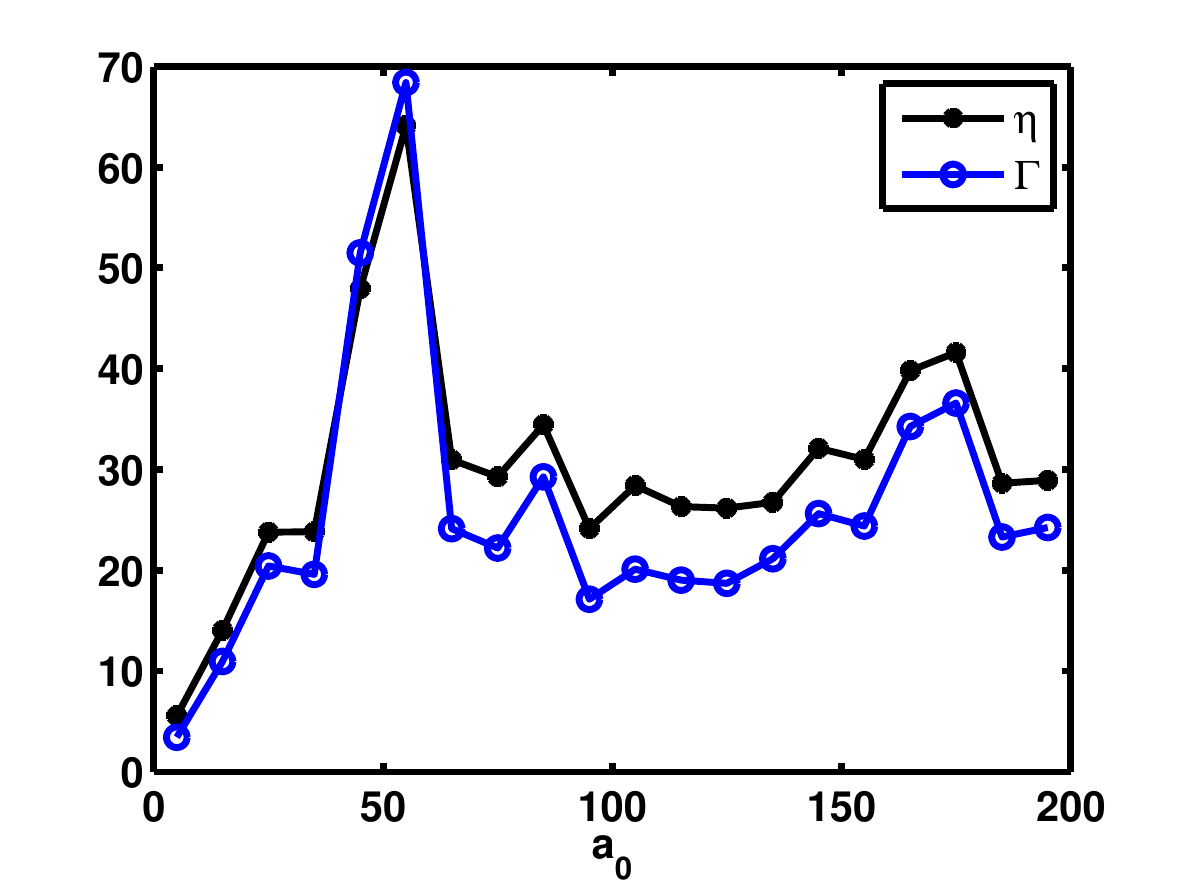}
\par\end{centering}

\caption{\selectlanguage{english}%
\label{fig:new_model_019_eta}\foreignlanguage{british}{Dependence
of the intensity boost $\eta=\max(E_{r}^{2})/\max(E_{i}^{2})$ and
the pulse compression $\Gamma=(\omega_{0}\tau)^{-1}$, where $\tau$
is the FWHM width of the attosecond intensity peak in the reflected
radiation, on $a_{0}$. The laser amplitude $a_{0}$ is varied between
5 and 195 in steps of 10. Other parameters are the same as in Fig.~\ref{fig:cse-generation}b.}\selectlanguage{british}
}
\end{figure}

In figure~\ref{fig:new_model_019_eta} the two parameters $\eta$
and $\Gamma$ are shown in dependence of $a_{0}$. Except for the
variation of $a_{0}$, the parameters chosen are the same as in Figs.~\ref{fig:CSE-new_model_time_spectra},
\ref{fig:cse-generation}(b) and \ref{fig:vy}.

First of all we notice, that for all simulations in this series with
$a_{0}\gg1$, we find $\eta\gg1$. Thus, Eq.~\eqref{eq: ARP_ansatz}
is violated in all cases. Since we also notice $\Gamma\gg1$ and $\Gamma\sim\eta$,
we know, that the radiation is emitted in the shape of attosecond
peaks with an efficiency of the order 1. This indicates, that we can
describe the radiation as CSE. The perhaps most intriguing feature
of Fig.~\ref{fig:new_model_019_eta} is the strongly pronounced peak
of both curves around $a_{0}=55$. We think that because of some very
special phase matching between the turning point of the electron bunch
and of the electromagnetic wave, the electron bunch experiences an
unusually high compression at this parameter settings. This is the
case that was introduced in subsection~\ref{sub:nanobunching-process}.

\begin{figure}
\begin{centering}
\includegraphics[width=4in]{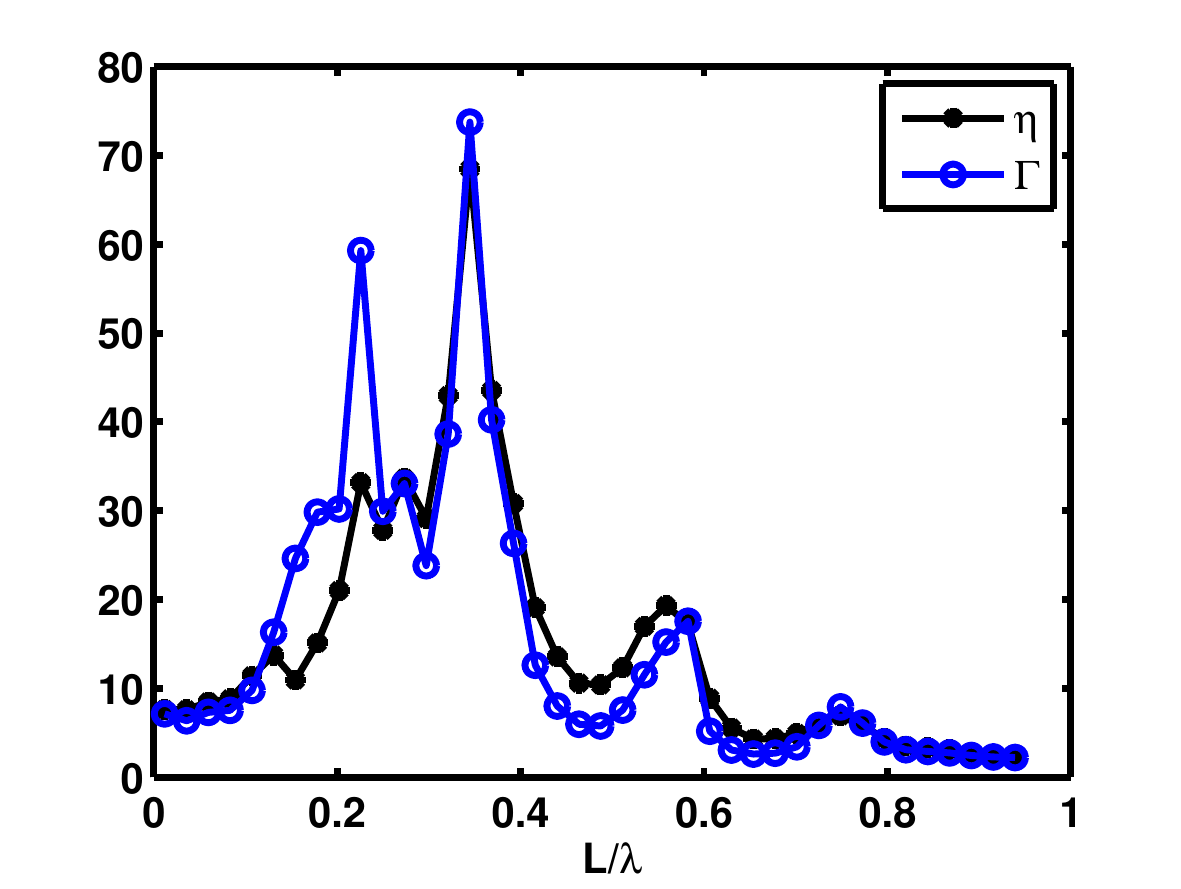}
\par\end{centering}

\caption{\selectlanguage{english}%
\label{fig:new_model_024_eta}\foreignlanguage{british}{Dependence
of the intensity boost $\eta=\max(E_{r}^{2})/\max(E_{i}^{2})$ and
the pulse compression $\Gamma=(\omega_{0}\tau)^{-1}$, on the plasma
scale length in units of the laser wavelength $L/\lambda$ in the
lab frame. Except for the plasma scale length, parameters are the
same as in Fig.~\ref{fig:cse-generation}b. The plasma ramp is again
an exponential one $\propto\exp(x/L)$.}\selectlanguage{british}
}
\end{figure}

Figure~\ref{fig:new_model_024_eta} shows the two parameters $\eta$
and $\Gamma$ as functions of the plasma gradient scale length $L$.
It is seen that both functions possess several local maxima. Further,
$\eta$ and $\Gamma$ behave similar apart from one runaway value
at $L=0.225\lambda$, where the FWHM peak duration is extremely short,
but the intensity boost is not as high. A look at the actual field
data tells us that in this case the pedestal of the attosecond peak
is broader, consuming most of the energy. This deviation might e.g.
be caused by a different, non-Gaussian shape of the electron nanobunch.

The maximum of both functions lies around $L=0.33\lambda$, the parameter
setting analyzed in detail before. In the limit of extremely small
scale lengths $L\lesssim0.1\lambda$, $\eta$ and $\Gamma$ become
smaller, but they remain clearly bigger than one. Thus the reflection
in this parameter range can still not very well be described by the
ARP boundary condition. For longer scale lengths $L>0.8\lambda$,
both key values approach 1, so the ARP boundary condition can be applied
here. This is a possible explanation for why the BGP spectrum \eqref{eq:BGP_spec}
could experimentally be measured at oblique incidence \cite{Dromey2006}.

\subsection{\label{sub:forward_harmonics}Harmonics emission in forward direction}

Up to now, we have discussed the harmonics emitted at the front side
of an overdense foil, propagating in backward direction together with
the reflected light. If however the foil used for HHG is sufficiently
thin, harmonics are also emitted in the forward direction, albeit
to a weaker extent \cite{u.2004harmonic,r.1997highorder,p.1997plasmadensity}.

Two main mechanisms \cite{george2009mechanisms} can be made responsible
for the production of harmonics at the back side of the foil:
\begin{enumerate}
\item Fast (Brunel) electrons which are transmitted through the foil may
trigger the emission of harmonics not only at the front side of the
foil, but also at its backside. Here, harmonics are produced up to
the maximum plasma frequency as is characteristic for the sub-relativistic
regime. This kind of radiation depends strongly on the properties
of the density gradient on the backside. For a too steep gradient,
only the very weak transition radiation is generated, whereas for
the right scale length, harmonics are observed due to the CWE mechanism.
\item Compressed electron bunches at the front side can emit high frequency
synchrotron radiation in both directions. Radiation with frequencies
above the maximum plasma frequency is transmitted through the foil,
resulting in forward emission of high harmonic orders. We focus on
this second mechanism here, as it is dominant in the relativistic
regime. Obviously, the {}``oscillating mirror'' model has no business
here, since we are talking about the transmitted, not the reflected
part of the light.\\
Let us also note, that the characteristics of the harmonics emitted
in the forward direction may well be different from the ones emitted
in backward direction, as two different $\gamma$-spikes are responsible
for their generation. Whereas high frequencies in the backward direction
are emitted, when the electrons are moving with maximum velocity away
from the surface, they are emitted in forward direction when the electrons
are moving with maximum velocity towards the surface. Thus the generation
efficiency of forward harmonics depends on the existence of such a
$\gamma$-spike, and on the compression of the electron bunch during
this instance.
\end{enumerate}
\begin{figure}
\begin{centering}
\includegraphics[width=4in]{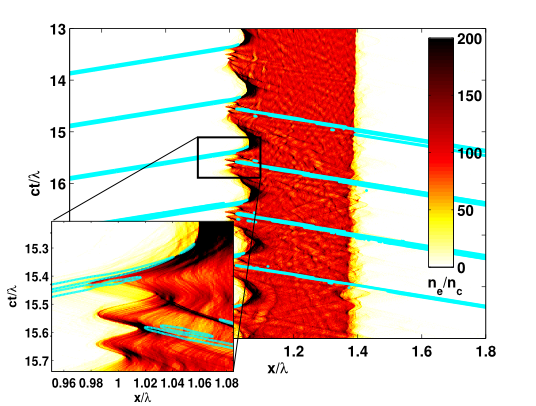}\caption{\label{fig:forward_001_generation}Electron density and contour lines
(cyan) of the emitted high frequency radiation ($\omega>\omega_{p}$).
Relativistic harmonics emission is observed in both forward and backward
direction, both generated at the front side. Simulation parameters
are: laser amplitude $a_{0}=10$, $45^{\circ}$ p-polarized incidence,
maximum plasma density: $n_{e}=100\, n_{c}$, exponential density
profile at the front side with scale length $L=0.06\,\lambda$ and
steep edge at the rear side.}

\par\end{centering}

\end{figure}

An example of the second mechanism is shown in Fig.~\ref{fig:forward_001_generation}.
It is seen, that in this case both forward and backward harmonics
are generated at the front side of the target. Further we observe
(see detail in Fig.~\ref{fig:forward_001_generation}) that they
are generated by two distinct electron bunches. The backward harmonics
are mainly emitted by a bunch propagating away from the surface and
becoming dispersed when returning into the plasma. In contrast, the
forward harmonics are generated by a second bunch that achieves its
highest compression when travelling into the plasma.

\begin{figure}
\centering{}\includegraphics[width=4in]{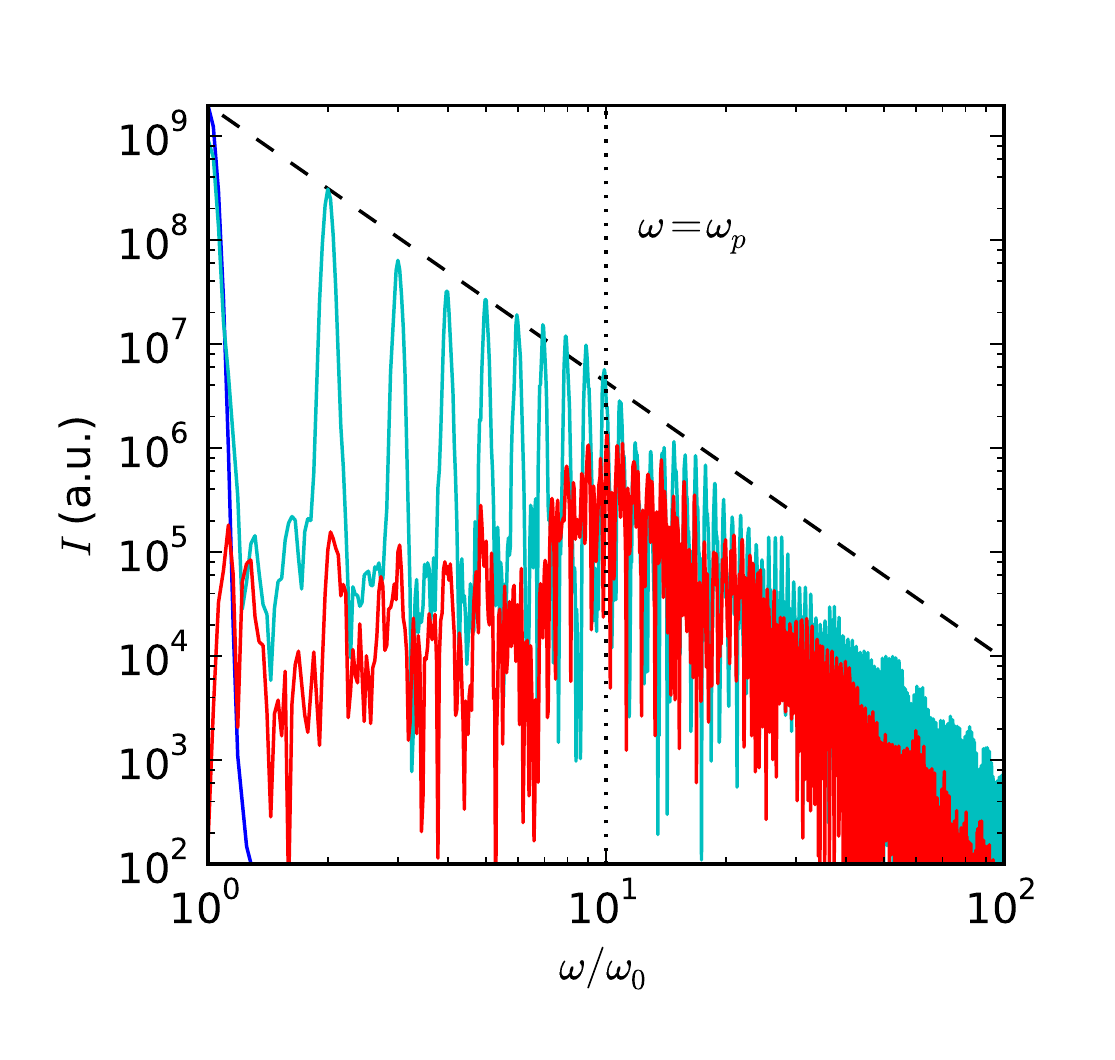}\caption{\label{fig:forward_001_spectra}Spectra of radiation emitted in forward
(red) and backward (cyan) direction, taken from the same simulation
as Fig.~\ref{fig:forward_001_generation}. The spectrum of the incident
laser (blue), the plasma frequency (black dotted) and an 8/3-power
law (black dashed) are also charted for reference.}
\end{figure}

Fig.~\ref{fig:forward_001_spectra} displays the spectra of the radiation
emitted in forward and backward direction from the same simulation
run. We notice, that the spectrum of the forward harmonics does not
contain much radiation at low frequencies $\omega<\omega_{p}$. This
is easily understood due to the fact that the harmonics generated
at the front side have to propagate through the plasma. Harmonics
generation at the rear side due was suppressed in our simulation by
the sharp density edge at this side. Further it is seen that the spectra
of forward and backward emitted harmonics fall off at different rates.
This is no surprise, since they are generated at different instants
and even by different electron bunches, as seen in the detail graph
inside Fig.~\ref{fig:forward_001_generation}. At their respective
$\gamma$-spikes, they possess distinct compression, shapes and energy,
leading to different spectral slopes.

\section{\label{sec:Relativistic-Harmonics-Radiation}Relativistic Harmonics
Radiation as Attosecond Pulse Train}

\subsection{\label{sub:attopulse-characterization}Characterization of the generated
attosecond pulses}

As the calculations above have shown, the emitted harmonics are phase
locked. Therefore, they form a train of extremely short pulses. The
shortest possible pulse duration can be estimated by $T\sim1/\omega_{r}$,
wherein $\omega_{r}$ is the characteristic roll-off frequency of
the harmonics generation process. If the spectral roll-off is due
to the relativistic $\gamma$-factor and the $\gamma$-spike is of
the first order, this means that\begin{equation}
T\sim\frac{1}{\omega_{0}\gamma^{3}}.\label{eq:attopulse_duration_estimate}\end{equation}

Thus for the typical values of $\gamma$, the achievable pulse duration
is in the order of a few attoseconds or even less. Note that the cubic
scaling in $\gamma$ exceeds the possible pulse compression by the
simple Doppler effect, which yields a duration not shorter than $T_{\textrm{Doppler}}\sim1/(4\gamma^{2})$.

The $1/\gamma^{3}$-scaling can physically be understood by having
another look at a characteristic $\gamma$-spikes. For the ROM-process,
this is $\gamma_{\arp}(t)$ as defined in Sec.~\ref{sub:spectrum-ARP}
and for the coherent synchrotron emission this is the $\gamma$-factor
corresponding to the longitudinal bunch velocity component $x_{el}(t)$.
Assuming a $\gamma$-spike of first order here, we have $v(t)\approx v_{0}-\alpha\omega_{0}^{2}t^{2}$
around the maximum. Consequently, the $\gamma$-factor can be written
as\begin{equation}
\gamma(t)\approx\frac{\gamma_{0}}{\sqrt{1+\gamma_{0}^{2}\alpha\omega_{0}^{2}t^{2}}}.\end{equation}

Evaluating the temporal width of the spike in $\gamma(t)$ at $t=0$
yields $\Delta t\sim1/(\omega_{0}\gamma_{0}^{3}\alpha^{1/2})$. Since
the high order harmonics are produced only during the $\gamma$-spike,
the duration of the corresponding attosecond pulses are in the same
order of magnitude, in agreement with \eqref{eq:attopulse_duration_estimate}.

\subsubsection{Attosecond pulses from ROM harmonics}

\global\long\def\high{\mathrm{high}}
\global\long\def\low{\mathrm{low}}
\global\long\def\atto{\mathrm{atto}}

In order to unravel the attosecond pulses contained in the ROM harmonics
radiation, it is required to filter out the lower harmonic orders.
The high-frequency cutoff of the power-law spectrum defines the shortest
pulse duration that can be achieved this way.

Assuming that the harmonics are emitted coherently and in phase, we
expect their duration $\tau_{\atto}$ to be roughly the inverse of
the absolute spectral width (ASW). So as to estimate the pulse duration
achieved by a certain filter we therefore calculate the ASW $\Delta\omega\equiv\left(\left\langle \omega^{2}\right\rangle -\left\langle \omega\right\rangle ^{2}\right)^{1/2}$.
This can be done analytically for a typical BGP spectrum. The spectral
high-pass filter can be introduced as a sharp low-frequency cutoff
beyond $\omega_{\low}$. We write $I(\omega)=I_{0}\,\omega^{-8/3}\,\exp(-\omega/\omega_{r})\,\theta(\omega-\omega_{\low})$,
substituting the Airy function in Eq.~\eqref{eq:BGP_spec} by an
exponential for the sake of simplicity. As shown in Ref.~\cite{bgp-theory2006},
this is a reasonable approximation here. Then we obtain for the ASW:\begin{equation}
\left(\Delta\omega\right)^{2}=\omega_{r}^{2}\,\frac{\Gamma(-\frac{5}{3};\, x)\Gamma(\frac{1}{3};\, x)-\left[\Gamma(-\frac{2}{3};\, x)\right]^{2}}{\left[\Gamma(-\frac{5}{3};\, x)\right]^{2}}\label{eq:asw_analytic}\end{equation}

wherein $\Gamma(s;\, x)\equiv\int_{x}^{\infty}t^{s-1}\exp(-t)\, dt$
is the upper incomplete gamma-function and $x\equiv\omega_{\low}/\omega_{r}$.

\begin{figure}
\centering{}\includegraphics[width=3in]{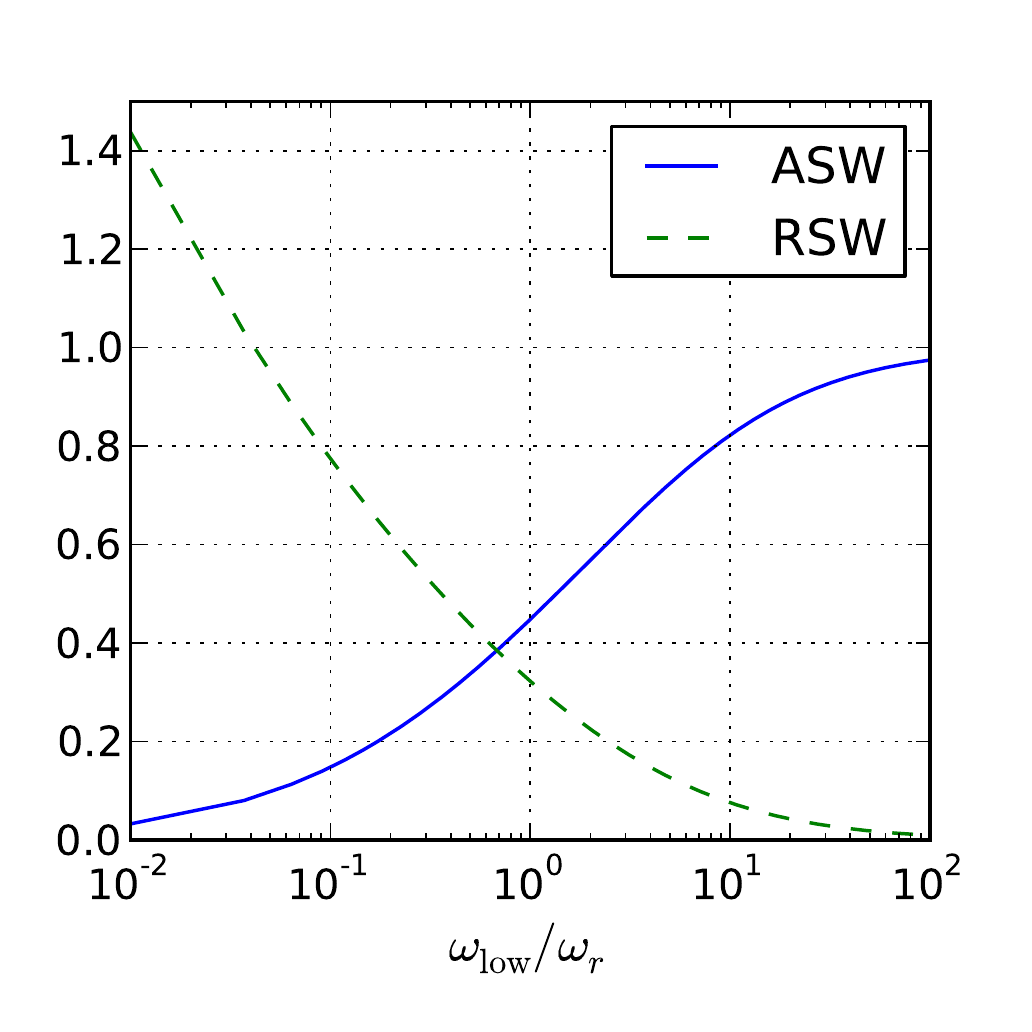}\caption{\label{fig:relative_spectal_width}Absolute spectral width $\mbox{ASW}\equiv\Delta\omega/\omega_{r}$
normalized to $\omega_{r}$ and relative spectral width $\mbox{RSW}\equiv\Delta\omega/\left\langle \omega\right\rangle $
of attosecond pulses from a high-pass filtered BGP spectrum $I\propto\omega^{-8/3}\,\exp(-\omega/\omega_{r})\,\theta(\omega-\omega_{\low})$.}
\end{figure}

The function, normalized to $\omega_{r}$, is plotted in Fig.~\ref{fig:relative_spectal_width}.
For very low filter frequencies $\omega_{\low}\ll\omega_{r}$, the
ASW is still much smaller than $\omega_{r}$, consequently the observed
attosecond pulses are not as short as they can be according to Eq.~\eqref{eq:attopulse_duration_estimate}.
If the filter frequency is increased up to $\omega_{r}$ or even beyond,
the ASW tends towards $\Delta\omega\rightarrow\omega_{r}$, therefore
the generated pulses approach the duration given by Eq.~\eqref{eq:attopulse_duration_estimate}.

Another interesting property can be extracted solely by looking at
the spectrum. If the filter frequency is well below the critical roll-off
frequency $\omega_{\low}\ll\omega_{r}$, the power law part of the
spectrum dominates, resulting in a large relative spectral width (RSW)
$\Delta\omega/\omega\gtrsim1$. Consequently, single-cycle pulses
are to be expected. If on the other hand $\omega_{\low}\gg\omega_{r}$,
the spectrum decays exponentially above the filter frequency, leading
to a relatively small RSW $\Delta\omega/\omega\ll1$. Analytically
we obtain for a BGP spectrum:\begin{eqnarray}
\left(\frac{\Delta\omega}{\left\langle \omega\right\rangle }\right)^{2} & = & \frac{\Gamma(-\frac{5}{3};\, x)\,\Gamma(\frac{1}{3};\, x)}{\left[\Gamma(-\frac{2}{3};\, x)\right]^{2}}-1.\label{eq:rsw_analytic}\end{eqnarray}

Again, the function is plotted in Fig.~\ref{fig:relative_spectal_width}.
It is seen that $\Delta\omega/\omega\rightarrow0$ for $x\rightarrow\infty$,
so for high filter frequencies the spectrum becomes increasingly monochromatic,
yielding many-cycle pulses. For comparatively low filter frequencies
$x\rightarrow0$ we find the asymptotic approximation $\left(\Delta\omega/\omega\right)^{2}\approx4\Gamma(\tfrac{1}{3})/(15\, x^{1/3})-1$,
leading to an increasing relative spectral width in agreement with
our previous thoughts.

\begin{figure}
\begin{centering}
\includegraphics[width=5in]{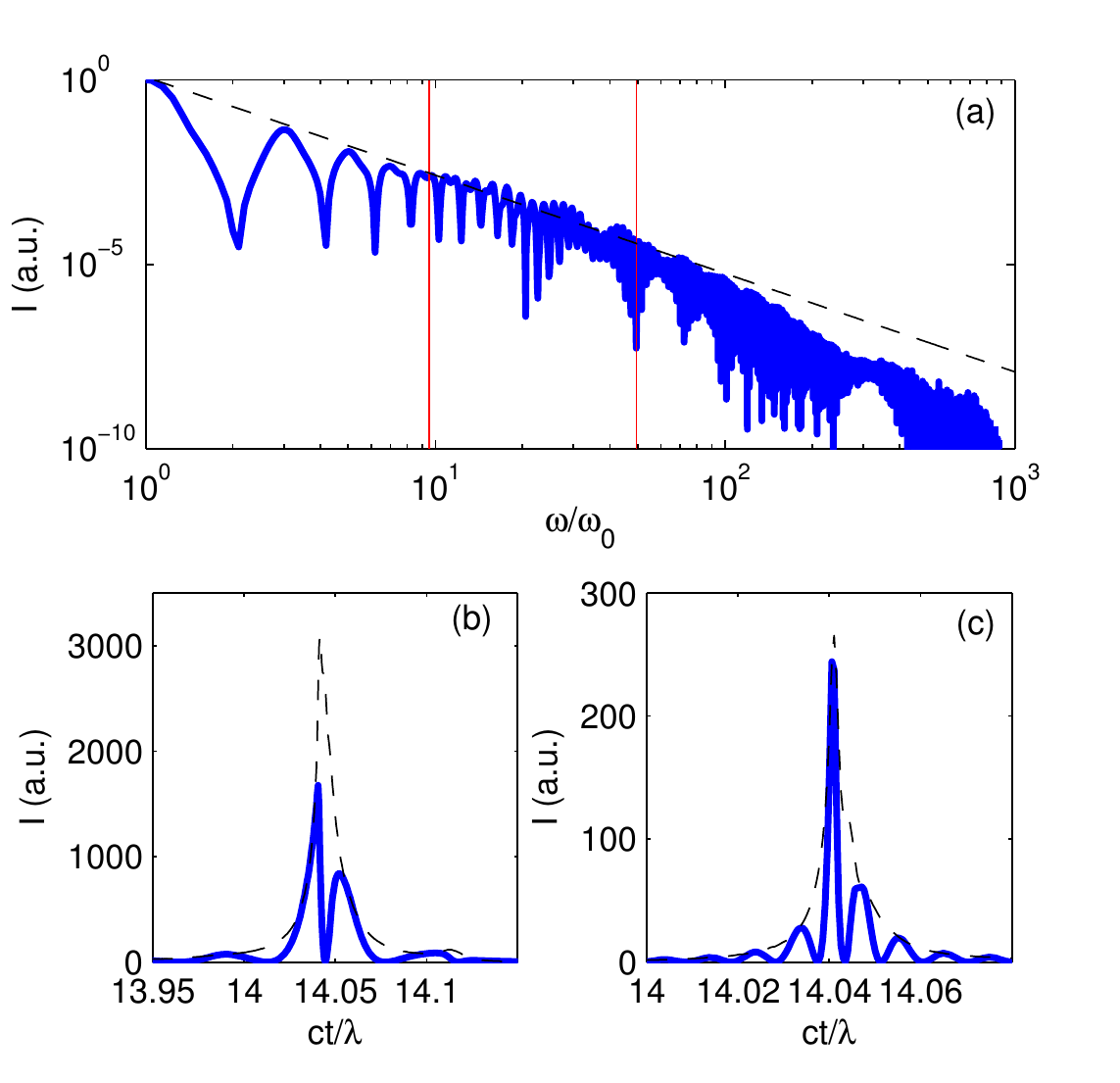}
\par\end{centering}

\caption{\label{fig:attopulse_filtering}Attosecond pulses in ROM harmonics
radiation can be revealed by spectral high-pass filtering. (a) shows
the spectrum of the radiation in a logarithmic representation, along
with the used filter frequencies as red lines, the black dashed line
represent the filter frequencies. (b) and (c) display the resulting
attosecond pulses for the filter frequencies $\omega_{\low}=9.5\,\omega_{0}$
respectively $\omega_{\low}=49.5\,\omega_{0}$. Here, the black dashed
line denote an envelope computed as the absolute square of the corresponding
analytic signal.}
\end{figure}

These analytical results compare well to the numerical ones shown
in Fig.~\ref{fig:attopulse_filtering}. For $\omega_{\low}=9.5\,\omega_{0}\approx0.1\,\omega_{r}$
the pulse duration is approximately $\tau_{\atto}\approx0.05\,\lambda/c=5\,\tau_{r}$,
where $\tau_{r}\equiv2\pi/\omega_{r}=0.01\,\lambda/c$. This agrees
reasonably with what can be expected from Eq.~\eqref{eq:asw_analytic}:
$\Delta\omega\approx0.15\,\omega_{r}$ for $x=0.1$. Further notice
that the pulse is single-cycle, in agreement with the large RSW $\Delta\omega/\left\langle \omega\right\rangle \approx0.77$,
as given by Eq.~\eqref{eq:rsw_analytic}.

Things look differently for the higher filter frequency $\omega_{\low}=49.5\,\omega_{0}\approx0.5\,\omega_{r}$.
Now the pulse duration is about $\tau_{\atto}\approx0.02\,\lambda/c=2\,\tau_{r}$,
roughly a little more than twice as short as before. This is in agreement
with $\Delta\omega\approx0.34\,\omega_{r}$ for $x=0.5$ from Eq.~\eqref{eq:asw_analytic}.
We also see that the pulse now contains slightly more than one optical
cycle, indicating the somewhat smaller RSW $\Delta\omega/\left\langle \omega\right\rangle \approx0.44$
given by Eq.~\eqref{eq:rsw_analytic}.

\subsubsection{\label{sub:nanobunch-radiation-properties}Attosecond pulses from
electron nanobunches}

As we see in Fig.~\ref{fig:CSE-new_model_time_spectra}(a), the CSE
radiation is emitted in the form of a single attosecond pulse whose
amplitude is significantly higher than that of the incident pulse.
This pulse has a FWHM duration of $0.003$ laser periods, i.e. $9\,\text{as}$
for a laser wavelength of $800\,\text{nm}$. This is very different
from emission of the ROM harmonics, which need to undergo diffraction
(see also Sec.~\ref{sec:3d_harmonics}) or spectral filtering \cite{bgp-theory2006}
before they take on the shape of attosecond pulses.

When we apply a spectral filter in a frequency range $\left(\omega_{\mathrm{low}},\omega_{{\scriptstyle \mathrm{high}}}\right)$
to a power-law harmonic spectrum with an exponent $q$, so that $I(\omega)=I_{0}(\omega_{0}/\omega)^{q}$,
the energy efficiency of the resulting attosecond pulse generation
process is

\begin{eqnarray}
\eta_{\text{atto}} & = & \int_{\omega_{\mathrm{low}}}^{\omega_{\mathrm{high}}}I(\omega)\, d\omega\nonumber \\
 & = & \frac{I_{0}\omega_{0}}{q-1}\left[\left(\frac{\omega_{0}}{\omega_{\text{low}}}\right)^{q-1}-\left(\frac{\omega_{0}}{\omega_{\text{high}}}\right)^{q-1}\right]\label{eq:cse_eff}\end{eqnarray}
 The scaling \eqref{eq:cse_eff} gives $\eta_{\text{atto}}^{\text{ROM}}\sim(\omega_{0}/\omega_{\text{low}})^{5/3}$
for the BGP spectrum with $q=8/3$. For unfiltered CSE harmonics with
the spectrum \textit{$q=4/3$} the efficiency is close to $\eta_{\text{atto}}^{\text{CSE}}=1$.
This means that almost the whole energy of the original optical cycle
is concentrated in the attosecond pulse. Note that absorption is very
small in the PIC simulations shown; it amounts to 5\% in the run corresponding
to Fig.~\ref{fig:CSE-new_model_time_spectra} and is even less in
the run corresponding to Fig.~\ref{fig:ROM-new_model_time_spectra}.

The ROM harmonics can be considered as a perturbation in the reflected
signal as most of the pulse energy remains in the fundamental. On
the contrary, the CSE harmonics consume most of the laser pulse energy.
This is nicely seen in the spectral intensity of the reflected fundamental
for the both cases {[}compare Figs.~\ref{fig:ROM-new_model_time_spectra}(b)
and \ref{fig:CSE-new_model_time_spectra}(b){]}. As the absorption
is negligible, the energy losses at the fundamental frequency can
be explained solely by the energy transfer to high harmonics. We can
roughly estimate this effect by $I_{0}^{BGP}/I_{0}^{CSE}\approx\int_{1}^{\infty}\omega^{-8/3}\, d\omega/\int_{1}^{\infty}\omega^{-4/3}\, d\omega=5$.
This value is quite close to the one from the PIC simulations: $I_{0}^{(\text{Fig. 1})}/I_{0}^{(\text{Fig. 5})}=3.7$.

Further, we can estimate amplitude of the CSE attosecond pulse analytically
from the spectrum. Since the harmonic phases are locked, for an arbitrary
power law spectrum $I(\omega)\propto\omega^{-q}$ and a spectral filter
$\left(\omega_{\mathrm{low}},\omega_{{\scriptstyle \mathrm{high}}}\right)$
we integrate the amplitude spectrum and obtain:

\begin{eqnarray}
E_{\text{atto}} & \approx & \frac{2\sqrt{\left.I\right|_{\omega=\omega_{1}}}}{q-2}\left[\left(\frac{\omega_{0}}{\omega_{\text{low}}}\right)^{\frac{q}{2}-1}-\left(\frac{\omega_{0}}{\omega_{\text{high}}}\right)^{\frac{q}{2}-1}\right]\label{eq:atto_amplitude}\end{eqnarray}
 Apparently, when the harmonic spectrum is steep, i.e. \textit{$q>2$},
the radiation is dominated by the lower harmonics $\omega_{\mathrm{low}}$.
This is the case of the BGP spectrum \textit{$q=8/3$}. That is why
one needs a spectral filter to extract the attosecond pulses here.
The situation changes drastically for slowly decaying spectra with
\textit{$q<2$} like the CSE spectrum with $q=4/3$. In this case,
the radiation is dominated by the high harmonics $\omega_{{\scriptstyle \mathrm{high}}}$.
Even without any spectral filtering the radiation takes on the shape
of an attosecond pulse. As a rule of thumb formula for the attosecond
peak field of the unfiltered CSE radiation we can write: \begin{equation}
E_{\text{atto}}^{\text{CSE}}\approx\sqrt{3}\left(m_{c}^{1/3}-1\right)E_{0}\label{eq:atto_peak}\end{equation}
 Using $m_{c}=\omega_{c}/\omega_{0}=225$, the lower of the two cut-off
harmonic numbers used for comparison with the PIC spectrum in Fig.~\ref{fig:CSE-new_model_time_spectra}(b),
we obtain $E_{\text{peak}}=8.8\, E_{0}$. This is in nice agreement
with Fig.~\ref{fig:CSE-new_model_time_spectra}(a).

\subsection{Isolation of single attosecond pulses by polarization gating}

Many applications in the field of imaging and control of quantum dynamics
on the attosecond timescale \cite{s.2010attosecond,ferenc2009attosecond}
require single attosecond pulse instead of a pulse train. The single
pulse can in principle be produced using a phase-stabilized single
cycle laser. However, relativistic harmonics require a laser pulse
intensity $I\gg10^{18}\,\mathrm{W/cm^{2}}$ and pulses in this intensity
range usually are several cycles long, leading to the production of
longer attosecond pulse trains. Therefore, we are in need of a method
to isolate single attosecond pulses from the pulse train.

\begin{figure}
\begin{centering}
\includegraphics[width=4in]{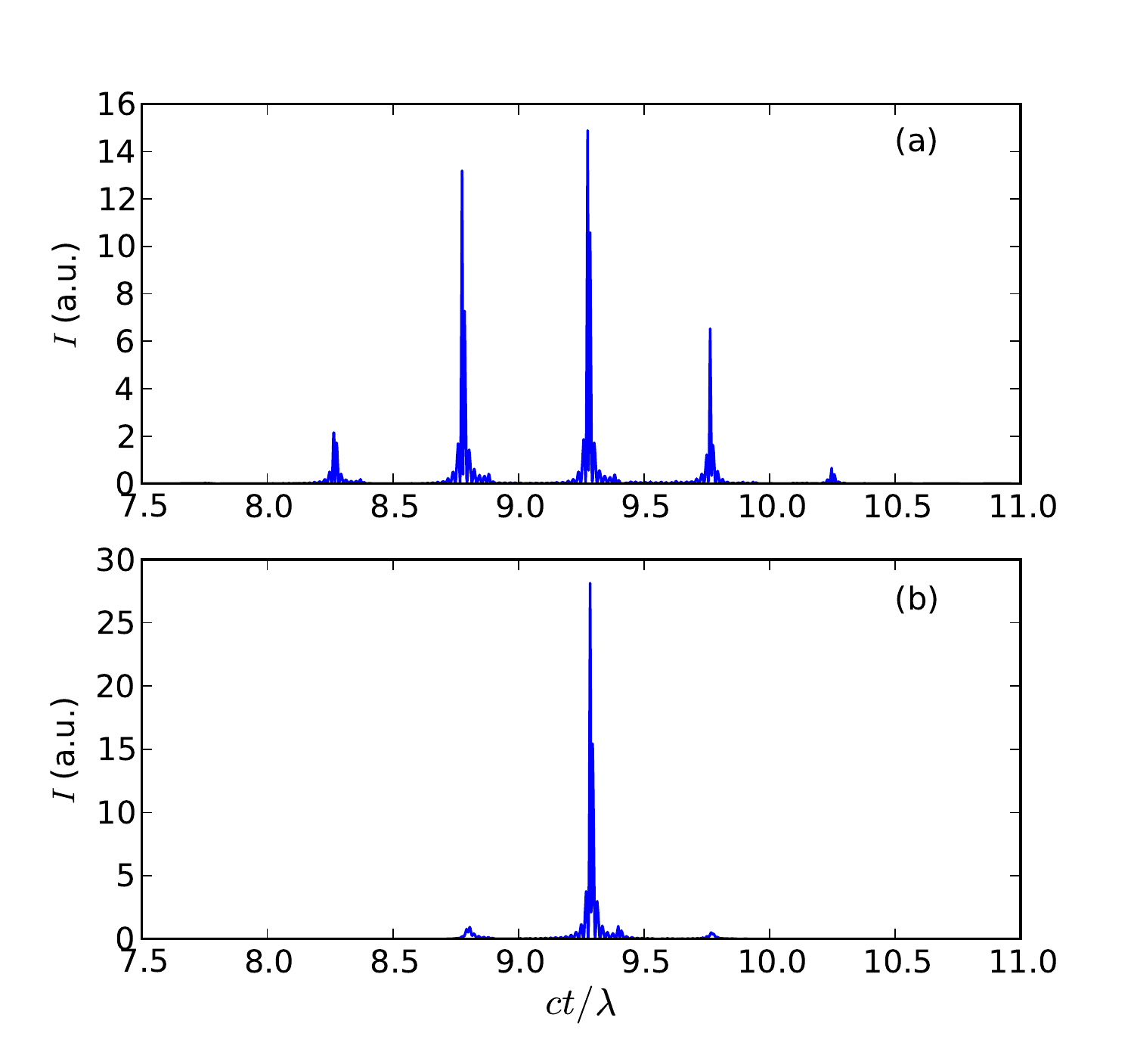}\caption{\label{fig:polarization_gating}Attosecond pulse train (a) without
and (b) with the use of polarization gating technique. Parameters
are: $a_{0}=20$, plasma density $N_{e}=90\, N_{c}$ with a sharply
defined surface. In (b) there is a second pulse with amplitude $a_{1}=6$
and polarization direction orthogonal to the first one, detuned to
$\omega_{1}=1.25\,\omega_{0}$ and dephased by $\Delta\phi=\pi/8$. }

\par\end{centering}

\end{figure}

It was shown above that the attosecond pulse are emitted when the
tangential components of the surface electron momentum vanish. This
property can be used to control the HHG and to gate a particular attosecond
pulse out of the train, see also Ref.~\cite{baeva2006relativistic}.
In the 1D geometry, the transverse generalized momentum is conserved:
$\mathbf{p}_{\perp}=e\mathbf{A}_{\perp}/c+\mathbf{p}_{\perp,0}$,
where $\mathbf{p}_{\perp}$ and $\mathbf{A}_{\perp}$are the tangential
components of the electron momentum $\mathbf{p}$ and the vector potential
$\mathbf{A}$. Consequently, the attosecond pulses are emitted when
the vector potential is zero. If the vector potential vanishes at
several moments, there are several $\gamma$-spikes and correspondingly,
several short pulses are observed in the reflected radiation, see
Fig.~\ref{fig:polarization_gating}(a). To select a single attosecond
pulse, we must ensure that the vector potential $\mathbf{A}_{\perp}$turns
zero exactly once. Since $\mathbf{A}_{\perp}$has two components,
how often it vanishes depends on its polarization. For linear polarization
under normal incidence it vanishes twice per laser period, while for
elliptic polarization it never equals zero. A laser pulse with time-dependent
polarization can be prepared in such a way that its vector potential
turns zero just once. A pulse of time-dependent polarization can be
equivalently represented as a superposition of two perpendicularly
polarized pulses, driving and controlling pulse, with slightly different
frequencies and phases. Our PIC simulations suggest that a controlling
signal with a small fraction of the driver intensity is sufficient
to manage the HHG, if the phase difference between the two laser pulses
is chosen carefully.

An example of this is shown Fig.~\ref{fig:polarization_gating}(b).
In addition to the main pulse, which is the same as in panel (a) of
the figure, a smaller controlling signal was used at an amplitude
of $a_{1}=6$. Frequency and phase of the controlling pulse were slightly
detuned in comparison to the main pulse. It is clearly seen that all
attosecond peaks except for the main pulse at $ct\approx9.3\,\lambda$
are strongly attenuated. Only two distinctly smaller side peaks are
left at $ct\approx8.8\,\lambda$ and $ct\approx9.8\,\approx\lambda$.

We conclude that it is possible to isolate single pulses in relativistic
HHG via a polarization gating technique, even if the driver pulse
contains more than one optical cycle.

\section{\label{sec:Fine-Structure-Harmonics}Line Structure in Relativistic
Harmonics Spectra}

In section~\ref{sec:Harmonics-Theory}, theoretical models of surface
HHG were discussed. From these models we were able to compute the
envelope of the harmonic spectrum, but they do not tell anything about
the structure of the individual harmonic lines. The line structure
provides additional details about the laser-plasma interaction on
the femtosecond timescale and thus may serve as a useful diagnostic.
However, to utilize it, a thorough understanding is needed at first.
This section aims to provide this understanding.

In subsection~\ref{sub:fine-structure-moderate-rel}, we briefly
discuss the line structure occurring in the moderately relativistic
regime. In the highly relativistic regime, the spectral line structure
is closely related to the phase of individual attosecond pulses inside
the generated pulse train. Therefore, we examine the dependence of
this phase on laser amplitude and plasma density in subsection~\ref{sub:Attosecond-phase}.
Next, we relate this to the chirp of the relativistic harmonics (Sub.~\ref{sub:Evidence-of-harmonic-chirp})
and calculate its spectral footprint (Sub.~\ref{sub:Spectral-footprint-chirp}),
which is well accessible in experiments. Such experiments have been
conducted at the ARCTURUS facility in Dusseldorf. In subsection~\ref{sub:chirp-Experimental},
we report about how they substantiate the presented theory.

\subsection{\label{sub:fine-structure-moderate-rel}Spectral line structure in
the moderately relativistic regime}

At moderate intensities, modulations in the spectral line structure
such as half integer harmonics are mainly caused by parametric instabilities
in the underdense part of the pre-plasma.

Parametric instabilities, such as stimulated Raman scattering and
the two plasmon decay in the underdense pre-plasma lead to creation
of plasmons at about half the laser frequency \cite{kruer}. These
plasmons can then recombine with the laser or harmonics photons via
sum frequency mixing, leading to side bands or spectral lines at half-integer
multiples of the fundamental \cite{PhysRevE.68.026410,DadB04}. This
mechanism is prevalent for moderate intensities $a_{0}\sim1$, longer
pulse durations $c\tau\gg\lambda$ and extended pre-plasmas.

Moderate broadening of the harmonic lines may also be caused by the
inherent chirp of the CWE process, see Ref.~\cite{quere2008phaseproperties}.
This chirp arises due to the dependence of the excursion times of
the Brunel electrons. For higher intensities, the excursion times
are longer, thus the attosecond pulses are emitted with a longer delay.
Assuming a bell shaped temporal profile of the laser pulse, this leads
to a negative (blue to red) chirp.

At higher intensities $a_{0}\gg1$, the relativistic ponderomotive
force of the laser sweeps away all electrons from the underdense plasma
regions. Therefore, parametric instabilities play no important role
anymore. Also, the CWE mechanism looses importance as the relativistic
effects take over. However, for these pulses, there is again a mechanism
that leads to a variation of the phase of the attosecond pulses depending
on the temporal variation of the laser intensity. This can lead to
heavy broadening and modulation of the harmonic lines, particularly
for extremely short pulses $c\tau\gtrsim\lambda$. Let us now go on
to discuss this mechanism in detail. We begin by numerically computing
the dependence of the phase of the attosecond peaks on the laser intensity
and other parameters.

\subsection{\label{sub:Attosecond-phase}Attosecond peak phase in the highly
relativistic regime}

This subsection is divided into the investigation of normal incidence
and the investigation of s- and p-polarized oblique incidence.

\subsubsection{Universal phase relation in normal incidence}

We start by examining the case of normal incidence on a perfectly
steep plasma boundary. To begin with, a suitable definition of the
{}``phase of the attosecond pulse'' is needed.

\begin{figure}[htp]
\begin{centering}
\includegraphics[width=3in]{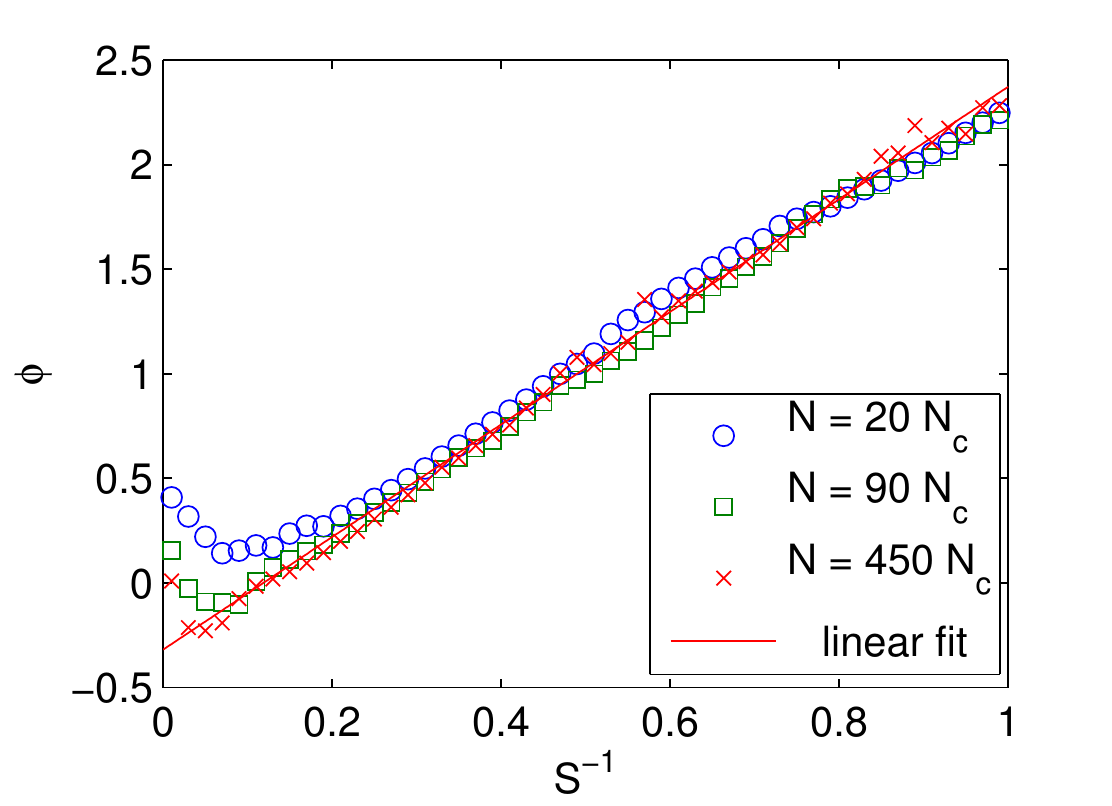}
\par\end{centering}

\caption{\label{fig:universal_phase_normal}Phase dependence of the attosecond
peak on the inverse $S$-parameter, $S^{-1}=a_{0}N_{c}/N$, under
normal incidence.}
\end{figure}

Having another look at Fig.~\ref{fig:ROM-new_model_time_spectra}(a),
showing a quite typical case of the reflected electric field in normal
incidence HHG, tells us what to do. Due to the discontinuities in
the function $E_{r}(t)$, the time derivative possesses clearly pronounced
peaks. Therefore, we define the {}``attosecond phase'' $\phi$ as
the position of the maximum of the time derivative of the reflected
electric field $\partial_{t}E_{r}$. Later on (Sec.~\ref{sec:3d_harmonics})
we will see, that $\partial_{t}E_{r}$ also happens to play an important
role in the computation of the far field. $\phi$ is normalized in
a way, that $\phi=0$ if there is only the Guoy phase shift in the
case of simple non-relativistic reflection from an infinitely dense
surface. With this definition, we measured $\phi$ for a huge range
of densities $N=20\ldots450\, N_{c}$ and laser amplitudes $a_{0}=0\ldots450$.
The result is displayed in Fig.~\ref{fig:universal_phase_normal}.

Very short pulses ($\tau=2\pi/\omega$) were used, so that the surface
remained intact during the interaction even for high intensities.
The phase of the incoming laser pulse is chosen in a way so that $E_{i}=0$
at the maximum of the envelope, thus the attosecond peak is located
close to the maximum of the envelope. The attosecond phase $\phi$
is plotted against the inverse $S$-parameter (see Ref.~\cite{gordienko:043109})
$S^{-1}=a_{0}N_{c}/N$. For all simulations in the highly relativistic
regime $a_{0}\gg1$, we find an excellent agreement with the fit

\begin{equation}
\phi=2.7\, S^{-1}-0.32,\label{eq:phase_dep}\end{equation}
while in the low intensity limit $S^{-1}\rightarrow0$ the phase shift
tends to the value $\phi=\textrm{acot}((N/N_{c}-2)/(2\sqrt{N/N_{c}-2}))$,
which is expected from non-relativistic optics, approving the correctness
of the PIC calculations once again.

Physically, the phase shift $\phi$ can be understood as a consequence
of the electron surface being pushed inside the plasma by the laser.
If the electron surface is pushed in to a depth of $\Delta$, we expect
the phase to experience an additional shift $\propto\Delta$. Let
us devise a rough model in order to understand the linear scaling
of $\phi$ with $S^{-1}$. Therefore we assume that there is a pressure
balance between the ponderomotive force $f_{pond}\propto a_{0}^{2}$
of the laser and the electrostatic restoring force $f_{stat}=qE\propto N^{2}\Delta^{2}$
of the plasma. Equalizing both terms yields $\Delta\sim a_{0}/N$
and consequently, a linear dependence of $\phi$ on $S^{-1}$.

Note further, that in the ultrarelativistic regime the function $\phi$
is indeed completely independent of the absolute plasma density. This
is the clearest footprint of the $S$-similarity \cite{gordienko:043109}
in laser-overdense plasma interaction observed so far.

\subsubsection{Phase behaviour at oblique incidence}

When considering oblique incidence, the polarization is crucial. For
s-polarized oblique incidence, we retain a behaviour similar to the
one observed under normal incidence. For p-polarized incidence, the
behaviour changes in many ways. We analyze both cases using 1D PIC
simulations in a Lorentz transformed frame (see App.~\ref{sec:Bourdier}).

\begin{figure}
\begin{centering}
\includegraphics[width=3in]{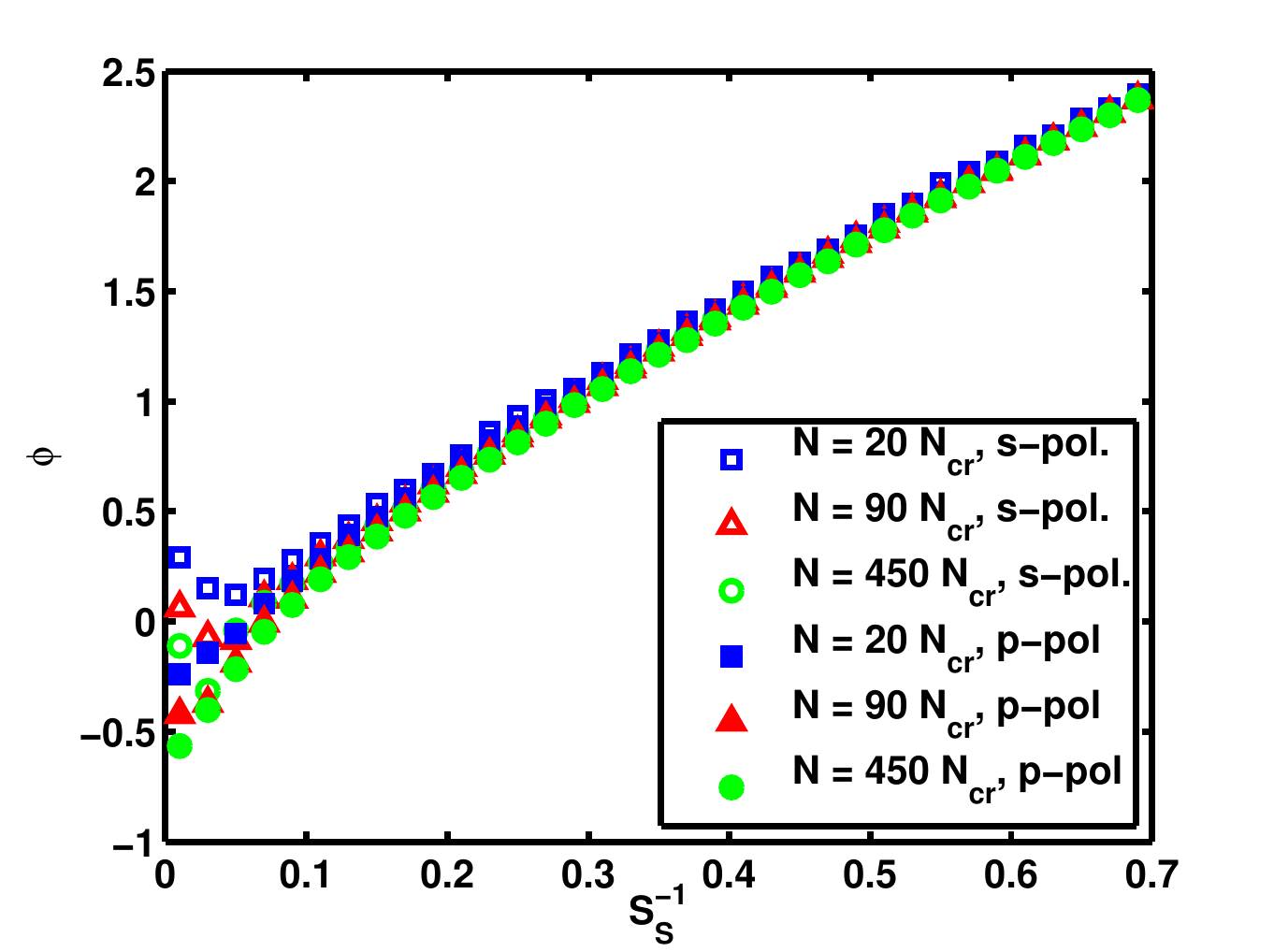}
\par\end{centering}

\caption{\label{fig:Phase-dependence-s-density}Phase dependence of the attosecond
peak on the inverse $S$-parameter in the simulation frame, $S_{S}^{-1}=a_{0}N_{c}^{(S)}/N^{(S)}$,
under s-polarized oblique incidence; density $N$ and laser amplitude
$a_{0}$ are varied. In relation to the laboratory frame $S$-parameter,
$S_{S}$ scales as $S_{S}=S_{L}/\cos^{3}\theta$. Both the phase of
s-polarized and p-polarized generated harmonics is displayed. The
angle of incidence is $\theta=45\text{\textdegree}$}
\end{figure}

Consider Fig.~\ref{fig:Phase-dependence-s-density}. As in the case
of normal incidence, we confirm the dependence on the $S$-parameter
with high accuracy in the ultrarelativistic regime. If the density
is varied, but the ratio $S^{-1}=a_{0}N_{c}/N$ is kept constant,
there is no change in the attosecond phase. We also see that there
is virtually no difference between the phase of the p-polarized and
the s-polarized generated harmonics. This is evidence that they both
are generated due to the same physical mechanism. They are not generated
at separate phases as are CWE and ROM pulses in the weakly relativistic
regime \cite{quere:125004}.

Further, as in the normal incidence case, an approximately linear
dependence on $S^{-1}$ is found. This can be understood, as the mechanism
leading to the indention of the electron plasma surface is basically
the same as for normal incidence: There is a pressure balance between
the ponderomotive light pressure and the electrostatic force.

In the laboratory frame however, the ponderomotive light pressure
is expected to be weaker compared to normal incidence, since the laser
does not hit the surface head on, but under an angle $\theta$. Seen
in the simulation frame, the ions and the electrons possess currents
in opposite directions. This generates a magnetic repulsion, counteracting
the electrostatic restoring force. Effectively, it leads to a mitigation
of the electrostatic force by a factor of $1/\gamma$. Therefore we
expect, that the scaling in s-polarized oblique incidence should remain
independent of the angle $\theta$ if we consider it a function of
$S_{\textrm{eff}}\equiv S_{S}/\gamma=S_{L}/\cos^{2}\theta$.

\begin{figure}
\centering{}\includegraphics[width=3in]{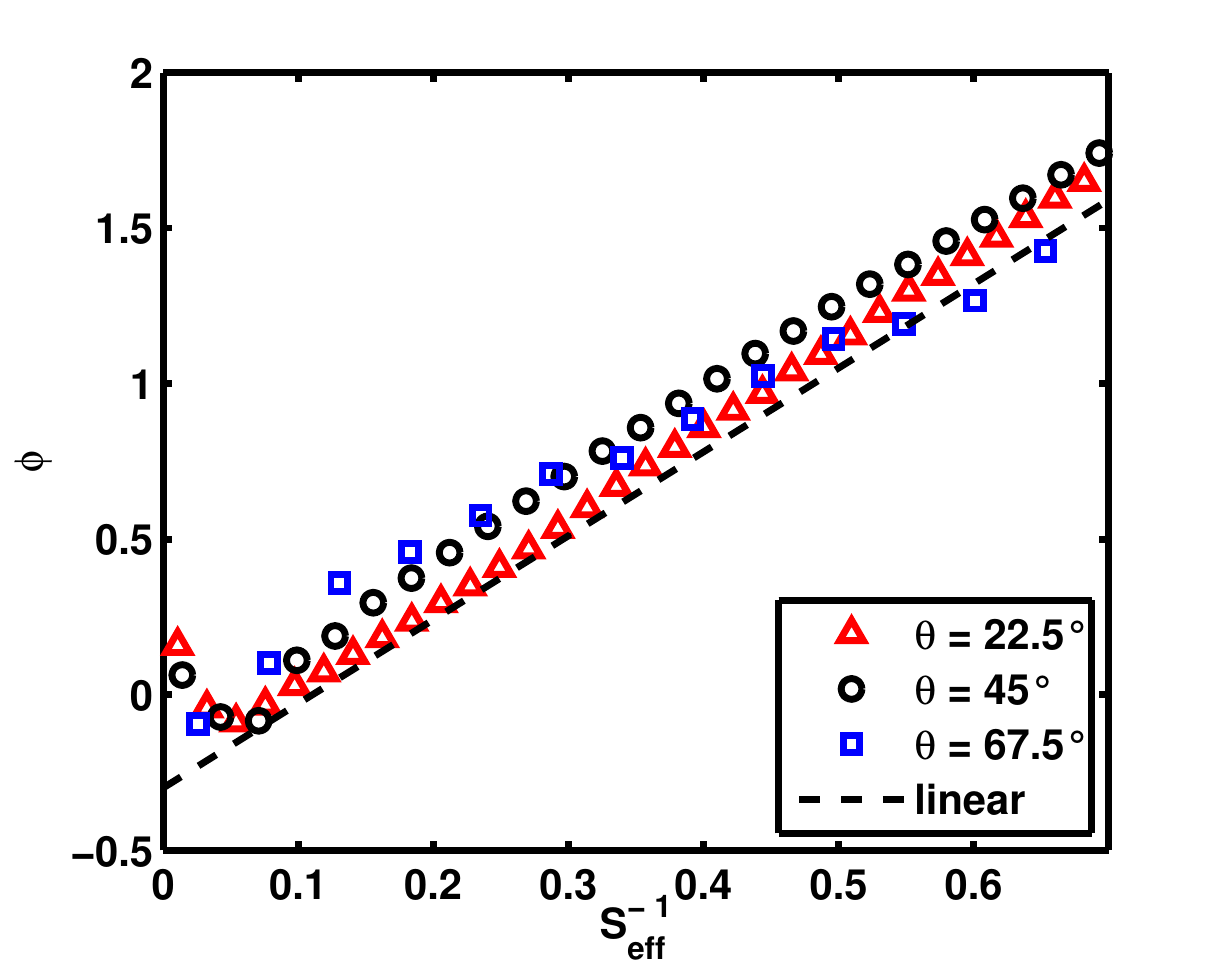}\caption{\label{fig:Phase-dependence-s-angle}Phase dependence of the attosecond
peak on the inverse effective $S$-parameter under s-polarized incidence,
angle of incidence $\theta$ and laser amplitude $a_{0}$ varied.
The effective $S$-parameter is defined as $S_{\textrm{eff}}\equiv S_{L}/\cos^{2}\theta=S_{S}/\gamma$
to bring out the consistent linear dependence. Here, only the phase
of the s-polarized generated harmonics is displayed since the one
for the p-polarized harmonics almost agrees (see Fig.~\ref{fig:Phase-dependence-s-density}).}
\end{figure}

This can well be confirmed by the numerical results depicted in Fig.~\ref{fig:Phase-dependence-s-angle}.
We conclude, that for s-polarized incidence, the phase of both the
s-polarized and the p-polarized fraction of the generated harmonics
is determined \emph{only} by the effective $S$-parameter $S_{\textrm{eff}}=a_{0}N_{c}/(N\,\cos^{2}\theta)$
and does \emph{not} depend on $a_{0}$, $N$ and $\theta$ separately.

\begin{figure}
\begin{centering}
\includegraphics[width=3in]{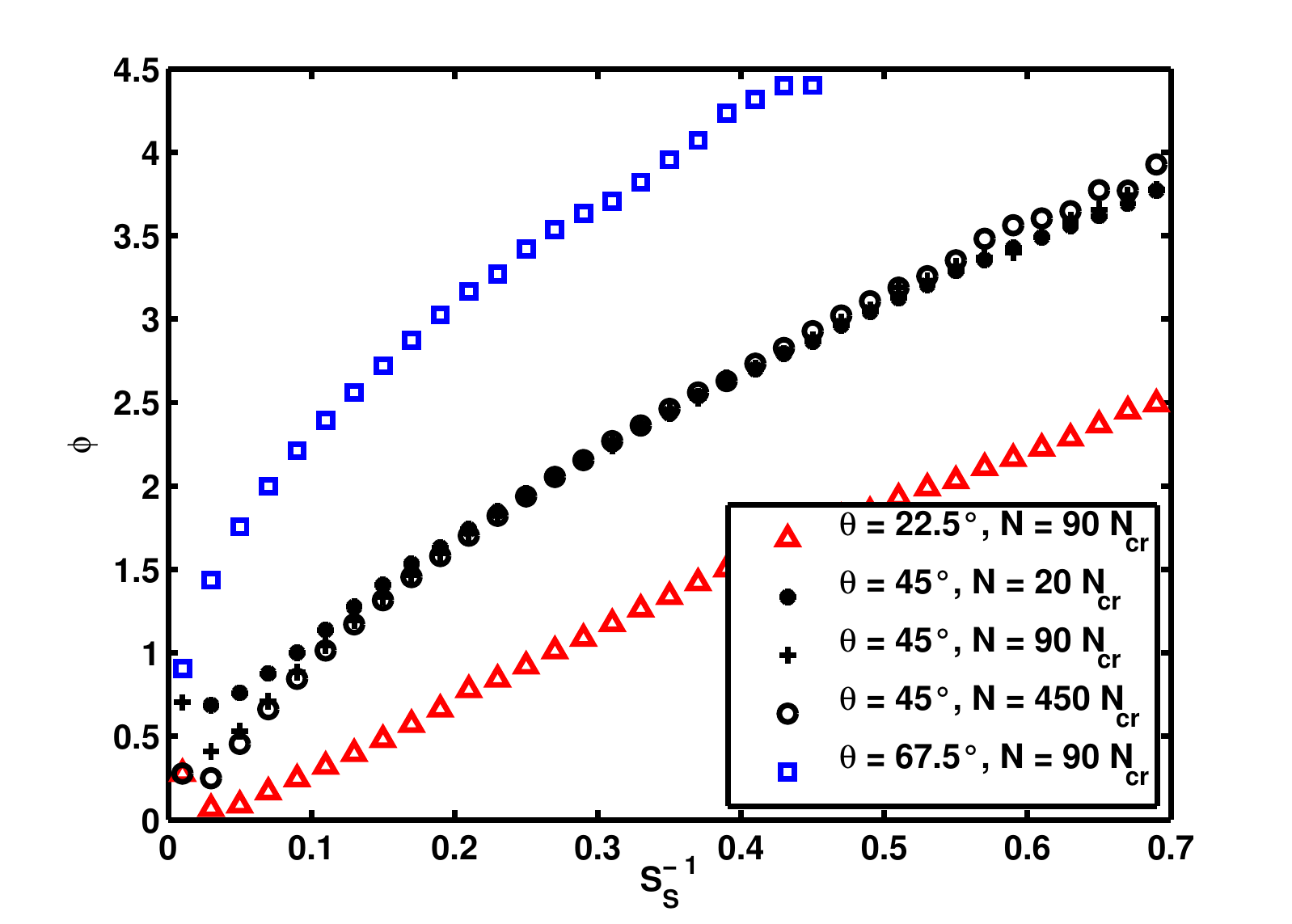}
\par\end{centering}

\caption{\label{fig:Phase-dependence-p-angle}Phase dependence of the attosecond
peak on the inverse $S$-parameter in the simulation frame under p-polarized
incidence, angle of incidence $\theta$, density $N$ and laser amplitude
$a_{0}$ are varied.}
\end{figure}

For p-polarized incidence, matters are more complex. In addition to
the ponderomotive force, the surface is also pushed in and pulled
out directly by the longitudinal electric field component of the laser.
Numerical results are shown in Fig.~\ref{fig:Phase-dependence-p-angle}.

Despite of the highly complex interaction, the attosecond phase $\phi$
again depends only on the $S$-parameter, not on $a_{0}$ and $N$
separately. The slope is however not linear anymore. As we can see
from Fig.~\ref{fig:Phase-dependence-p-angle}, the non-linearity
increases with the angle of incidence $\theta$.

Note further that in the case of p-polarized incidence, the duration
of the pulse may also play an important role. Oblique p-polarized
incidence can lead to the generation of very strong quasi-static magnetic
fields close to the surface. Therefore, memory effects are present
and $\phi$ is not just a function of the instantaneous intensity
but a functional of the whole history of the incident field. In section~\ref{sub:chirp-Experimental},
we will show an example of this highly interesting effect.

\subsection{\label{sub:Evidence-of-harmonic-chirp}Evidence of harmonic chirp
in PIC simulation}

As we have just seen, the phase of the attosecond pulses generated
from overdense plasmas depends on the $S$-parameter of the interaction.
Because of the relativistic radiation pressure of the laser pulse,
the electrons are pushed inside the plasma during the rising edge
of the laser pulse, causing an initial red shift of the reflected
light. Later, the electron fluid will return to its original position
and therefore cause a blue shift. This shifting of frequencies is
called \emph{harmonic chirp}%
\footnote{In addition to the Doppler shift due to the physical displacement
of the electron surface, for an extended density gradient there is
also a {}``virtual'' Doppler shift due to the motion of the reflecting
surface because of the variation in relativistic transparency, compare
Ref.~\cite{DadB04}. However, we do not intend to further distinguish
these phenomena here, since their effect on the reflected radiation
is essentially the same.%
}.

\begin{figure}
\begin{centering}
\includegraphics[width=3in]{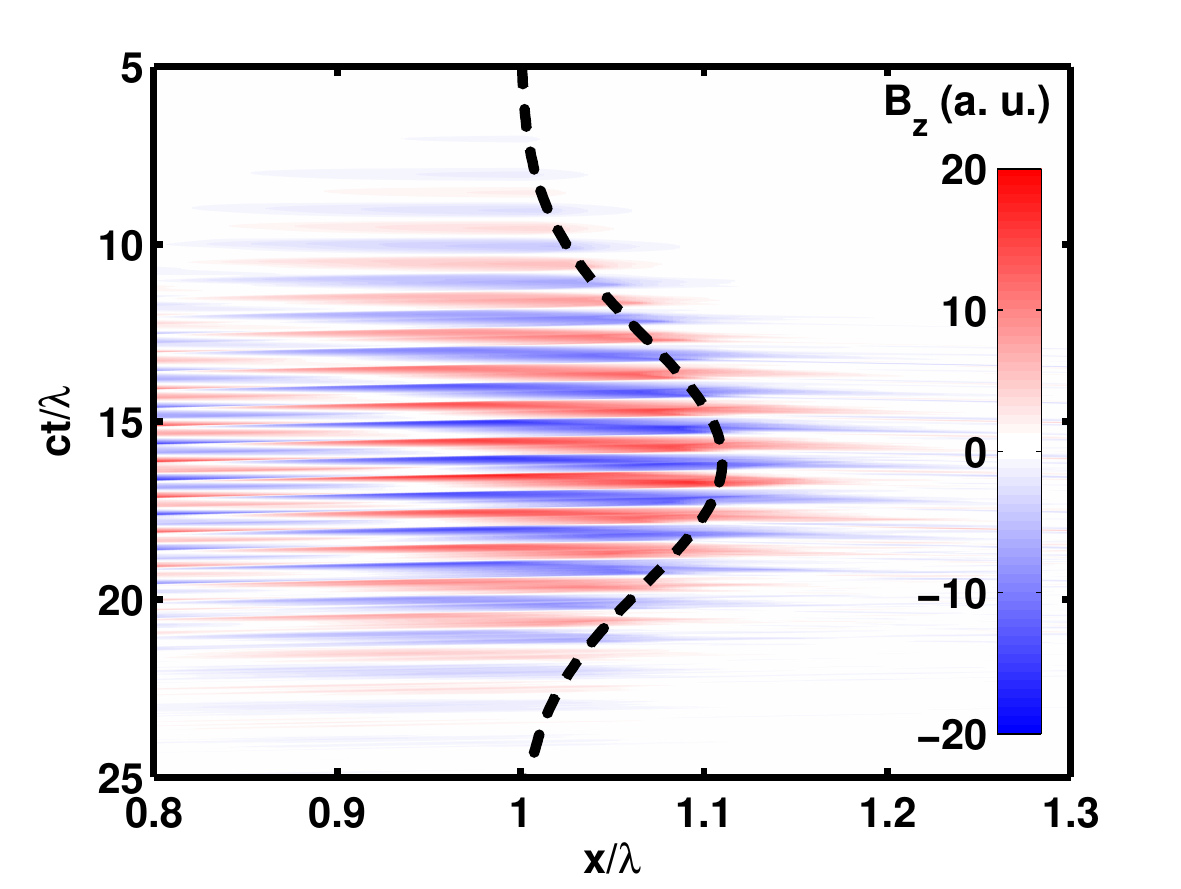}
\par\end{centering}

\caption{\label{fig:chirp_001_bz}Transverse magnetic field component $B_{z}$
of a laser reflecting and generating harmonics at an overdense plasma
surface at normal incidence. The dashed line highlights the motion
of the reflecting surface, fitted by a Gaussian function, corresponding
to Eq.~\eqref{eq:attoseq_dir_lin_respons} with $\alpha=1.35$. Parameters
are: laser amplitude $a_{0}=10$, duration $c\tau=5\lambda$; sharp
edged plasma with density $n_{e}=20\, n_{c}$ starting at $x=1\,\lambda$,
fixed ions.}
\end{figure}

The motion of the reflecting surface can be followed in an $x$-$t$-colourscale
image of the transverse magnetic field component. Figure~\ref{fig:chirp_001_bz}
shows such an image for a PIC simulation of normal laser incidence
on a perfectly sharp plasma boundary. Realistic cases with oblique,
p-polarized incidence on a plasma with a finite density gradient will
be discussed in subsection~\ref{sub:chirp-Experimental}. In our
simple case we see that the motion of the surface is well described
by a Gaussian function, i.e. the surface displacement is proportional
to the instantaneous laser amplitude. This agrees with the observations
made in subsection~\ref{sub:Attosecond-phase} with even shorter
laser pulses.

As a result, the reflected radiation contains a positive chirp. This
chirp can be made visible in a time-frequency image (or spectrogram).
To compute the spectrogram, the time-series data from the PIC simulation
is multiplied with a bell-shaped window function that is gradually
moved over the data. Then, spectra of the products are calculated,
yielding the spectrogram.

\begin{figure}
\begin{centering}
\includegraphics[width=4in]{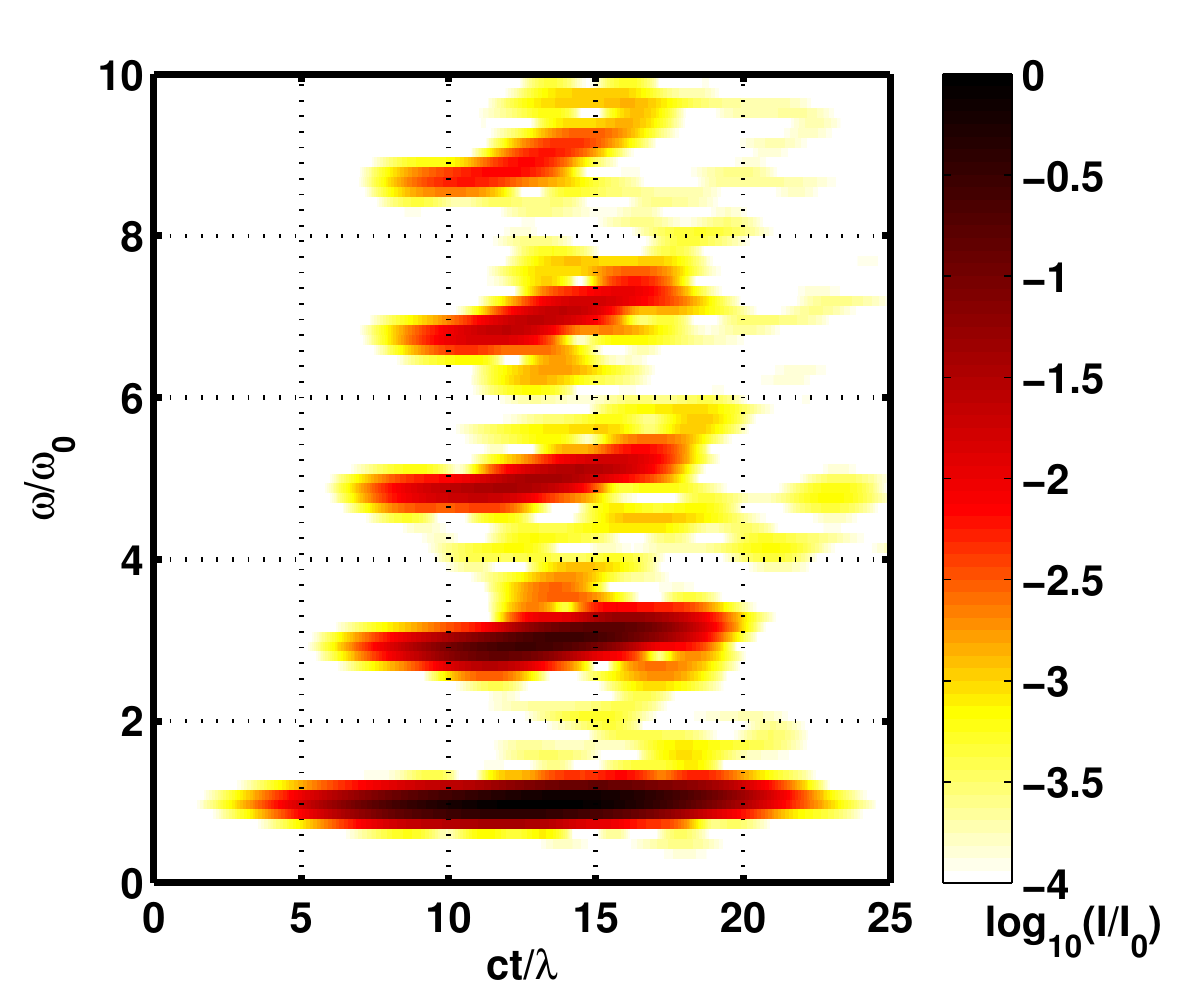}
\par\end{centering}

\caption{\label{fig:chirp_001_specgram}Spectrogram with logarithmic colourscale
of the reflected radiation. Simulation parameters are the same as
in figure~\ref{fig:chirp_001_bz}. For the computation of the spectrogram,
a Blackman-Harris window with a width of about eight laser periods
was used.}
\end{figure}
In Fig.~\ref{fig:chirp_001_specgram} such a spectrogram is shown.
The data stems from the same simulation as the previous figure. The
chirp from red to blue is clearly visible in the fundamental and all
harmonics shown. The higher the harmonic number, the more pronounced
is the chirp. This is due to the nature of the Doppler effect. Because
the relative Doppler frequency shift $\Delta\omega/\omega$ is constant,
the absolute shift $\Delta\omega$ is proportional to the frequency.
Further note that the \emph{positive} chirp (from red to blue) observed
here distinguishes this relativistic regime from the CWE regime, where
a \emph{negative} chirp (from blue to red) is observed~\cite{quere2008phaseproperties}.

\subsection{\label{sub:Spectral-footprint-chirp}Spectral footprint of harmonic
chirp}

In the time integrated spectrum the chirp is visible in the line structure.
Due to the dependence of the Doppler shift on $\omega$, we also expect
the line structure to vary according to the spectral region.

To describe this structure analytically, assume the radiation is given
as a sequence of attosecond pulses emitted at the times $t_{k}$ with
identical shape $f(t)$ but possibly different amplitudes $E_{k}$,
corresponding to the laser envelope: \begin{equation}
E(t)=\sum_{k}E_{k}\, f(t-t_{k}).\label{eq:attopulse_sequence_time_domain}\end{equation}

Now we Fourier transform Eq.~\eqref{eq:attopulse_sequence_time_domain}
and take the absolute square to arrive at the spectrum:\begin{equation}
I(\omega)=|\tilde{f}(\omega)|^{2}\underbrace{\left|\sum_{k}E_{k}e^{-i\omega t_{k}}\right|^{2}}_{J(\omega)},\label{eq:attopulse_sequence_spectrum}\end{equation}
where $\tilde{f}(\omega)$ denotes the Fourier transformation of the
attosecond pulse shape function $f(t)$. Its absolute square $|\tilde{f}(\omega)|^{2}$
corresponds to the spectral envelope that has been discussed in Sec.~\ref{sec:Harmonics-Theory}.
Here, we concentrate on the second factor $J(\omega)$ that represents
the spectral line structure.

In the trivial case of equidistant pulses with constant intensities,
i.e. $E_{k}\equiv1$ and $t_{k}=kT_{0}=2\pi k/\omega_{0}$, the result
is a sequence of sharp harmonic lines at multiples of the fundamental
frequency $\omega_{0}$. Such a spectrum occurs for harmonics generated
by comparatively long laser pulses (picosecond range) with moderate
intensities. Early experiments on surface harmonics generation worked
with such pulses and obtained spectra close to this prediction, compare
e.g. Ref.~\cite{PhysRevA.24.2649}.

The spectrum changes as pulses become shorter and more intense. As
described above, the harmonics move from a red-shifted to a blue-shifted
phase due to the Doppler effect of the averaged surface motion. To
get a first impression of the effect on the spectra, let us consider
two trains of pulses with a slightly different periodicity $T_{1}$
and $T_{2}$. Both pulse trains will produce a train of harmonic lines
corresponding to their repetition frequencies $\omega_{i}=2\pi/T_{i}$
($i\in\{1,2\}$). The harmonic lines will then interfere with each
other. Provided they possess a finite linewidth $\delta\omega$ and
the difference between the two interfering frequencies is small in
the sense $\Delta\equiv\omega_{2}-\omega_{1}\ll\delta\omega$, we
can calculate the frequency period $\Omega$ of the occurring interference
pattern by setting $\Omega=n\omega_{2}=(n+1)\omega_{1}$ and therefore
$\Omega\approx\omega_{0}^{2}/\Delta$, where $\omega_{0}=(\omega_{1}+\omega_{2})/2$
is the centre frequency.

\begin{figure}
\begin{centering}
\includegraphics[width=4in]{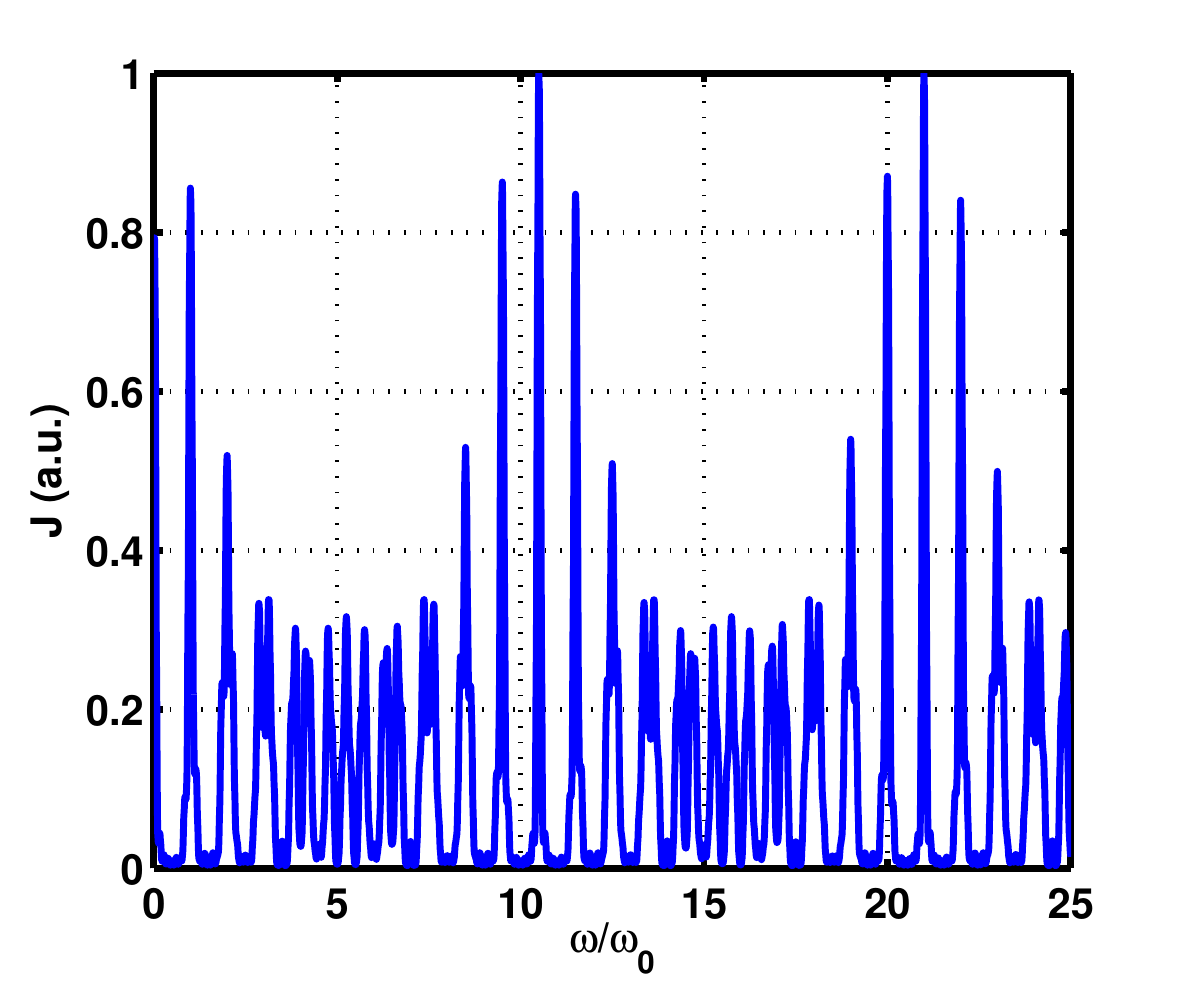}
\par\end{centering}

\caption{\label{fig:spectrum_two_attotrains}Spectral line structure of as-pulse
train with two different frequencies $\omega_{1}=0.95\,\omega_{0}$
and $\omega_{2}=1.05\,\omega_{0}$. The pulse train contains 9 pulses.}
\end{figure}

Fig.~\ref{fig:spectrum_two_attotrains} shows the actual spectrum
$J(\omega)$. The figure confirms the analytically evaluated modulation
frequency of $\Omega=10\,\omega_{0}$. The first few lines clearly
pronounce multiples of the fundamental frequencies, then the lines
become broader and at around $5\,\omega_{0}$, we observe two lines
per harmonic. Around $10\,\omega_{0}$, we observe one peak per harmonic
again, but this time at half integer frequencies. After that, the
structure repeats, shifting back to integer harmonics around $20\,\omega_{0}$.

Certainly, in reality the period of the as-pulses does not change
abruptly, but continuously. Let us therefore consider a linearly chirped
train of attosecond pulses:

\begin{eqnarray}
t_{k} & = & \frac{2\pi}{\omega_{0}}\left(k-\frac{\beta}{m}k^{2}\right),\label{eq:attoseq_lin_chirp}\end{eqnarray}
where $k=-m\ldots m$. Note that the parameter $\beta$ represents
the maximum cycle averaged velocity acquired by the reflecting surface.
Inserting \eqref{eq:attoseq_lin_chirp} into \eqref{eq:attopulse_sequence_spectrum}
we arrive at \begin{equation}
\sum_{k}E_{k}\exp(-i\omega t_{k})=1+2\sum_{k=1}^{m}\cos\left(\frac{2\pi k\omega}{\omega_{0}}\right)\exp\left(i\frac{2\pi k^{2}\omega\beta}{m\omega_{0}}\right).\label{eq:attoseq-model-sum}\end{equation}

\begin{figure}
\begin{centering}
\includegraphics[width=6in]{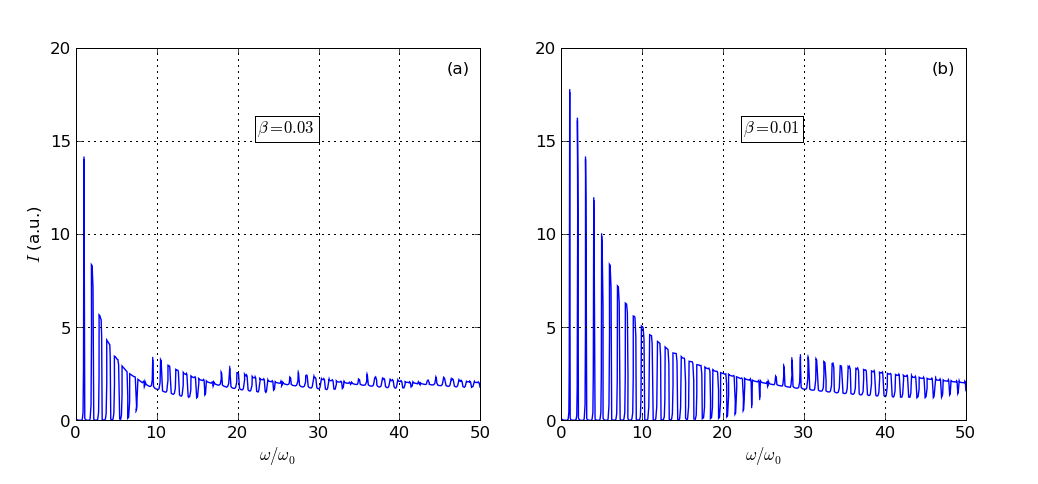}
\par\end{centering}

\caption{\label{fig:attopulse_model_spectra_subplots}The model from Eqs.~\eqref{eq:attopulse_sequence_spectrum}
and \eqref{eq:attoseq_lin_chirp}, using $|\tilde{f}(\omega)|\equiv1$
and $m=64$ throughout. Spectra are smoothed by convolution with a
Gaussian function of FWHM $\Delta\omega=0.1\,\omega_{0}$.}
\end{figure}

The resulting spectra are depicted in Fig.~\ref{fig:attopulse_model_spectra_subplots}.
Again, we see a periodic modulation of the spectrum. Because of the
continuously changing period lines also broaden, especially in the
high frequency range. This leads to a decay of the modulation amplitude.
Further, a quasi-continuum is observed at the spectral regions where
two harmonics per period were observed with the discrete frequency
model. The modulation period corresponds to the frequency difference
$\Delta$ between the extreme ends of the linear chirp and therefore
is proportional to the maximum surface velocity. Thus, it is possible
to extract physical information from the line structure. If the reflecting
surface attains a velocity of $\beta$ (in units of $c$), we can
expect large scale modulations in the line structure with a period
$\Omega$ given by \begin{equation}
\Omega\approx\frac{\omega_{0}}{4\beta}.\label{eq:large_scale_modulation_harm_spectrum}\end{equation}

Let us now reconsider the example from the previous subsection~\ref{sub:Evidence-of-harmonic-chirp}.
From Sec.~\ref{sub:Attosecond-phase} and also Fig.~\ref{fig:chirp_001_bz}
of this section, we learn that the phase of the reflected radiation
depends roughly linear on the laser field. Therefore, we can directly
relate $E_{k}$ and $t_{k}$ to the envelope $g(t)$ of the laser.
At normal incidence we further expect two attosecond pulses per period
with alternating sign. This leads us to the model: \begin{eqnarray}
t_{k} & \approx & \pi k/\omega_{0}+\alpha\, g(\pi k/\omega_{0})\nonumber \\
E_{k} & \approx & (-1)^{k}\, g(\pi k/\omega_{0}).\label{eq:attoseq_dir_lin_respons}\end{eqnarray}

Inserting Eqs.~\eqref{eq:attoseq_dir_lin_respons} into Eq.~\eqref{eq:attopulse_sequence_spectrum}
yields the structure of the spectral lines. For $S\equiv n_{e}/a_{0}n_{c}\gtrsim1$,
the model parameter $\alpha$ can be determined from the linear slope
in Fig.~\ref{fig:universal_phase_normal} (or Eq.~\eqref{eq:phase_dep}).
Therefore, $\alpha=1.35$ corresponds to $S=2$ and $\alpha=0.54$
to $S=5$.

\begin{figure}
\begin{centering}
\includegraphics[width=5in]{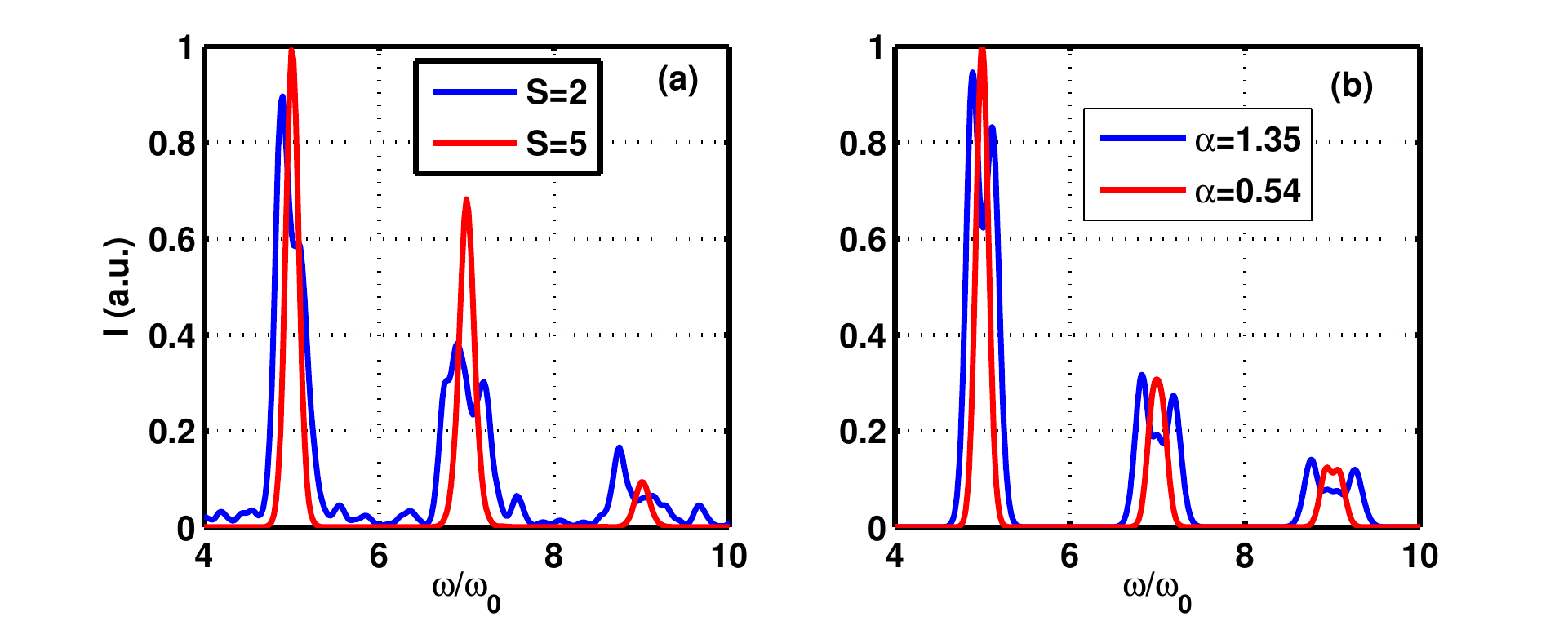}
\par\end{centering}

\caption{\label{fig:chirp_spectra_PIC_vs_model}Comparison of the line structure
from the fifth to the ninth harmonic (a) from PIC data and (b) due
to Eqs.~\eqref{eq:attopulse_sequence_spectrum} and \eqref{eq:attoseq_dir_lin_respons}.
Simulation parameters corresponding to the blue line in (a) are the
same as in Figs.~\ref{fig:chirp_001_bz} and \ref{fig:chirp_001_specgram},
the red line in (a) corresponds to the same set of parameters except
for $n_{e}=50\, n_{c}$.}
\end{figure}
 Now we can compare the simple model to spectra obtained from PIC
data. As shown in Fig.~\ref{fig:chirp_spectra_PIC_vs_model}, this
comparison shows good qualitative agreement. Remaining differences
can arguably be attributed to the non-linear dependence of the attosecond
phase and the harmonics intensity in the only moderately relativistic
interaction at the edges of the pulse, which are not included in the
simple model.

We conclude, that the line broadening observed in relativistic harmonics
spectra can to a large extent be explained by the chirp due to unequal
spacing of the attosecond pulses. It does \emph{not} imply a loss
in coherency of the individual attosecond pulses.

\subsection{\label{sub:chirp-Experimental}Experimental confirmation of harmonic
chirp}

\begin{figure}
\begin{centering}
\includegraphics[width=3in]{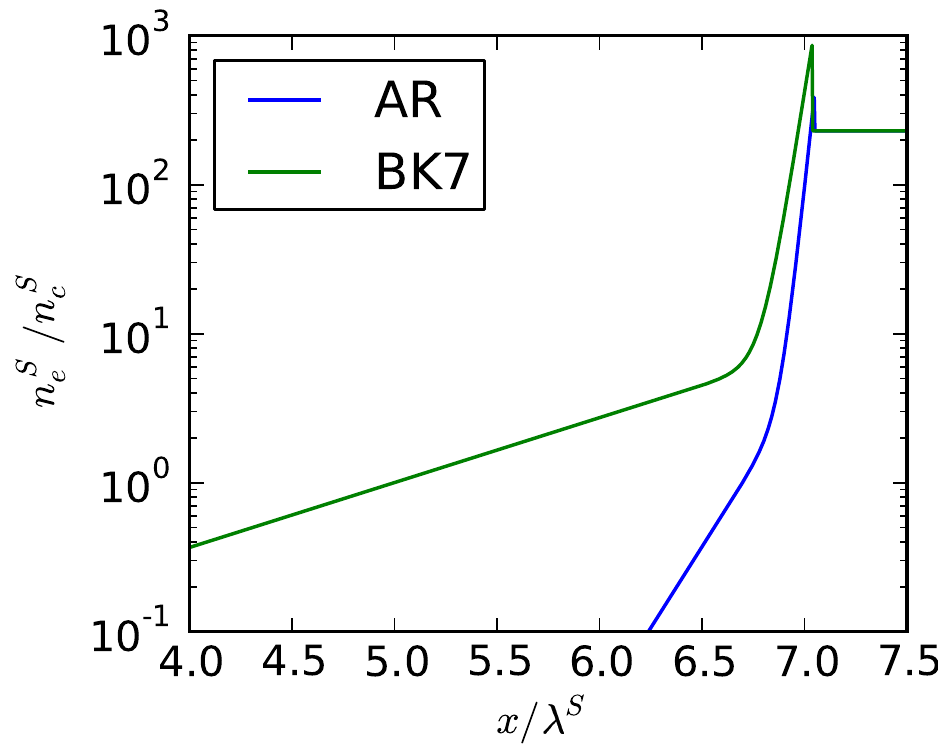}
\par\end{centering}

\caption{\label{fig:ha_exp_013_014_density_prof}Plasma density profiles corresponding
to experiments at the Arcturus laser facility using AR and BK7 type
plasma mirrors, as used in the PIC simulation. All magnitudes are
shown in the Lorentz transformed simulation frame (see App.~\ref{sec:Bourdier}).}
\end{figure}

Let us now have a look at a set of parameters derived from real experiments
carried out at the Düsseldorf ARCTURUS laser facility. The laser is
obliquely incident under an angle of 45\textdegree{} and the light
is p-polarized with an estimated peak amplitude of about $a_{0}=8$.
Two different kinds of plasma mirrors were used to improve the laser
contrast ratio: the AR (anti-reflex coated, 0.1\% reflectivity) mirror
leading to a high contrast and an extremely steep density gradient
and the polished BK7 glass (\textasciitilde{}4\% reflectivity) yielding
a medium contrast and a little less steep density gradient. In the
PIC simulation, {}``double exponential'' density profiles of the
type\begin{equation}
n_{e}(x)=\begin{cases}
\exp\left[a(x-x_{0})\right]+\exp\left[b(x-x_{1})\right] & (x<x_{2})\\
n_{0} & (x>x_{2})\end{cases}\end{equation}
were chosen as shown in Fig.~\ref{fig:ha_exp_013_014_density_prof}.
These profiles closely resemble the ones in the experiment, which
could be estimated by simulations of the hydrocode Multi-FS\cite{ramis1988multitextemdash},
conducted by Michael Behmke and Jens Osterholz.

\begin{figure}
\begin{centering}
\includegraphics[width=4in]{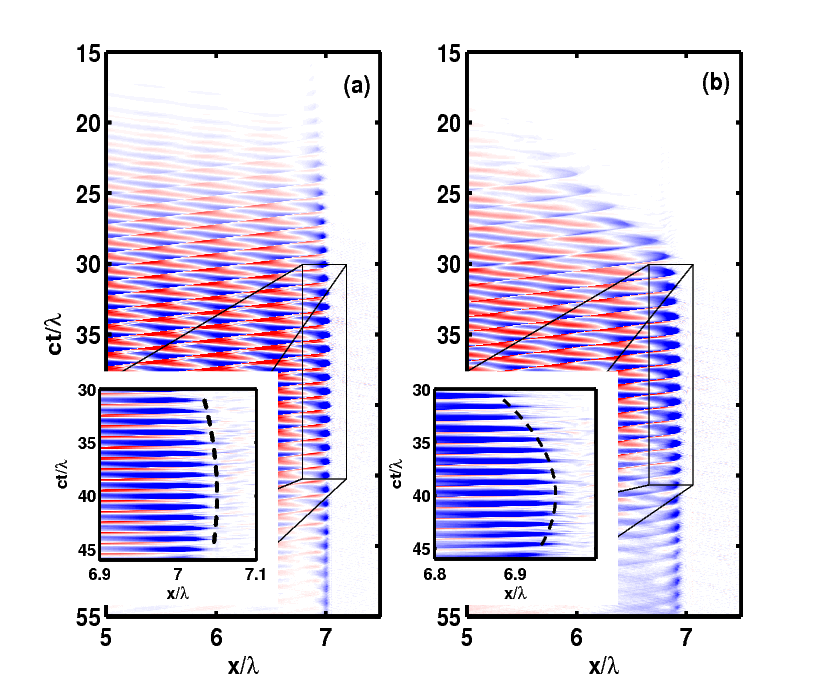}
\par\end{centering}

\caption{\label{fig:ha_exp_013_014_bz_overview}Colourscale image of the transverse
magnetic field component $B_{z}$ as a function of time $t$ and space
$x$, overview and detail. The dashed lines in the detail graphs denote
the part of the surface motion used to model the spectral chirp. The
data stems from simulations with a realistic set of parameters: The
incident laser has a Gaussian temporal profile $a=a_{0}\exp(-t^{2}/\tau^{2})$
with amplitude $a_{0}=8.1$ and pulse duration $\tau=10\,\lambda/c$
for both subfigures. The density profiles used are displayed in Fig.~\ref{fig:ha_exp_013_014_density_prof},
here (a) corresponds to the AR and (b) to the BK7 plasma mirror.}
\end{figure}
Figure~\ref{fig:ha_exp_013_014_bz_overview} shows the transverse
magnetic field from the interaction. Compare this to Fig.~\ref{fig:chirp_001_bz}.
The strong temporal asymmetry is conspicuous: Instead of instantly
returning to its initial position as in the normal incidence case
(Fig.~\ref{fig:chirp_001_bz}), the electron surface remains indented.
The static magnetic field, created by the current of the Brunel electrons,
holds the electrons inside.

Because of this temporal asymmetry and the non-linear dependence of
the phase on the amplitude, Eq.~\eqref{eq:attoseq_dir_lin_respons}
ceases to apply here. To reproduce the exact spectral shape in our
model would therefore require to exactly trace the surface motion
with a non-elementary function. The purpose of our simple model is
however not to exactly reproduce the spectrum, but to extract some
crucial features. Our aim is to provide clear evidence that the modulations
in the experimental and PIC spectra are caused by the unequal spacing
between the attosecond peaks and to show, which information can be
gained from the spectra. We therefore design the model as plain as
possible, leaving only two free parameters $\beta$ and $m$. We concentrate
on the main phase of harmonic generation $t=30\ldots45\,\lambda/c$
and approximate the surface motion during this phase by a parabola,
corresponding to a linear chirp. Also, the temporal asymmetry is ignored,
taking the sum in Eq.~\eqref{eq:attoseq-model-sum} always from $-m$
to $m$.

\begin{figure}
\begin{centering}
\includegraphics[width=5in]{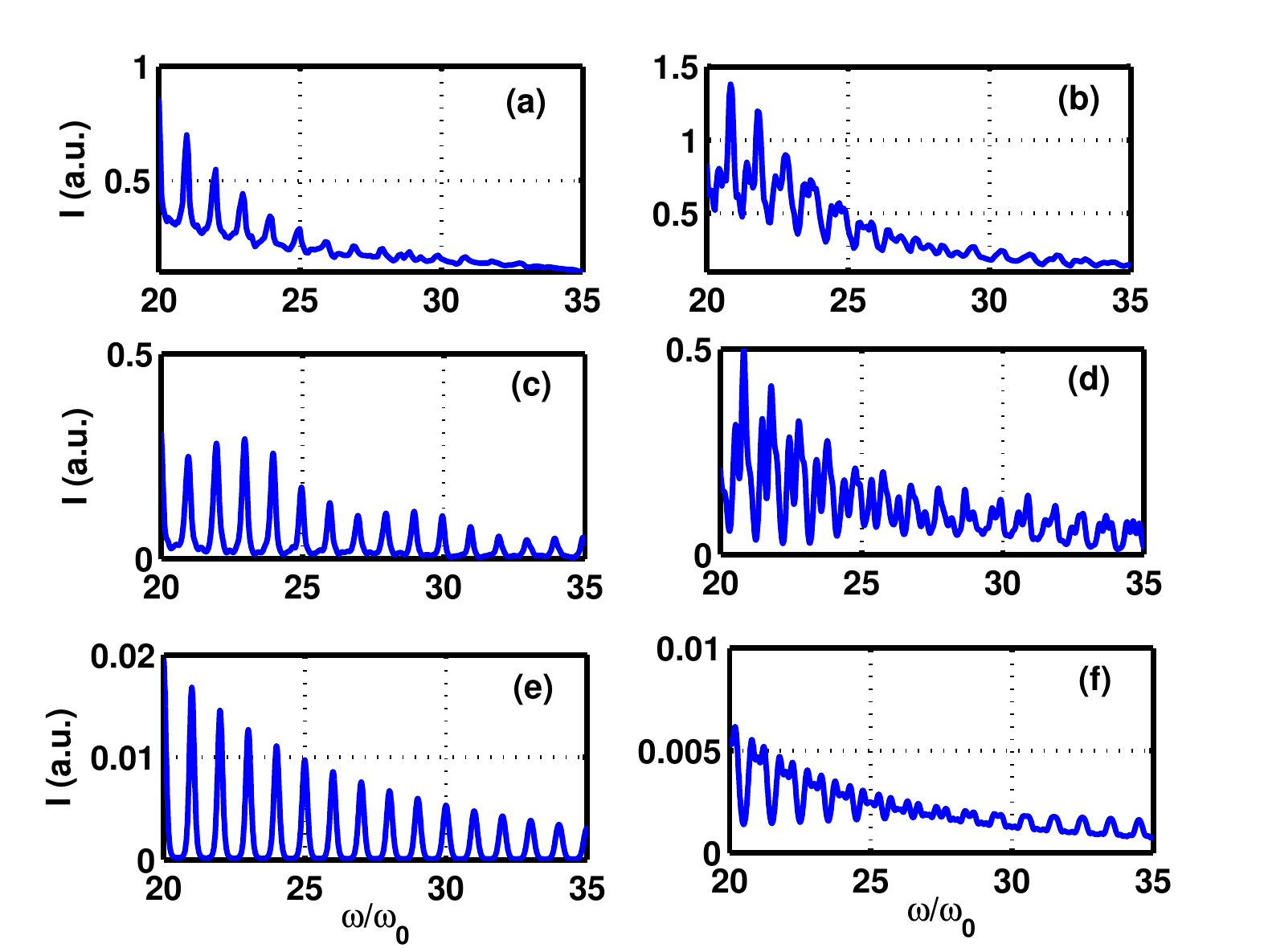}
\par\end{centering}

\caption{\label{fig:ha_exp_013_014_spectra_compare}Comparison of (a)-(b) experimental,
(c)-(d) PIC and (e)-(f) model spectra according to Eqs.~\eqref{eq:attopulse_sequence_spectrum}
and \eqref{eq:attoseq_lin_chirp}. The left column represents the
experiment conducted with the AR plasma mirror (high contrast, very
short pre-plasma) and the corresponding simulations, the right column
represents an experiment conducted with the BK7 plasma mirror (medium
contrast, slightly longer pre-plasma) and the corresponding simulations.
Parameters in (c) and (d) are the same as in Fig.~\ref{fig:ha_exp_013_014_bz_overview}.
Parameters chosen in model: (e) $\beta=0.0028$, (f) $\beta=0.011$,
$|\tilde{f}(\omega)|=\omega^{-8/3}$ and $m=7$ for both. PIC and
model spectra were smoothed by convolution with a Gaussian function
of FWHM $\Delta\omega=0.15\,\omega_{0}$ for the sake of better comparability
with the spectrometer data. Intensity units within one row are comparable,
but non in between the rows.}
\end{figure}

Considering the complexity of the actual, highly non-linear process
and the simplicity of the model, we observe an excellent agreement
between the experimental spectra, the PIC spectra and the analytically
calculated ones, compare Fig.~\ref{fig:ha_exp_013_014_spectra_compare}.
From the conjunction between the plasma motion in Fig.~\ref{fig:ha_exp_013_014_bz_overview}
and the model spectra from Fig.~\ref{fig:ha_exp_013_014_spectra_compare},
it becomes clear that the modulations in the spectrum are caused by
the unequal spacing between the attosecond pulses within the generated
pulse train.

Another interesting detail is the slight redshift to be observed in
experimental {[}Fig.~\ref{fig:ha_exp_013_014_spectra_compare}(b){]}
and PIC {[}Fig.~\ref{fig:ha_exp_013_014_spectra_compare}(d){]} data,
in particular between the 15th and the 25th harmonic. This overall
redshift is a footprint of the aforementioned temporal asymmetry in
the femtosecond plasma dynamic. Therefore, it witnesses the self-generated
static magnetic field.

Let us now estimate the surface velocity from the experimental spectra,
employing Eq.~\eqref{eq:large_scale_modulation_harm_spectrum}. In
the {}``medium contrast'' case, we observe a transition from integer
harmonics in the region up to the 25th order to half-integer harmonics
in the region beyond the 30th order. In between, the lines merge into
a quasi-continuum. Thus, the modulation cycle is about $\Omega\sim27\omega_{0}$,
and the cycle averaged surface velocity is in the order of $0.01\, c$.

We conclude that the harmonic spectrum is rich in information about
the plasma dynamics on the femtosecond timescale. In the presented
experiment, the influence of the laser contrast on the pre-plasma
scale is clearly reflected in the harmonics chirp and thus, in the
spectral line structure. From modulations in the spectrum, we can
estimate the cycle averaged velocity of the electron plasma surface
during its interaction with the main pulse.

\section{\label{sec:3d_harmonics}Relativistic Harmonics in 3D Geometry}

Up to now, we studied the theory of surface HHG in a simplified 1D
geometry. This chapter is dedicated to the investigation of harmonics
spectra and the attosecond pulses in realistic 3D geometry. There
are two new groups of phenomena to be considered:
\begin{enumerate}
\item Due to the extremely broad spectrum of the generated radiation, diffraction
will exert a major influence on the spectrum in the far field. Although
the photon flux through a surface surrounding the whole solid angle
of $2\pi$ in front of the plasma remains constant for each frequency,
we expect the radiation field to be inhomogeneous so that the spectrum
changes as a function of position. Firstly, because of the different
diffraction lengths of the harmonics, and secondly, because the harmonics
field distribution at the plasma surface may differ in intensity and
phase from that of the driving laser. It is our main concern to investigate
these phenomena since they may lead to very useful applications.
\item In particular for very tightly focused laser pulses ($\sigma\sim\lambda$),
3D effects may play a role in the physics of the harmonics generation
itself, so that 1D theory is not applicable anymore. We discuss the
transition to this regime in section~\ref{sub:3d-focusing-scalings}.
\end{enumerate}
Two works \cite{PhysRevLett.94.103903,Naumova:04} precede the study
of this section, which is based on the author's own work Ref.~\cite{adBPuk2007,pukhov2009relativistic}.
The letter \cite{PhysRevLett.94.103903} by Gordienko \emph{et al.}
studies the coherent focusing of the surface harmonics radiation under
strongly idealized conditions, and Naumova \emph{et al. }\cite{Naumova:04}
concentrate on extremely tightly focused ({}``$\lambda^{3}$-regime'')
pulses. Here, we present a broader overview of 3D phenomena that play
a role in HHG experiments.

We begin with some analytical estimations about how the harmonics
spectrum changes due to diffraction in vacuum (subsection~\ref{sub:3d_estimate_diffraction}).
Then (subsection~\ref{sub:3d-LI-generation}), we verify the assumption
of {}``locally independent'' HHG, to see down to which focal spot
sizes 1D theory is still applicable for the generation process itself.
We propose to exploit diffraction effects as {}``spatial spectral
filters'' (subsection~\ref{sub:3d-SSF}) - an alternative or supplement
to spectral transmission filtering to unveil the attosecond pulses
in the harmonics radiation. The coherent focusing of harmonics (CHF),
as proposed in Ref.~\cite{PhysRevLett.94.103903}, is a promising
pathway towards extremely strong fields, perhaps reaching up to intensities
that should allow us to observe exotic effects such as non-linear
vacuum polarization \cite{schwinger1951ongauge,brezin1970pairproduction},
predicted by quantum electrodynamics (QED). In subsection~\ref{sub:3d-focusing-scalings},
we discuss it, thereby considering a more realistic setup compared
to the original proposition in Ref.~\cite{PhysRevLett.94.103903}.

\subsection{\label{sub:3d_estimate_diffraction}Harmonics spectrum changes due
to vacuum propagation}

Vacuum propagation exerts an intriguing influence on high harmonics
radiation generated from solid surfaces. We start with simple analytical
estimations in order to illustrate this. Further, we aim to give an
idea of how these effects might be harnessed to work in our favour.

We begin by considering a linearly polarized Gaussian laser pulse
normally incident onto a planar and sharp-edged overdense plasma surface.
The generated harmonic spectrum can be approximated by a power law
everywhere close to the surface, so that $I_{\mbox{near}}(r,\,\omega)\approx I_{0}(r)\,(\omega/\omega_{0})^{-p}$
for $\omega\gg\omega_{0}$, where the exponent $p$ depends on the
exact HHG mechanism, e.g. $p=8/3$ in the BGP case (see section~\ref{sub:spectrum-ARP}).
If we neglect the intensity dependence of absorption, $I_{0}(r)$
is proportional to the intensity of the incoming beam, and therefore
Gaussian.

Let us at first regard the idealized case that the phase of the generated
harmonics does \emph{not} depend on $r$. Since the frequency $\omega$
is of course very different for distinct harmonic orders but the focal
spot size $\sigma$ is the same for all of them, we easily find that
the beam divergence $\theta=\theta_{0}\omega_{0}/\omega$ is inversely
proportional to the harmonic order. The high orders are emitted into
a much smaller cone than the lower orders. On the optical axis, the
spectrum in the far-field is therefore expected to show a much stronger
pronunciation of high harmonic orders compared to the spectra obtained
within 1D models.

Let us therefore evaluate the development of the spectrum on the optical
axis. We are also interested in including the intensity-dependent
frequency cut-off in our estimate, so we start with: \begin{equation}
I_{\mbox{near}}(r,\omega)=I_{0}(r)\,\left(\frac{\omega}{\omega_{0}}\right)^{-p}\,\theta\left(\left[a(r)\right]^{q}\omega_{c}-\omega\right).\label{eq:Near-Field}\end{equation}
Here, the cut-off frequency is assumed to have a power law dependence
on the vector potential amplitude $a(r)$ at a given point of the
surface.

The general starting point for our computations of the far-field is
the Kirchhoff formula:

\begin{equation}
\psi(\vec{r}')=\frac{1}{4\pi}\oint\vec{dA}\cdot\left(G(\vec{r},\vec{r}')\nabla\psi(\vec{r}')-\psi(\vec{r})\nabla G(\vec{r},\vec{r}')\right),\label{eq:Kirchhoff1}\end{equation}

\noindent with $G(\vec{r},\vec{r}')=\exp(i\omega|\vec{r}-\vec{r}'|)/|\vec{r}-\vec{r}'|$.
Specialized to the geometry of a beam focused onto a planar surface,
assuming cylindrical symmetry and using $F_{-}=0.5(E_{y}-B_{z})$
this becomes

\begin{equation}
\left.F_{-}(x,t)\right|_{r=0}=\frac{1}{x}\int_{0}^{R_{mx}}r\, dr\,\partial_{t}F_{-}\left(x=0,r,t-\frac{\sqrt{x^{2}+r^{2}}}{c}\right),\label{eq:Kirchhoff2}\end{equation}

\noindent where $|x|\gg R_{mx}\sim\sigma$ is assumed. Equation~\eqref{eq:Kirchhoff2}
is specialized to planar surfaces, but the results of the following
calculations may easily be re-interpreted for curved surfaces, as
we are going to see in section~\ref{sub:3d-focusing-scalings}.

\noindent Applying \eqref{eq:Kirchhoff2} to the harmonics generated
by a Gaussian laser pulse as given by Eq.~\eqref{eq:Near-Field}
and no phase dependence on $r$ included, we find the far field spectrum
to be\begin{eqnarray}
I(x,\omega) & = & I_{0}\frac{\left(\frac{\omega}{\omega_{0}}\right)^{-p+2}}{\left(\frac{x}{x_{Rl}}\right)^{2}+\left(\frac{\omega}{\omega_{0}}\right)^{2}}\left(1-\frac{1}{a_{0}}\sqrt[q]{\frac{\omega}{\omega_{c}}}\right)^{2}\nonumber \\
 & \underset{x,\, a_{0}\rightarrow\infty}{\approx} & \frac{I_{0}x_{Rl}^{2}}{x^{2}}\left(\frac{\omega}{\omega_{0}}\right){}^{-p+2},\label{eq:theoretspec}\end{eqnarray}

\noindent wherein $x_{Rl}=\pi\sigma^{2}/\lambda$ is the Rayleigh
length of the fundamental.

\noindent Eq.~\eqref{eq:theoretspec} shows explicitly how vacuum
propagation influences the harmonics spectrum on-axis. Just by picking
the right point in space in front of the harmonics-generating surface,
we can find a spectrum decaying two powers slower than the spectrum
predicted by 1D theory, i.e. (using the BGP exponent $p=8/3$) $I\propto\omega^{-2/3}$.
Physically, the reason for this is the much stronger collimation of
the higher harmonic orders.

Another interesting detail is that the sharp spectral cut-off in the
near-field yields a soft roll-off in the far-field. Provided the far
field spectrum can be measured accurately, conclusions on the constants
$q$ and $\omega_{c}$, which determine the general intensity dependence
of the harmonics cut-off (see Eq. \eqref{eq:Near-Field}) in the near-field
are possible in principle.

However, the above calculation presumed that there is no phase dependence
on the distance from the optical axis $r$ in the near field. In section~\ref{sub:Attosecond-phase},
we have seen that the attosecond phase $\phi$ depends on the $S$-parameter
of the interaction. Therefore, to produce the effect explained above,
it is necessary to keep the local $S$-parameter constant along the
surface. In section~\ref{sub:3d-SSF}, we discuss methods to achieve
this, employing PIC simulation to substantiate our proposal.

Now let us estimate the consequences of the variation of the local
$S$-parameter along the surface in the case of a Gaussian laser pulse,
normally%
\footnote{For s-polarized oblique incidence, the results remain valid when replacing
$S_{0}$ by the effective $S$-parameter $S_{\textrm{eff}}=S_{0}/\cos^{2}\theta$.%
} incident on a surface with steep density gradient up to a constant
density. Due to the curved phase surface, we expect the harmonics
pulse to self-focus. Applying Eq.~\eqref{eq:phase_dep}, we can calculate
the curvature of the generated phase surface and consequently, the
self-focusing distance $x_{sf}$:

\begin{equation}
x_{sf}=\frac{S_{0}}{2.7}x_{Rl}.\label{eq:self_focusing_normal_incidence-1}\end{equation}

Due to this self-focusing, the divergence angle of the individual
harmonics is not simply proportional to the harmonic wavelength anymore.
Equation~\eqref{eq:phase_dep} allows us to derive an expression
for the divergence angle:\begin{equation}
\theta_{\lambda}=\theta_{0}\,\sqrt{\left(\frac{\lambda}{\lambda_{0}}\right)^{2}+\left(\frac{2.7}{S_{0}}\right)^{2}},\label{eq:divergence_normal_incidence}\end{equation}
wherein $\theta_{0}$ is the solid angle, from which the laser itself
is focused.

\subsection{\label{sub:3d-LI-generation}Checking the assumption of locally independent
generation}

In the above calculations we have assumed, that the generation process
itself can be described by the 1D models discussed in section~\ref{sec:Harmonics-Theory}.
Only to investigate the diffraction of the emerging radiation, we
consider the real 3D geometry. In other words, we presumed that the
harmonics are generated \emph{locally independently} at each point
of the surface. This means that the radial field gradient has no influence
on the harmonics spectrum and phase at a certain point. There is no
transverse energy transfer. Mathematically this condition can be expressed
as\begin{equation}
F_{-}(y,z,t)=F_{-}[\left.F_{+}\right|_{y'=y,\, z'=z,\, t'\leq t}]\label{eq:locallyindependtly}\end{equation}
where $F_{-}$ stands for the reflected field and $F_{+}$ for the
field of the incoming radiation.

Note that this assumption also allows us to perform 1D instead of
3D PIC simulations, even if we are interested in the far-field of
a realistic 3D geometry. Assuming the validity of Eq.~\eqref{eq:locallyindependtly},
we can merge the results of a series of 1D simulations, utilizing
Eq.~\eqref{eq:Kirchhoff2} to obtain the far-field. 1D simulations
are computationally much cheaper and so they can be performed with
a higher resolution in the same amount of time on the same computer.

Let us now check in which parameter region the condition \eqref{eq:locallyindependtly}
is satisfied. The reflected field is of course generated by plasma
electrons. The electrons are driven by the electromagnetic field of
the laser pulse. In the ultra-relativistic case, the size of the electron
orbits is on the order of $\lambda$. The scale length, on which the
radiation intensity at the surface changes in radial direction is
the beam waist $\sigma$. Therefore, if $\sigma\sim\lambda$, the
electrons might mediate between regions of different intensities,
endangering the validity of Eq.~\eqref{eq:locallyindependtly}. If
$\sigma\gg\lambda$, the electrons are not able to travel this distance
and one expects that \eqref{eq:locallyindependtly} is fulfilled.

We now compare results of 1D and 3D PIC simulations to verify this.
3D simulations were performed using the spot sizes $\sigma=5,\,2,\,1,\,0.5\,\lambda$,
all with the dimensionless laser amplitude $a_{0}=30$ and the plasma
density $n_{e}=90\, n_{c}$, where $n_{c}=\omega_{0}^{2}m/4\pi e^{2}$
is the critical density. The laser pulse is linearly polarized in
$y$-direction. The reflected field in the 3-dimensional PIC simulations
was always recorded at a distance of $1\lambda$ to the originally
sharp-edged plasma surface. This field is directly compared to the
result of 1D PIC simulations with the same parameters ($\delta_{1D,3D}$).
Anyway, for the most tightly focused pulses the field recording distance
is in the order of the fundamental Rayleigh length so that a direct
comparison with 1D PIC is pointless. Instead of this, it can be compared
to the far field calculated with equation \eqref{eq:Kirchhoff2} from
1D results. For $\sigma=1\,\lambda$, the distance is already too
big to compare directly but yet too small to use \eqref{eq:Kirchhoff2},
so the value is missing here.

Yet two more comparisons were performed. The absolute far field ($x\rightarrow\infty$)
calculated by Eq.~\eqref{eq:Kirchhoff2} from 1D and 3D PIC data
is compared ($\delta_{1D,3D}^{\mathrm{far}}$). This method verifies
the accuracy of our quasi-1D calculations of the far-field directly.

The third comparison ($\delta_{y,z}$) concerns the radial symmetry
of the reflected pulse. In the 3D geometry, it can in principle be
broken because of the linear polarization of the incoming laser pulse,
but it obviously cannot be broken as long as \eqref{eq:locallyindependtly}
holds. Thus this symmetry check provides another indirect criterion
for verifying \eqref{eq:locallyindependtly}. The fields compared
are $F_{-}(x=1\lambda,\, y=\sigma\sqrt{\ln2},\, z=0,\, t)$ and $F_{-}(x=1\lambda,\, y=0,\, z=\sigma\sqrt{\ln2},\, t)$,
both obtained from the 3D PIC simulations.

\begin{table}
\begin{tabular}{cccc}
\hline
$\sigma/\lambda$ & $\delta_{1D,3D}$ & $\delta_{1D,3D}^{\mathrm{far}}$ & $\delta_{y,z}$\tabularnewline
\hline
5.0 & 0.011 & 0.018 & 0.016\tabularnewline
2.0 & 0.011 & 0.019 & 0.036\tabularnewline
1.0 & - & 0.036 & 0.115\tabularnewline
0.5 & 0.196 & 0.092 & 0.240\tabularnewline
\hline
\end{tabular}\caption{\label{tab: limits}Deviations from the assumption of locally independent
harmonic generation according to the measure \eqref{eq:deviation}.}
\end{table}

The results of all these comparisons are collected in Tab.~\ref{tab: limits}.
The relative deviations are measured using \begin{equation}
\delta_{1,2}=\frac{\int dt\:|f_{1}(t)-f_{2}(t)|^{2}\ }{\int dt\:|f_{1}(t)|^{2}+|f_{2}(t)|^{2}}\label{eq:deviation}\end{equation}

This measure is 0 if $f_{1}$and $f_{2}$ are identical functions
and 1 if they are completely uncorrelated.

It can be seen that deviations are very small for not too tiny focal
spots such as $\sigma=5\lambda$. Thus, models based on Eq.~\eqref{eq:locallyindependtly}
may be used for the vast majority of today's HHG experiments. Even
in the case $\sigma=\lambda$, Eq.~\eqref{eq:locallyindependtly}
still holds as a rough approximation. Our studies of even tinier focal
spots have shown, that the deviations from Eq.~\eqref{eq:locallyindependtly}
are generally not favourable for the generation of attosecond pulses.

We now go on to discuss diffraction effects, assuming the validity
of Eq.~\eqref{eq:locallyindependtly}, which is correct for not too
small focal spots in the sense explained above.

\subsection{\label{sub:3d-SSF}Self-focusing and spatial spectral filtering using
Super-Gaussian pulses or Constant-$S$ surfaces}

In this subsection we would like to present the results of some numerical
experiments. These experiments were carried out with the 1D version
of the VLPL (Virtual Laser Plasma Laboratory) PIC (Particles In Cells)
code in combination with a 3D cylindrical geometry numerical propagator
based on Eq.~\eqref{eq:Kirchhoff2} to obtain the far-field on the
optical axis from each series of 1D simulations. The simulations were
made with HHG at a planar surface in mind, but in subsection~\ref{sub:3d-focusing-scalings}
we are going to see that all results can easily be re-interpreted
to suit HHG at a curved surface in confocal geometry. This is crucial
in regard of the exciting possibilities opened up by coherent focusing
of the harmonics radiation.

We start with a simulation of a Gaussian laser pulse, given by $a(x,r,t)=a_{0}\,\textrm{Re}\left[\exp\left(i\,\omega_{0}(x/c-t)-(t/\tau)^{2}-(r/\sigma)^{2}\right)\right]$,
normally incident onto a surface with a steep and constant density
profile. The laser parameters are: $a_{0}=20$, $\sigma=5\lambda$,
$\tau=2\pi/\omega_{0}$. The surface density is $n_{e}=90\, n_{c}$,
so that the ultra-relativistic similarity parameter \cite{gordienko:043109}
$S_{0}=n_{e}/(a_{0}n_{c})=4.5$ at the maximum of the laser pulse.

\begin{figure}
\begin{centering}
\includegraphics[width=5in]{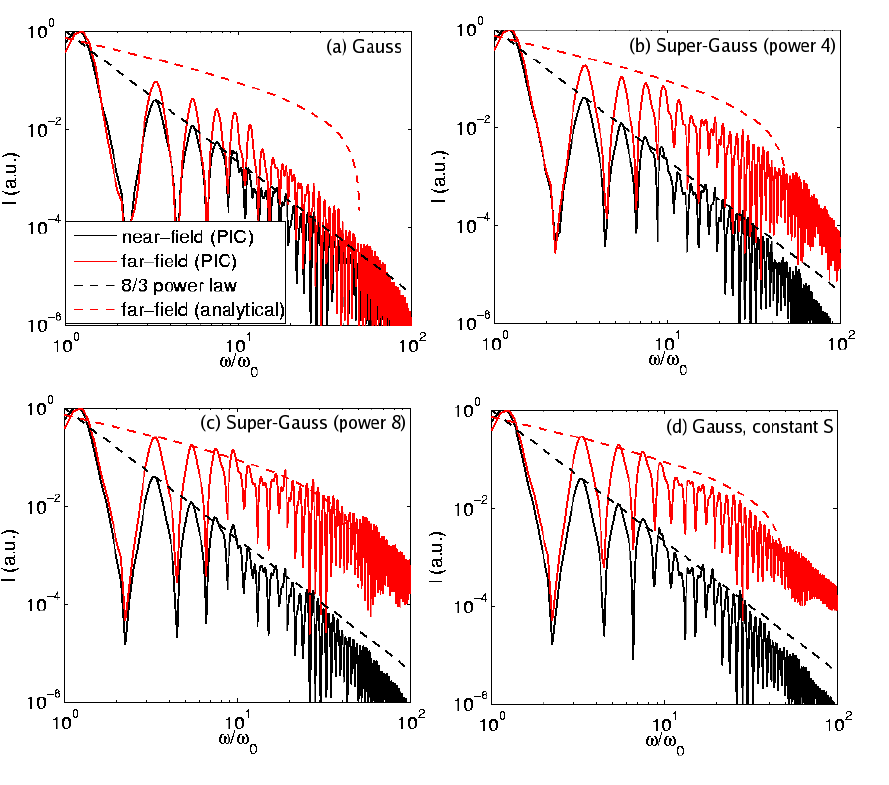}
\par\end{centering}

\caption{\label{fig:spectra}Spectra, near- and far-field ($x=\infty$) for
different types of laser pulses, normalized to the intensity of the
fundamental. ($a_{0}=20$)}
\end{figure}

The resulting spectra for the near and the far-field on the optical
axis in comparison to the analytical estimate \eqref{eq:theoretspec}
disregarding the phase variation are shown in Fig.~\ref{fig:spectra}(a).
It is seen, that there is still a big difference in the slope of the
spectrum compared to Eq.~\eqref{eq:theoretspec}. Although the spectrum
in the far field decays slightly slower than the spectrum close to
the surface, the improvement is far behind from what we could expect
if the attosecond phase remained constant along the surface.

\begin{figure}
\begin{centering}
\includegraphics[width=4in]{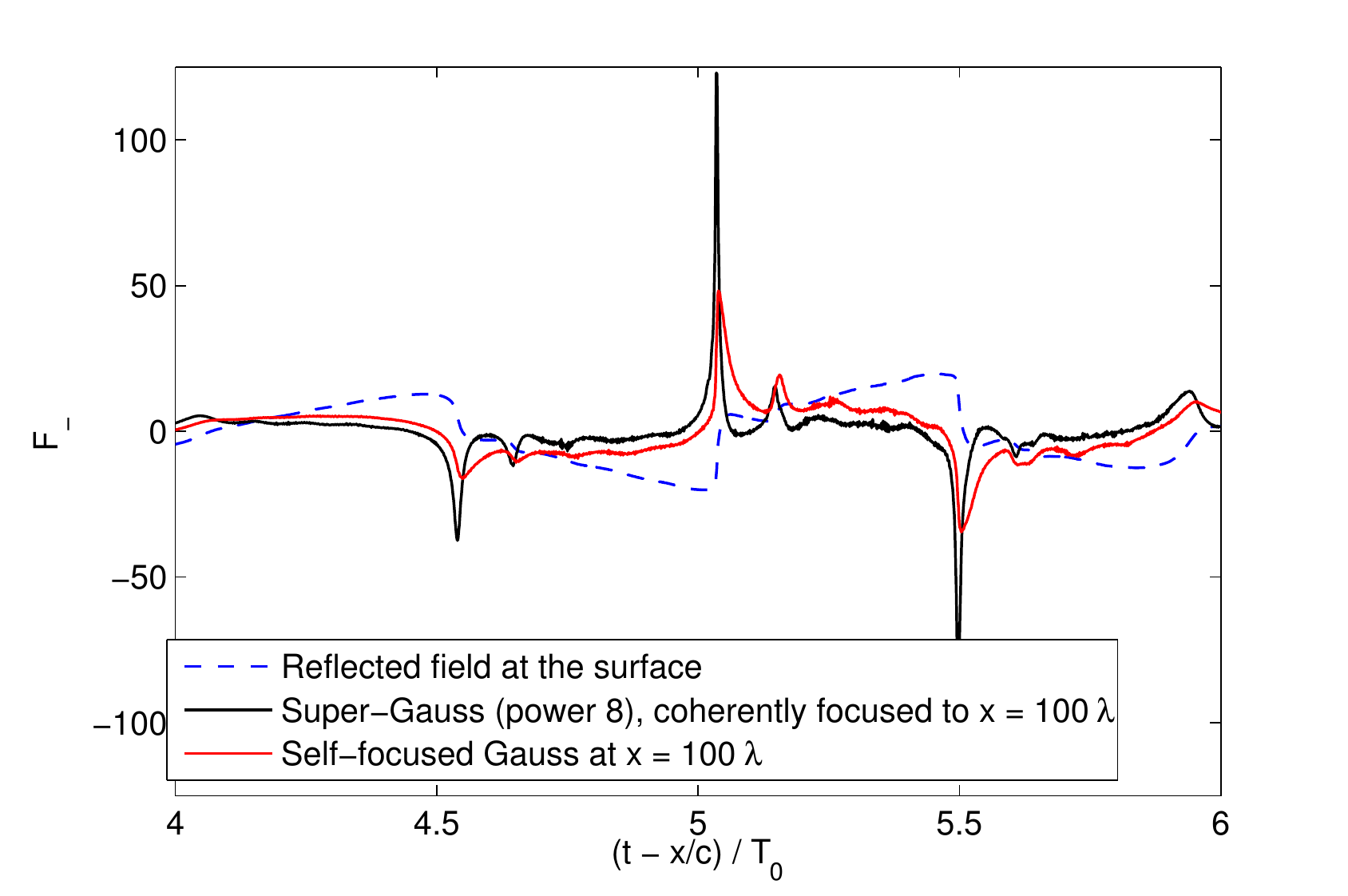}
\par\end{centering}

\caption{\label{fig:self focus}Self-focusing of a Gaussian pulse ($a_{0}=20$,
$\sigma=5\lambda$) in front of a planar surface compared to focusing
of a Super-Gaussian pulse (same amplitude and power) to the same distance
using a curved surface. The time unit $T_{0}=2\pi/\omega_{0}$ is
just the laser period.}
\end{figure}

In Fig.~\ref{fig:self focus} we see what happens at an intermediate
distance from the surface. According to Eq.~\eqref{eq:self_focusing_normal_incidence-1},
the pulse should be self-focused at a distance of around $x_{sf}\approx130\,\lambda$.
Indeed, we observe that the pulse is self-focused, and the self-focusing
length is in reasonable agreement with the analytical estimate. The
achieved intensity is about four times the laser intensity, and it
is reached at a distance of about $100\lambda$ from the surface.
Even attosecond peaks can be seen, yet the contrast ratio is quite
poor.

In order to improve the quality of these pulses, we should aim to
keep the attosecond phase $\phi$ - and thus the relativistic $S$-parameter
- constant alongside the surface. Basically, there are two possibilities
to achieve this:
\begin{enumerate}
\item The use of laser pulses with sufficiently flat intensity distributions
across the focal spot, e.g. Super-Gaussian. This causes the part of
the laser pulse that contributes considerably to the HHG to be at
a nearly constant intensity level.
\item Varying the surface density in a way, so that $S(r)=n_{e}(r)/(a(r)\, n_{c})=\mathrm{const}$.
\end{enumerate}
For testing these ideas, we perform additional simulation runs: some
of them using Super-Gaussian laser pulses $I\propto\exp\left[-(r/\sigma)^{4}\right]$
or $I\propto\exp\left[-(r/\sigma)^{8}\right]$ and some of them using
a conventional Gaussian laser pulse, but a surface with radially varied
density so that the local similarity parameter remains constant $S(r)=4.5$.
In the simulations with the Super-Gaussian laser profile, $\sigma$
was chosen in a way so that the laser power and amplitude are the
same as in the corresponding simulations with the Gaussian pulse.

The spectra obtained from these simulations are depicted in Fig.~\ref{fig:spectra}(b)-(d).
We see a great improvement compared to the unoptimized case Fig.~\ref{fig:spectra}(a).
Evidently, the spectra decay much slower in the far-field, reaching
close to the ideal $I\propto\omega^{-2/3}$ line that was expected
from the analytical estimate in subsection~\ref{sub:3d_estimate_diffraction}.
The lower frequencies are filtered out by diffraction in space. Therefore
we may refer to these schemes as {}``spatial spectral filters''.

\begin{figure}
\begin{centering}
\includegraphics[width=4in]{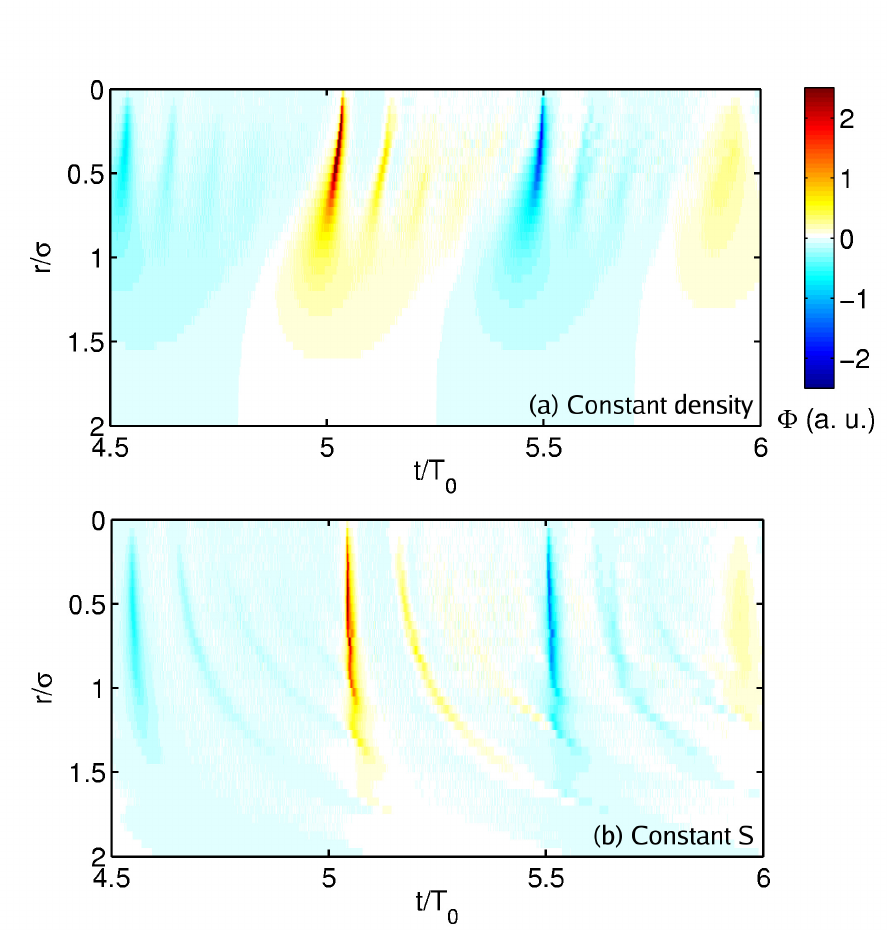}
\par\end{centering}

\caption{\label{fig:integrand}The integrand $\Phi(r,t)=r\,\partial_{t}F_{-}(x=0,r,t)$
from Eq.~\eqref{eq:Kirchhoff2}. The colour scale is the same for
both plots. Gaussian laser pulses with $a_{0}=20$ are used.}
\end{figure}

The advantage of the constant-$S$ surface can nicely be seen in Fig.~\ref{fig:integrand},
showing the integrand from which the far-field is calculated. The
integration to obtain the far-field is carried out along the path
$t'=t-\sqrt{x^{2}+r^{2}}/c$. Thus for the very far field in front
of a planar surface, the integration path becomes a straight line.
The intensity of the integral becomes maximal if the integrand is
big over the whole integration path. Therefore we see, that self-focused
or defocused attosecond pulses are represented by curved lines in
our diagram, whereas non-self-focused attosecond pulses are represented
by straight lines. Further, longer lines lead to higher peak intensities.
With this knowledge, the advantage of a constant-$S$ surface can
be easily understood from Fig.~\ref{fig:integrand}. Notice also
that for the constant-$S$ surface the side peaks are strongly defocused,
yielding a better contrast ratio to the main peaks. For Super-Gaussian
pulses the image would look similar to Fig.~\ref{fig:integrand}(a),
except for that the upper part is stretched.

\begin{figure}
\includegraphics[width=0.05\columnwidth]{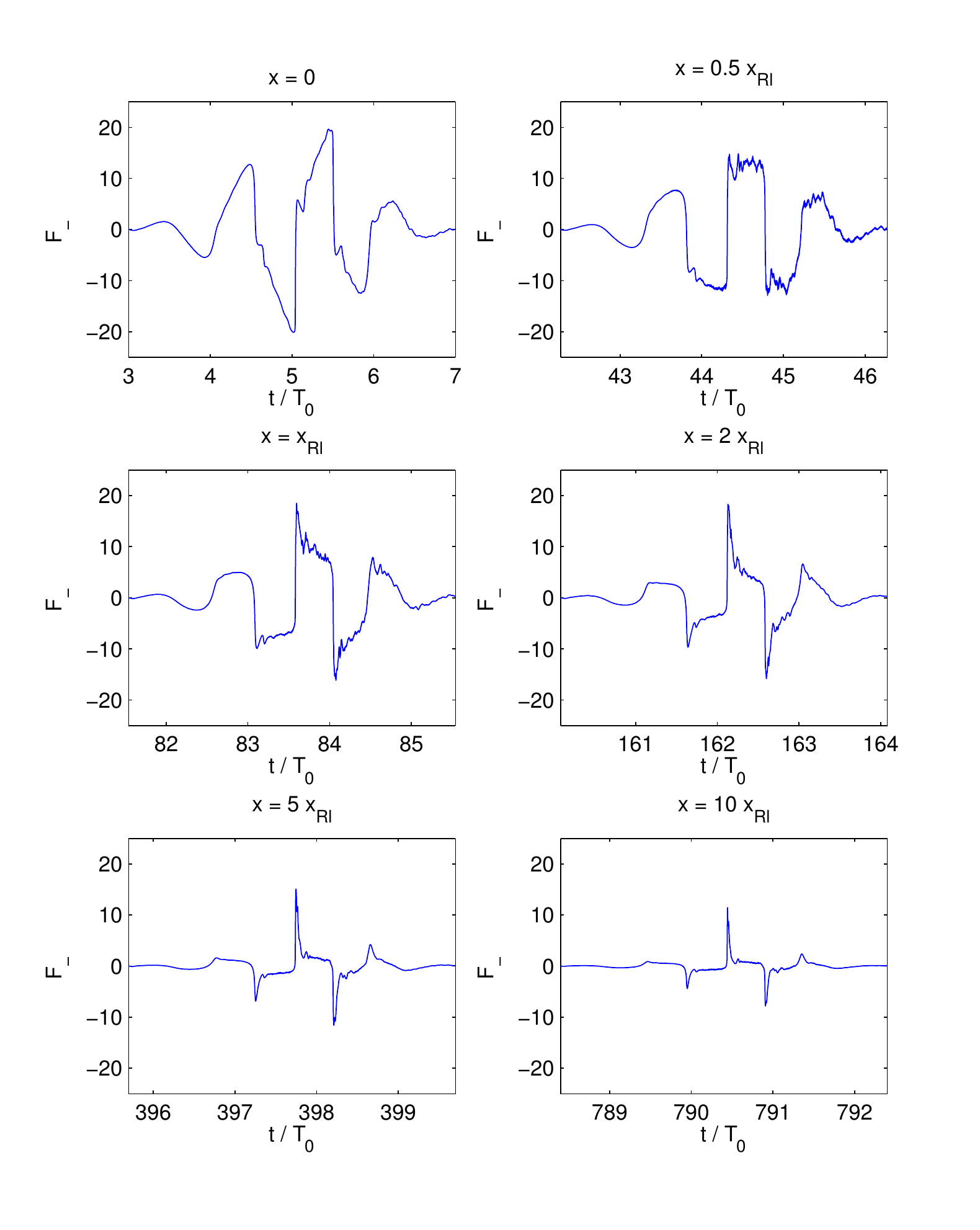}

\caption{\label{fig:move away}The reflected field on the optical axis, observed
at different distances $x$ from the surface. Data from PIC simulation
using a Gaussian laser pulse with $a_{0}=20$ focused on a planar
constant-$S$ surface.}
\end{figure}

To liven up the picture of how the attosecond pulses emerge, we take
a look at the on-axis field at different distances $x$ from the surface.
For Fig.~\ref{fig:move away} we choose a simulation with a constant-$S$
surface, because the process of {}``vacuum'' attosecond pulse generation
is most pronounced here. While we depart from the surface together
with the reflected radiation, we see how the attosecond pulses get
rectified and the whole rest of the radiation is simply diffracted
away from the optical axis.

Focusing these improved pulses using a confocal setting yields a much
better result than the self-focusing of a Gaussian pulse in front
of a planar surface as can be seen from Fig.~\ref{fig:self focus}.
Here a Super-Gaussian pulse was chosen, but the use of a constant-$S$
surface leads to a similar effect, as shown further below.

\begin{figure}
\includegraphics[width=6in]{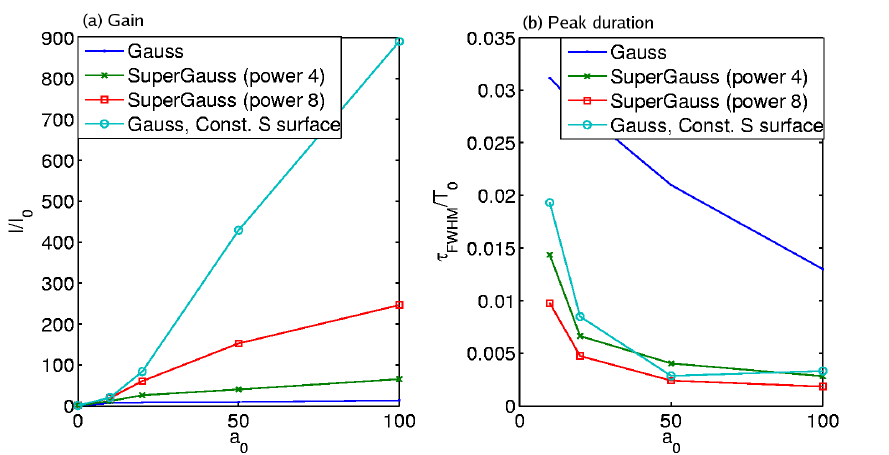}

\caption{\label{fig:Comparison atto}Comparison of the intensity and duration
(full width half maximum) of the far-field attosecond pulses for different
laser intensities, radial laser shapes and plasma surfaces. The intensity
unit $I_{0}=(a_{0}x_{Rl}/x)^{2}$ is the peak intensity value we would
get at the same distance from the surface for a Gaussian laser pulse
with the same power and maximum vector amplitude \emph{without} the
generation of harmonics. The time unit $T_{0}=2\pi/\omega_{0}$ is
the laser period. The constant $S$-line (turquoise) in panel (a)
also denotes the intensity gain that can be expected from the focusing
scheme depicted in Fig.~\ref{fig:chf_schematic}. Here, $S=4.5$
was used.}
\end{figure}

Let us now have a look at a broader range of parameters. In Fig.~\ref{fig:Comparison atto}
the intensity and duration of the attosecond pulses in the far-field
is compared for different laser amplitudes $a_{0}$ and all the proposed
schemes. As shown in the following subsection~\ref{sub:3d-focusing-scalings},
the results can be applied for planar surfaces as well as for focusing
geometries. We compare the attosecond pulses in the far-field of a
planar surface or at the focal spot in front of a spherically curved
surface. We notice once again, that the Super-Gaussian laser pulse
focal spots and the constant-$S$ surfaces yield a clear advantage
for the attosecond pulse generation.

In earlier works \cite{tsakiris2006routeto,bgp-theory2006}, transmission
filtering has been suggested as a technique to improve the quality
of the attosecond pulses. Let us compare this to our method of spatial
spectral filtering via shaping of the laser pulse focal spot.

Applying transmission filters directly to the results of 1D PIC calculations,
as it has been done in previous works, yields a somewhat unrealistic
picture, since the filters have to be placed inside the far-field
in a real experiment. In this work we consider the 3D geometry and
apply optical filters to the far-field radiation.

\begin{figure}
\includegraphics[width=5in]{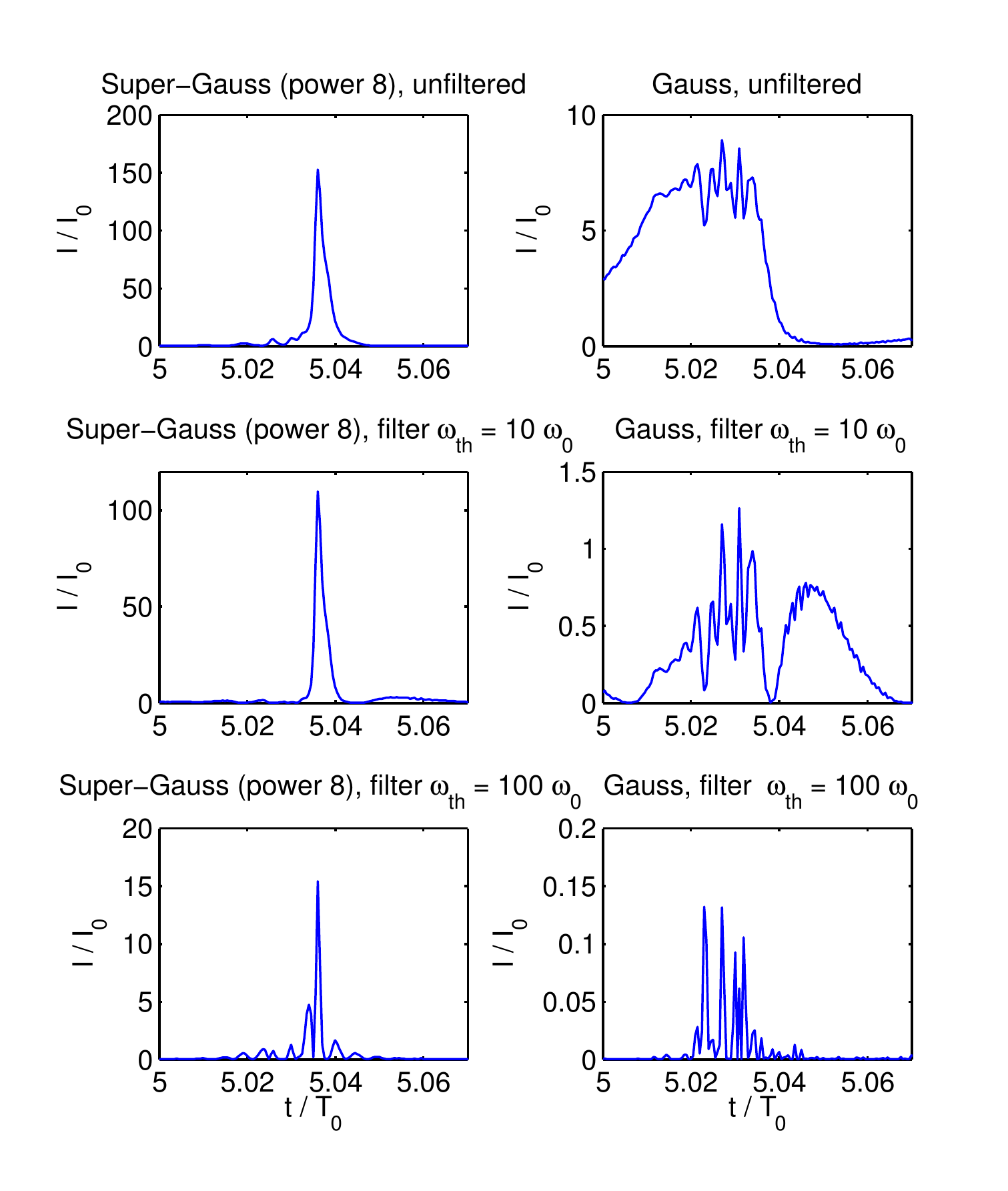}

\caption{\label{fig:filtered}The temporal structure of the attosecond pulses
in the far-field of a Supergaussian (left column) and a Gaussian (right
column) laser-pulse, both with the same power and peak amplitude $a_{0}=50$
in the focus. The intensity is normalized so that the maximum in case
of no harmonics and no absorption for the Gaussian Laser would be
at $I/I_{0}=1$. In the second and third lines high-pass filters are
applied. The filter function is a simple step like function with a
linear transition of the width $\delta=\omega_{0}$, the threshold
frequency is $\omega_{th}$.}
\end{figure}

We compare attosecond pulses generated by lasers with a Gaussian and
a Super-Gaussian focal spot, see Fig.~\ref{fig:filtered}. The first
thing to notice is that filtering influences the temporal structure
of the attosecond pulses. While the attosecond pulses in the unfiltered
far-field are pure half-cycle pulses, optical filtering can generate
single- or multi-cycle pulses, depending on the filter frequency,
see also Sec.~\ref{sub:attopulse-characterization}. Then, unlike
transverse pulse shaping (see Fig.~\ref{fig:Comparison atto}), transmission
filtering naturally leads to a decrease of the attosecond peak intensity.
To obtain a significantly shortened pulse, one needs to use filters
with a very high threshold frequency, eating up most of the pulse
energy. Nevertheless, filtering leads to an improvement of the contrast
ratio by a factor of about 3 for $\omega_{th}=100\omega_{0}$.

To get the shortest possible pulse duration and the best contrast
ratio, we recommend to combine transverse pulse shaping with the use
of an optical filter. When attosecond pulses with a maximum peak intensity
are required, the transverse pulse shaping or constant-$S$ surfaces
combined with the proper focusing geometry are the best option.

\subsection{\label{sub:3d-focusing-scalings}Optical scalings for harmonics focusing}

In this subsection, we examine focusing geometries for the surface
harmonics radiation. In Ref.~\cite{PhysRevLett.94.103903} it has
been shown, that coherent harmonic focusing (CHF) has the potential
to produce unprecedentedly intense electromagnetic fields. The created
intensities may be so extreme, that they can be used to explore exotic
QED effects such as vacuum polarization or even electron-positron
pair creation \cite{schwinger1951ongauge,brezin1970pairproduction}.
However, the conditions under which the phenomenon was examined in
Ref.~\cite{PhysRevLett.94.103903} were strongly idealized: A perfect
spherical wave, uniformly illuminating a curved plasma surface with
a large solid angle of $\Omega=1$ and a tiny radius of $R=4\lambda$
was studied - something that is not achievable with a focused Gaussian
beam. Therefore, important effects such as the variation of the laser
intensity on the harmonics generating surface were not taken into
account. Here, we discuss CHF under more realistic conditions.

In order to better understand CHF, we start by assembling some optical
scaling laws for the broadband harmonics radiation. These laws are
immediate consequences of Eq.~\eqref{eq:Kirchhoff1}. As the geometries
and frequency spectra involved in CHF may be unusual, we also make
an effort to give conditions of validity for the scaling laws in the
cases when they are different from the ones for the fundamental Kirchhoff
integral~\eqref{eq:Kirchhoff1}.

Before we start considering curved surfaces, we have a look at what
happens, when the size of the focal spot is changed on a planar surface.
First, we consider the case when the laser field is focused onto a
planar surface and the focal spot size is varied but the maximum amplitude
of the vector potential $a_{0}$ is kept constant: \[
F_{-}^{(1)}(0,\, r,\, t)=F_{-}^{(\alpha)}(0,\,\alpha r,\, t),\]
wherein $\alpha$ is the dimensionless factor describing the focal
spot scaling. Now assuming $\pi\sigma^{4}/(4\lambda x^{3})\ll1$,
we can calculate that the reflected radiation scales like:

\begin{eqnarray}
F_{-}^{(1)}(x,\,0,\, t-x/c) & = & F_{-}^{(\alpha)}(\alpha^{2}x,\,0,\, t-\alpha^{2}x/c).\label{eq:spotsizes}\end{eqnarray}

Therefore, as long as Eq.~\eqref{eq:locallyindependtly} holds, a
variation in the focal spot size will yield an exactly similar field
structure in the far-field, just scaled in size.

\begin{figure}
\begin{centering}
\includegraphics[width=4in]{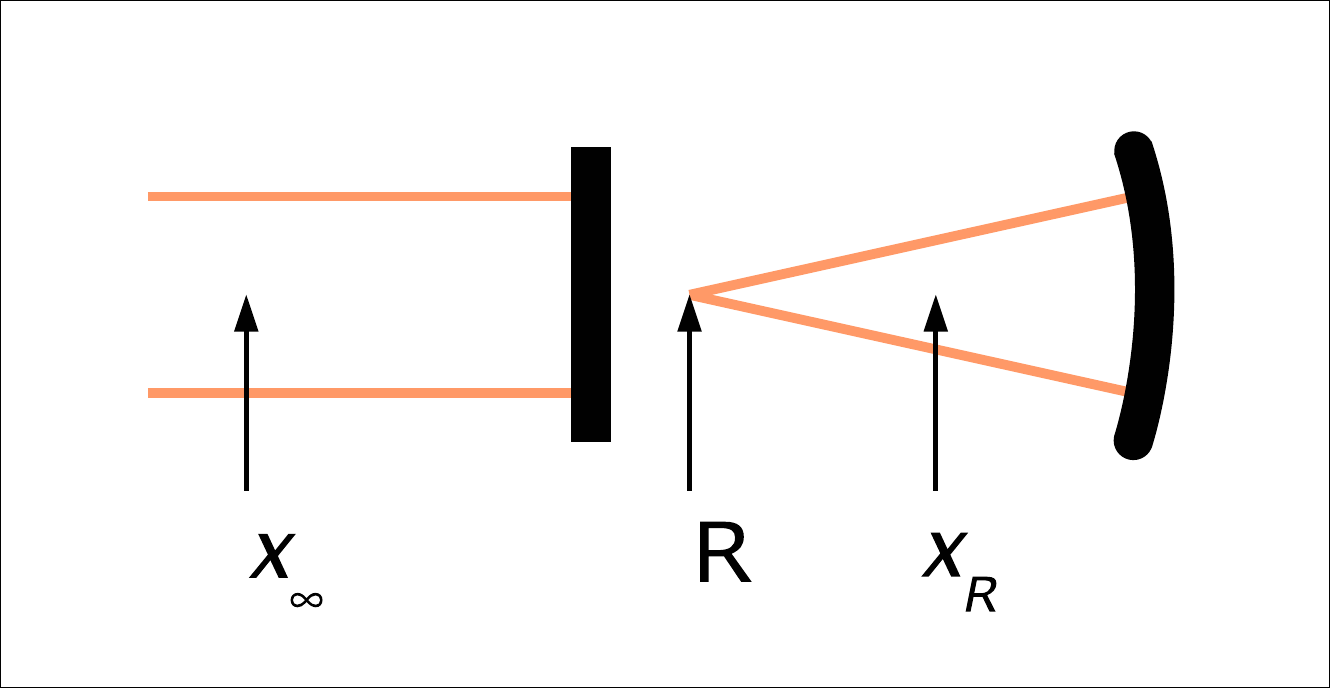}
\par\end{centering}

\caption{\label{fig:geometry}Illustration of the geometry underlying Eqs.~\eqref{eq:distances},~\eqref{eq:focusing}}
\end{figure}

Now we are interested in actively focusing the harmonics radiation.
The most straightforward way to do so is the use of a curved surface
for HHG, since it can do without relying on any optical components
suitable for the extremely broadband radiation. We consider a confocal
geometry, in the sense that the field distribution $F_{-}^{(R)}$
of the radiation on the curved surface is the same as for the focal
spot in the planar geometry $F_{-}^{(\infty)}$ - in both phase and
intensity. This can be written as:

\[
F_{-}^{(R)}(x=\sqrt{R^{2}-r^{2}}-R,\, r,\, t)=F_{-}^{(\infty)}(0,\, r,\, t),\]
where $R$ is the radius of the curved surface. We find that the field
in front of the planar surface at the distance $x_{\infty}$ is similar
to that in front of the curved surface at the distance $x_{R}$, where
$x_{\infty}$ and $x_{R}$ are related according to

\begin{equation}
\frac{1}{x_{R}}=\frac{1}{x_{\infty}}+\frac{1}{R}\label{eq:distances}\end{equation}

\noindent In this case, the field relation is

\begin{eqnarray}
F_{-}^{(R)}(x_{R},\,0,\, t-x_{R}/c) & = & \frac{x_{\infty}}{x_{R}}F_{-}^{(\infty)}(x_{\infty},\,0,\, t-x_{\infty}/c).\label{eq:focusing}\end{eqnarray}

This result becomes exact for small focusing solid angles, but also
in the most interesting limit $x_{R}\rightarrow R$, which corresponds
to the actual focal spot when there is no self-focusing present. The
scaling law allows us to simply re-interpret all results obtained
for a planar surface in subsection~\ref{sub:3d-SSF} to such for
a spherical; of course, as long as the condition \eqref{eq:locallyindependtly}
holds.

Next, we consider the variation of the focal distance, but keep the
field amplitude at the surface and the solid angle constant:

\begin{eqnarray}
F_{-}^{(\alpha)}(\alpha r,\,\theta,\, t) & = & F_{-}^{(1)}(r,\,\theta,\, t)\nonumber \\
\Rightarrow F_{-}^{(\alpha)}(0,\, t-\alpha r/c) & = & \alpha F_{-}^{(1)}(0,\, t-r/c).\label{eq:intensityscaling}\end{eqnarray}

So if the focal distance is varied, the intensity in the focal spot
increases proportionally to the input power. Since a higher intensity
at the surface generally creates a bigger number of harmonics and
therefore leads to more than linear amplification of the radiation
in the focal spot, we should aim for a focal distance as small as
possible.

If the solid angle of the mirroring surface is varied, but the field
amplitude and the focal distance are kept constant, the intensity
in the focal spot increases stronger than the input power:

\begin{eqnarray}
F_{-}^{(\alpha)}(R,\,\alpha\theta,\, t) & = & F_{-}^{(1)}(R,\,\theta,\, t)\nonumber \\
\Rightarrow F_{-}^{(\alpha)}(0,\, t) & = & \alpha^{2}F_{-}^{(1)}(0,\, t).\label{eq:anglescaling}\end{eqnarray}

This intensity gain is even stronger than the one achieved by the
temporal focusing gain due to HHG. Therefore, to maximize the focal
spot intensity with a constant laser power, it is first needed to
maximize the solid angle of CHF. Of course, this solid angle is limited
by the focusing geometry of the driving laser itself. Second, the
CHF distance should be minimized in order to make maximum use of the
temporal focusing gain due to surface HHG. As shown in subsection~\ref{sub:3d-SSF},
the use of a constant-$S$ surface would be ideal here, compare Fig.~\ref{fig:Comparison atto}.

\begin{figure}
\begin{centering}
\includegraphics[width=3in]{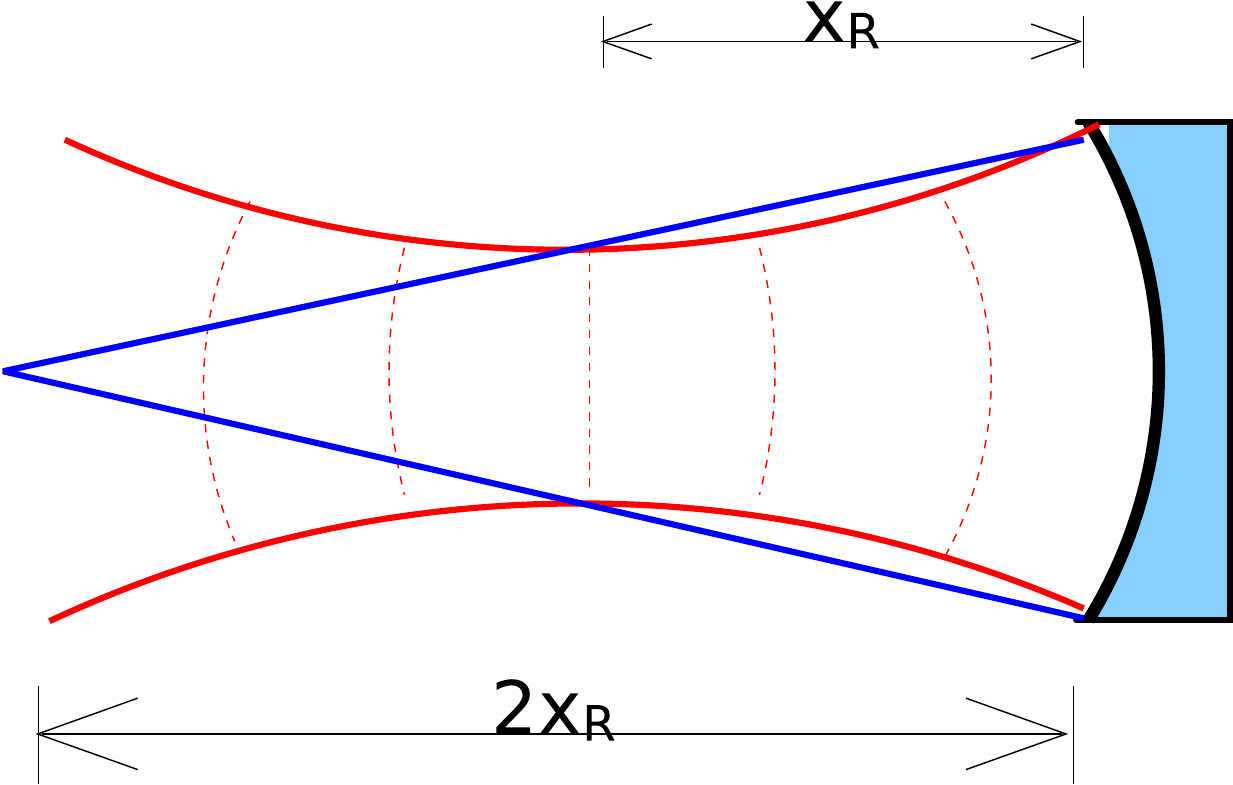}
\par\end{centering}

\caption{\label{fig:chf_schematic}Suggested geometry for coherent focusing
of the surface harmonics radiation, produced by a Gaussian laser beam.
Ideally, the plasma surface should be designed in a way, that the
$S$-parameter does not change with the distance from the optical
axis.}
\end{figure}

A focusing geometry that fulfils the above criteria is presented in
Fig.~\ref{fig:chf_schematic}. The laser is focused to a distance
of one Rayleigh length $x_{R}$ in front of the HHG surface. The plasma
surface is spherically curved with a radius of $2x_{R}$ and the density
of the surface is modulated in a way, that the $S$-parameter $S=n_{e}/(a_{0}n_{c})$
is constant everywhere. The harmonics radiation will then be coherently
focused to a distance of $2x_{R}$ in front of the curved surface.
\begin{table}
\begin{tabular}{|c|c|c|c|c|}
\hline
$I\lambda^{2}\,(\mathrm{W\, cm^{-2}\,\mu m^{2}})$ in focus & $a_{0}$ on surface & \multicolumn{3}{c|}{$\eta_{\mbox{CHF}}$}\tabularnewline
\hline
 &  & $S=2$ & $S=3$ & $S=4$\tabularnewline
\hline
$2.74\times10^{20}$ & 10 & 4.4 & 5.3 & 3.3\tabularnewline
\hline
\selectlanguage{english}%
$2.74\times10^{22}$\selectlanguage{british}
 & 100 & 337 & 346 & 187\tabularnewline
\hline
\selectlanguage{english}%
$2.74\times10^{24}$\selectlanguage{british}
 & 1000 & 1185 & 5250 & 5311\tabularnewline
\hline
\end{tabular}

\caption{\label{tab:chf_gain}Expected gain in temporal peak intensity for
coherent focusing of harmonics according to the scheme Fig.~\ref{fig:chf_schematic},
as computed by series of 1D-PIC simulations in combination with a
diffraction integral.}
\end{table}

The intensity gain achieved by this scheme is directly related to
the idealized gain displayed in Fig.~\ref{fig:Comparison atto}.
However due to the inevitable finite distances from the focus it is
reduced to\begin{equation}
\eta_{\textrm{CHF}}=\frac{1}{8}\,\eta\left(\frac{a_{0}}{\sqrt{2}}\right).\end{equation}

Table~\ref{tab:chf_gain} gives an overview of the gain to be expected
from the scheme for different intensities and surface $S$ parameters.
Note that in the case of $S=3$ and $a_{0}=1000$, as can be expected
for the European ELI project \cite{gerstner2007laserphysics}, an
output intensity of $I_{\mbox{CHF}}\sim10^{28}\,\mathrm{W\, cm^{-2}}$
is computed. This would be sufficient to access an entirely new regime
of physics in which the vacuum itself becomes non-linear due to exotic
QED effects.

In this section, we have discussed the effect of diffraction on the
HHG radiation under realistic conditions. To make maximum use of these,
techniques to shape the focal spot or the plasma surface have to be
implemented. With these techniques, it may be possible to produce
intensities that are more than a thousand times higher than the conventional
focusing intensity of the laser, opening up the possibility of verifying
vacuum QED effects with ultraintense laser systems.

\section{Conclusions}

The reflection of relativistic light at overdense plasma surfaces,
performing a strongly non-linear oscillation, is currently one of
the most promising candidates for the production of intense attosecond
pulses.

We have taken a fresh look at the theory of their generation. The
foundations of the supposedly well-known ROM model have been re-investigated,
yielding a clearer picture of the scope of application of the model.
Further, the model has been extended to higher order $\gamma$-spikes,
demonstrating the possibility of modulated spectral structures and
power law spectra $I\propto\omega^{-q}$ with exponents $q<8/3$ even
within the ROM model. The explicit formulation of the TROM model and
its comparison to the ROM model sheds additional light on the physics
of relativistic high harmonics generation.

A third model was motivated by numerical observations: amazingly dense
and narrow electron {}``nanobunches'' may form at the plasma surface,
emitting coherent synchrotron radiation efficiently. This nanobunching
regime of relativistic HHG is optimal for attosecond pulse generation
in the sense that the generated pulses bear almost the full energy
of the entire optical cycle of the driving laser. Here, we expect
a flat power law spectrum with $q\leq4/3$ up to a smooth cut-off
at a frequency which is determined either by the bunch relativistic
energy $\omega_{rs}\propto\omega_{0}\gamma^{3}$ or by the nanobunch
width $\omega_{rf}\propto c/\delta$ .

We have also studied carefully the phase properties of the relativistic
harmonics, something that has largely been neglected so far. The relation
of the electron surface motion to the spectral line structure has
been investigated. It has been found that the spectral line structure
can deliver valuable information on the motion of the electron surface
on a femtosecond timescale.

The phase properties are also crucial when considering the free space
propagation of the harmonic radiation. We have shown that diffraction
can be harnessed as a spatial spectral filter for the harmonics radiation
by designing the target surface or the laser pulse focal spot in the
right way. This way, attosecond pulses can be extracted efficiently
without the use of optical transmission filters.

Our study can further provide the basis for focusing schemes of the
harmonics radiation. CHF (coherent harmonic focusing) has the potential
to produce intensities that exceed the one of the driving laser by
several orders of magnitude. Our proposed focusing scheme was shown
to yield an intensity enhancement by a factor of 3000 with parameters
as expected for the ELI facility. Here, unlike previous studies, we
took a realistic Gaussian laser pulse geometry into account.

\appendix

\section{\label{sec:Stationary-Phase-Method}Stationary Phase Method}

In this section, the asymptotic evaluation of integrals via the stationary
phase method is explained. The method has e.g. been applied to diffraction
integrals and the calculation of synchrotron spectra and plays a vital
role in the theory of high harmonics generation at overdense plasma
surfaces. For a comprehensive introduction to this method and related
ones, consider e.g. the book by Wong \cite{wong2001asymptotic}. Here,
we describe the method briefly with the applications from section~\ref{sec:Harmonics-Theory}
in mind.

We are interested in integrals of the form\begin{equation}
F(\omega)=\int_{-\infty}^{\infty}g(t)\,\exp\left[i\,\omega\, f(t)\right]\: dt,\label{eq:oscillating_integral}\end{equation}
where $g(t)$ and $f(t)$ are assumed to be smooth functions. We want
to find an asymptotic approximation for $F(\omega)$ in the limit
of big $\omega$. Then, the rapidly oscillating integrand cancels
everywhere except for the regions of {}``stationary phase'' where
$df(t)/dt\approx0$.

In the simplest case, we can find a set of well separated points $\{t_{k}\}$
on the real axis where $df(t_{k})/dt=0$ and $d^{2}f(t_{k})/dt^{2}\neq0$.
Then $f(t)$ and $g(t)$ can be Taylor expanded around these points:
$f(t\approx t_{k})\approx a_{k}+b_{k}(t-t_{k})^{2}$ and $g(t\approx t_{k})\approx c_{k}$.
Now, the integral can be evaluated analytically:\[
{\displaystyle \begin{array}{ccc}
F(\omega) & \approx & {\displaystyle \sum_{k}c_{k}e^{i\omega a_{k}}\int_{-\infty}^{\infty}\exp\left[i\,\omega\, b_{k}(t-t_{k})^{2}\right]\, dt}\\
 & = & {\displaystyle \sqrt{\frac{\pi}{i\omega}}\,\sum_{k}b_{k}^{-1/2}c_{k}\, e^{i\omega a_{k}}}\end{array}.}\]

It can be seen that the behaviour of the integral depends on two factors.
The first factor scales $\propto\omega^{-1/2}$ and constitutes a
spectral envelope, whereas the sum determines if the contributions
from each stationary point interfere positively or negatively and
is thus responsible for the structure of the single {}``harmonic''
lines. This sort of behaviour also extends to the more complicated
cases discussed below, but the envelope factor varies sensitively
according to the exact structure of the stationary phase points. Let
us now go on to discuss the relevant cases.

\subsection{\label{sub:Two-Saddles}First order $\gamma$-spikes}

Taking a look at the integrals that we encounter in the models presented
in Section~\ref{sec:Harmonics-Theory}, we find that none of them
ever contains points where the condition $df(t)/dt=0$ is exactly
fulfilled. It is possible to understand this in terms of physics.
The phase functions $f(t)$ in the models are always connected to
the difference between the trajectory of a point, that is somehow
connected to the plasma motion and the motion of the emitted light
wave. Since the plasma cannot be faster than light, $f(t)$ is strictly
monotonic, consequently $df/dt\neq0$.

Still it is possible to apply the stationary phase method by considering
points where $df(t)/dt\approx0$. Technically, this can be viewed
as a region where two saddle points, that are located in the complex
plane slightly off the real axis, closely merge.

The ROM model, as discussed in Sec.~\ref{sub:harm_theo-ROM} leads
us to integrals of the sort

\begin{equation}
F_{\pm}(\omega)=\int\exp\left[i\left(t(\omega\pm\omega_{0})+x(t)(\omega\mp\omega_{0})\right)\right]\,\left(1+\dot{x}(t)\right)\, dt.\label{eq:ARP_basic_int}\end{equation}

This integral has to be handled with attention: Note that although
$\omega\gg\omega_{0}$ is assumed, we must not neglect $\omega_{0}$
in the exponent. The reason for this will become evident later.

The Taylor expansion of $x(t)$ around the velocity maximum can be
written down as $x(t)=-vt+\alpha t^{3}/3$. In the case $\alpha\neq0$,
we speak about a {}``$\gamma$-spike of the order 1''. In the case,
when $\alpha=0$, higher orders of the Taylor expansion have to be
considered. It is discussed in subsection~\ref{sub:higher-order-gamma}.

We shift the stationary phase point to $t_{0}=0$ without loss of
generality here, as we are not interested in absolute phase terms.
Using the abbreviation $\delta=1-v\approx1/(2\gamma^{2})$, we get
$F(\omega)=F_{1}(\omega)+F_{2}(\omega)$ with\begin{equation}
\begin{array}{ccc}
{\displaystyle F_{1}(\omega)} & {\displaystyle =} & {\displaystyle \delta\int\exp\left[i\left(t\left(\delta\omega\pm(2-\delta)\omega_{0}\right)+\alpha(\omega\mp\omega_{0})\frac{t^{3}}{3}\right)\right]\, dt}\\
{\displaystyle F_{2}(\omega)} & {\displaystyle =} & {\displaystyle \alpha\int t^{2}\,\exp\left[i\left(t\left(\delta\omega\pm(2-\delta)\omega_{0}\right)+\alpha(\omega\mp\omega_{0})\frac{t^{3}}{3}\right)\right]\, dt}\end{array}.\end{equation}
By now it should become clear why $F_{2}$ and the above mentioned
$\omega_{0}$-terms could not be neglected: We want to presume $\omega\gg\omega_{0}$,
but not $\delta\omega\gg\omega_{0}$. Later on, we may neglect $\omega_{0}$
compared to $\omega$ and $\delta$ compared to $1$.

$F_{1}$ and $F_{2}$ can now be expressed in terms of the well-known
Airy function $\textrm{Ai}(x)\equiv(2\pi)^{-1}\int_{-\infty}^{\infty}\exp\left(i\left(xt+t^{3}/3\right)\right)dt$:

\begin{eqnarray}
F_{1}(\omega) & = & \frac{2\pi\delta}{\sqrt[3]{\alpha\left(\omega\mp\omega_{0}\right)}}\,\textrm{Ai}\left(\xi\right)\\
F_{2}(\omega) & = & \frac{-2\pi\xi}{\omega\mp\omega_{0}}\,\textrm{Ai}\left(\xi\right).\end{eqnarray}
with $\xi=\left(\delta\omega\pm(2-\delta)\omega_{0}\right)/\left(\alpha\left(\omega\mp\omega_{0}\right)\right)^{1/3}$.
For the calculation of $F_{2}$, we made use of $\textrm{Ai}''(x)=x\textrm{Ai}(x)$.
After taking the sum of $F_{1}$ and $F_{2}$, the $\omega^{-1/3}$
terms cancel and only the $\omega^{-4/3}$ term remains, which represents
the leading order now:\begin{equation}
F_{\pm}(\omega)=\frac{\pm4\pi}{\sqrt[3]{\alpha}\left(\omega\mp\omega_{0}\right)^{4/3}}\,\textrm{Ai}\left(\xi\right).\label{eq:ARP_F_omg}\end{equation}

Taking the absolute square yields the famous $-8/3$-power law spectrum.

For the TROM model (section~\ref{sub:harm_theo-TROM}), the integral
looks a bit different:

\begin{equation}
F_{\pm}(\omega)=\int\exp\left[i\left(t(\omega\pm\omega_{0})+x(t)(\omega\mp\omega_{0})\right)\right]\,\left(1-\dot{x}(t)\right)\, dt.\label{eq:TROM_basic_int}\end{equation}

The integration works in complete analogy to the case shown above,
but in this case, the $\omega^{-1/3}$-terms do not cancel out. Therefore,
we obtain in highest order:\begin{eqnarray}
F_{\pm}(\omega) & = & \frac{4\pi}{\sqrt[3]{\alpha\left(\omega\mp\omega_{0}\right)}}\,\textrm{Ai}\left(\xi\right).\end{eqnarray}

In the case of CSE (section~\ref{sub:harm_theo-nanobunching}), the
integral is of the sort:\begin{equation}
F(\omega)=\int\dot{y}(t)\,\exp\left[-i\omega\left(t+x(t)\right)\right]\, dt.\label{eq:CSE_basic_int}\end{equation}

To get some meaningful result out of this, we need to make an assumption
about the relation between $y(t)$ and $x(t)$. We assume, that during
the time of harmonic generation, the absolute velocity $(\dot{x}^{2}+\dot{y}^{2})^{1/2}$
is approximately constant and close to the speed of light. This is
reasonable in the ultra-relativistic regime. With this assumption,
the stationary phase points are exactly the points, where $\dot{y}$
vanishes and the electrons move towards the observer. Now we can Taylor
expand $\dot{y}(t)=\alpha_{0}t$ and $x(t)=-vt+\alpha_{1}t^{3}/3$.
Substituting into Eq.~\eqref{eq:CSE_basic_int} yields:

\begin{equation}
F(\omega)=\alpha_{0}\int t\,\exp\left[i\left(-\omega\delta t-\omega\alpha_{1}\frac{t^{3}}{3}\right)\right]\, dt,\end{equation}
where $\delta=1-v$ as above. Again, the result can be expressed in
terms of the Airy function:\begin{equation}
F(\omega)=\frac{-2\pi\alpha_{0}i}{\left(\alpha_{1}\omega\right)^{2/3}}\,\textrm{Ai'}\left(\frac{\delta\omega^{2/3}}{\sqrt[3]{\alpha_{1}}}\right).\label{eq:CSE_G_omg}\end{equation}

\subsection{\label{sub:higher-order-gamma}Higher order $\gamma$-spikes}

In the previous subsection (Sec.~\ref{sub:Two-Saddles}), we dealt
with the case when the transverse velocity of the electrons pass zero.
Together with the assumption of ultrarelativistic motion this lead
us to the Taylor expansion $x(t)=-vt+\alpha t^{3}/3$. In this section
we deal with the possibility, that the transverse velocity does not
go through, but touches zero, so that the third order of $x(t)$ vanishes.

In general, if the first $2n$ orders of $x(t)$ vanish, it can be
written: $x(t)=-vt+\alpha t^{2n+1}/(2n+1)$. We refer to this case
as a {}``$\gamma$-spike of the order $n$''. Inserting this into
Eq.~\eqref{eq:ARP_basic_int} yields:

\begin{equation}
\begin{array}{ccc}
{\displaystyle F_{1}(\omega)} & {\displaystyle =} & {\displaystyle \delta\int\exp\left[i\left(t\left(\delta\omega\pm(2-\delta)\omega_{0}\right)+\alpha(\omega\mp\omega_{0})\frac{t^{2n+1}}{2n+1}\right)\right]\, dt}\\
{\displaystyle F_{2}(\omega)} & {\displaystyle =} & {\displaystyle \alpha\int t^{2n}\,\exp\left[i\left(t\left(\delta\omega\pm(2-\delta)\omega_{0}\right)+\alpha(\omega\mp\omega_{0})\frac{t^{2n+1}}{2n+1}\right)\right]\, dt}\end{array}.\label{eq:F1F2_multiple}\end{equation}
These integrals can now be expressed by a generalized Airy function,
which we define as $\gai(x)\equiv(2\pi)^{-1}\int_{-\infty}^{\infty}\exp\left[i\left(xt+t^{2n+1}/(2n+1)\right)\right]dt$.
Note that for $n=1$ we retain the Airy function and for $n=2$ we
obtain a special case of the canonical swallowtail integral \cite{1984Swallowtail}.
Since the $\gai(x)$ are not available in general purpose numerical
libraries, their numerical evaluation is explained in Sec.~\ref{sec:num-integration-gen-airy}.

In analogy to the Airy function, the $\gai(x)$ fulfil ODEs: $d^{2n}\gai(x)/dx^{2n}\,+\,(-1)^{n}x\,\gai(x)=0$.
Exploiting this, $F_{1}$ and $F_{2}$ become\begin{eqnarray}
F_{1}(\omega) & = & \frac{2\pi\delta}{\sqrt[2n+1]{\alpha(\omega\mp\omega_{0})}}\,\gai(\xi)\\
F_{2}(\omega) & = & \frac{-2\pi\xi}{\omega\mp\omega_{0}}\,\gai(\xi),\end{eqnarray}
where $\xi=\left(\delta\omega\pm(2-\delta)\omega_{0}\right)/\left(\alpha\left(\omega\mp\omega_{0}\right)\right)^{1/(2n+1)}$.
After again taking the sum of $F_{1}$ and $F_{2}$, the $\omega^{-1/(2n+1)}$
terms cancel and what remains is:

\begin{equation}
F_{\pm}(\omega)=\frac{\pm4\pi}{\sqrt[2n+1]{\alpha}(\omega\mp\omega_{0})^{\nicefrac{2n+2}{2n+1}}}\,\gai(\xi).\end{equation}

In complete analogy the TROM model, represented by the integral~\eqref{eq:TROM_basic_int},
yields:

\begin{equation}
F_{\pm}(\omega)=\frac{4\pi}{\sqrt[2n+1]{\alpha(\omega\mp\omega_{0})}}\,\gai(\xi).\end{equation}

Finally we calculate the CSE integral \eqref{eq:CSE_basic_int} for
arbitrary orders of the $\gamma$-spike. Here, this means $\dot{y}(t)=\alpha_{0}t^{n}$
and, consequently, $x(t)=-vt+\alpha_{1}t^{2n+1}/(2n+1)$. Since this
works in complete analogy to the hitherto discussed cases, we just
present the result:\begin{equation}
F(\omega)=\frac{-2\pi\alpha_{0}i^{n}}{(\alpha_{1}\omega)^{\nicefrac{n+1}{2n+1}}}\,\frac{d^{n}\gai(\xi)}{d\xi^{n}},\end{equation}
wherein $\xi=\omega^{2n/2n+1}\delta/\alpha_{1}^{1/(2n+1)}$.

\section{\label{sec:Bourdier}Lorentz Transformation to Describe Oblique Incidence
in 1D}

\global\long\def\lab{\mathcal{L}}
\global\long\def\simu{\mathcal{S}}

In laser-plasma theory, it is often convenient to describe things
in a one dimensional (1D) slab geometry, i.e. all spatial derivatives
perpendicular to the optical axis are neglected. For not too small
laser focal spot sizes, this is very often a reasonable approximation
and leads to great simplifications in numerical as well as analytical
theory. In numerical computations, the grid size can be reduced by
orders of magnitude, allowing for higher resolution in the critical
dimension. In analytical calculations, it sometimes enables us to
give closed form solutions and straightforward, comprehensible models.
Whereas it is obvious that the 1D treatment can be employed in situations
of normal laser incidence, it can also be extended to oblique laser
incidence. Therefore, as shown by Bourdier in Ref.~\cite{Bourdier1983},
a Lorentz transformation does the job.

From the lab frame $\mathcal{L}$, we transform to the inertial frame
$\mathcal{S}$, in which the laser is normally incident. Let the light
wave in $\lab$ be described by the frequency 4-vector $\left(\omega_{0}^{\lab},\, ck_{x}^{\lab},\, ck_{y}^{\lab},\,0\right)=\omega_{0}^{\lab}\left(1,\,\cos\alpha,\,\sin\alpha,\,0\right)$,
wherein $x$ denotes the direction normal to the surface and $\alpha$
is the angle of incidence. In $\simu$, we claim $ck_{y}^{\simu}\overset{!}{=}0$.
Thus, the Lorentz transformation is given by the matrix\begin{equation}
\Lambda=\left(\begin{array}{cccc}
\gamma & 0 & -\beta\gamma & 0\\
0 & 1 & 0 & 0\\
-\beta\gamma & 0 & \gamma & 0\\
0 & 0 & 0 & 1\end{array}\right),\end{equation}
with $\beta=\sin\alpha$ and $\gamma=\left(\cos\alpha\right)^{-1}$.
Now, we can derive all interesting magnitudes. In the frame $\simu$,
the plasma is streaming with the velocity $v_{y}^{\simu}=-c\beta$,
the laser wavelength is altered by $\lambda^{\simu}=\gamma\lambda^{\lab}$
and therefore the corresponding critical density changes to $n_{cr}^{\simu}=\gamma^{-2}n_{cr}^{\lab}$.
The electron density itself changes to $n^{\simu}=\gamma n^{\lab}$,
thus the \emph{normalized} density \begin{equation}
\bar{n}^{\simu}\equiv\frac{n^{\simu}}{n_{cr}^{\simu}}=\gamma^{3}\bar{n}^{\lab}\end{equation}
scales even with $\gamma^{3}$.

\section{\label{sec:num-integration-gen-airy}Numerical Calculation of the
Generalized Airy Function}

This section explains the numerical computation of the integral \begin{equation}
\gai(x)\equiv\frac{1}{2\pi}\int_{-\infty}^{\infty}e^{i\left(xt+\frac{t^{2n+1}}{2n+1}\right)}dt.\label{eq:gen_airy}\end{equation}

The unmindful immediate application of a trapezoidal formula would
fail here due to the rapidly oscillating integrand, which further
does not vanish at infinity. The integral only converges because of
the steadily decreasing oscillation period for $t\rightarrow\pm\infty$.

However, a simple trick can be applied to calculate the integral numerically:
We shift the contour of integration along the imaginary axis by a
margin of $a>0$. Due to Cauchy's integral theorem, this will not
change the results, since for large $t$ the contours connecting the
real axis to the new integration path do not contribute. Next, the
symmetry of the integrand can be exploited, so that the contour of
integration can be halved. We obtain: \begin{eqnarray}
\gai(x) & = & \frac{1}{\pi}\int_{0}^{\infty}\exp\left[-\textrm{Im}\left(f(x,t)\right)\right]\times\cos\left[\textrm{Re}\left(f(x,t)\right)\right]\, dt\label{eq:Gai_numerically}\\
f(x,t) & = & (t+ia)x+\frac{(t+ia)^{2n+1}}{2n+1}.\end{eqnarray}

Note that, since $a>0$, $f(x,\, t)$ possesses the highly desirable
property $\lim_{t\rightarrow\infty}\left[\textrm{Im}\left(f(x,t)\right)\right]=+\infty$.
This means, that the integrand vanishes exponentially for large $t$,
and we can approximate \eqref{eq:Gai_numerically} with a trapezoidal
formula.

Since $a$ does not affect the value of the integral, it can be chosen
in a way so as to minimize the computational effort. For too small
$a$, the integrand vanishes only slowly for $t\rightarrow\infty$
so that the numerical upper boundary would have to be very high. For
too large $a$, the integrand oscillates more rapidly, so that the
time step would have to be very small. Our experience showed, that
the best choice for $a$ depends mainly on the order $n$ of the function.
For $n\leq2$ we found $a=1$ a good choice, whereas for $n>2$, smaller
$a$ work better.

\section*{Acknowledgements}

This work has been partially funded within the DFG SFB Transregio
TR 18 and Graduiertenkolleg GRK1203.

\bibliographystyle{unsrt}

\end{document}